\documentclass[prd,preprint,preprintnumbers,superscriptaddress,nofootinbib]{revtex4}
\usepackage[a4paper, hdivide={1.91cm,,1.165cm}, vdivide={1.83cm,,3.6cm}]{geometry}

\usepackage{amstext,amssymb}
\usepackage{amsmath}
\usepackage{graphicx}
\usepackage[hyperfootnotes=true]{hyperref}
\usepackage{xspace}
\usepackage{color}
\usepackage{slashed}
\usepackage{multirow}
\usepackage{tikz}
\usetikzlibrary{snakes}
\usepackage[normalem]{ulem}
\usepackage[utf8]{inputenc}
\DeclareUnicodeCharacter{2212}{\ensuremath{-}}

\newcommand{\be}{\begin{equation}}
\newcommand{\ee}{\end{equation}}
\newcommand{\bea}{\begin{eqnarray}}
\newcommand{\eea}{\end{eqnarray}}
\def\bl#1{\textcolor{blue}{#1}}

\begin{document}

\title{Exploring two component doublet dark matter}

\author{Mariana Frank}
\email{mariana.frank@concordia.ca}
\affiliation{Department of Physics, Concordia University,7141 Sherbrooke St. West, Montreal, Quebec City H4B 1R6, Canada}

\author{Purusottam Ghosh}
\email{pghoshiitg@gmail.com}
\affiliation{Institute of Mathematical Sciences, Taramani, Chennai 600 113, India} 
\affiliation{Homi Bhabha National Institute, Anushakti Nagar, Mumbai 400094, India}

\author{Chayan Majumdar}
\email{c.majumdar@ucl.ac.uk}
\affiliation{Department of Physics and Astronomy, University College London, London WC1E 6BT, United Kingdom}

\author{Supriya Senapati}
\email{ssenapati@umass.edu}
\affiliation{Amherst Center for Fundamental Interactions, Department of Physics, University of Massachusetts Amherst, MA 01003, USA}

\begin{abstract}
We propose a two-component dark matter (DM) scenario by extending the Standard Model with two additional $SU(2)_L$ doublets, one scalar, and another fermion. To ensure the stability of the DM components, we impose a global $\mathcal{Z}_2 \times \mathcal{Z}_2^\prime$ symmetry. The lightest neutral states for both the scalar and fermion, which are non-trivially transformed under the extended symmetry, behave as stable two-component DM candidates. While single components are under-abundant due to their gauge interactions, in a mass region between $m_W$ and $525$ GeV for the scalar and a mass below $1200$ GeV for the fermion, and the fermion DM conflicts with direct detection limits over the whole parameter space,  having two components helps to saturate relic density in the regions with under-abundance. Compliance with direct detection constraints leads to two options, either introducing dim-5 effective operators, or embedding the scenarios into a complete UV theory, which reproduces a type II seesaw model, thus naturally including neutrino masses. We analyze the consequences of this scenario at the LHC.
\end{abstract}

\pacs{}
\maketitle

\section{Introduction}
\label{sec:intro}
The existence of dark matter (DM), a non-luminous and non-baryonic form of matter in the universe, is supported by astrophysical observations such as galaxy clusters \cite{2009GReGr..41..207Z}, galaxy rotation curves \cite{1970ApJ...159..379R}, and galaxy survey experiments \cite{Clowe:2006eq} that map its distribution based on gravitational lensing effects. Cosmological evidence suggests that around 26\% of the energy density of the present universe is in the form of DM, with a present DM abundance conventionally reported as $\Omega_{\text{DM}} h^2 = 0.120 \pm 0.001$ at 68\% C.L. \cite{Planck:2018vyg}. However, despite significant astrophysical and cosmological evidence, at present no experiments have detected DM particles. Direct detection (DD) experiments such as LUX-LZ 2024 \cite{LZ:2024zvo}, PandaX-II \cite{PandaX-II:2017hlx}, Xenon1T \cite{XENON:2018voc} and indirect detection (ID) such as FERMI LAT \cite{Fermi-LAT:2015att}, MAGIC \cite{MAGIC:2016xys} have produced no results. Theoretical problems also exist, as the Standard Model (SM) of particle physics cannot explain DM, and thus beyond the SM (BSM) proposals have been devised to explicitly include it \cite{Taoso:2007qk}. Of the numerous proposals, the weakly interacting massive particle (WIMP) paradigm is the most studied. WIMP interactions can obtain the correct relic density of DM but are increasingly in conflict with direct detection data. These particles can also give rise to DM particle production at the LHC \cite{Kahlhoefer:2017dnp}, but no results have been found. Indirect detection searches are ongoing to find an excess of antimatter, gamma rays, or neutrinos from DM annihilation or decay, but they also have observed no convincing DM signals yet. There are tight constraints on DM annihilation into SM particles \cite{MAGIC:2016xys}, especially charged ones, which can lead to an excess of gamma rays for WIMP-type DM.

The absence of detection of DM particles in previous experiments has not eliminated all possibilities for single particle DM models. However, it increases the possibility of the existence of a more complex DM sector, similar to the visible sector that consists of multiple types of particles. The multicomponent WIMP DM models proposed in recent years (as in \cite{Cao:2007fy, Chialva:2012rq, Heeck:2012bz, Bhattacharya:2013hva, Bian:2013wna, Esch:2014jpa, Karam:2016rsz, Bhattacharya:2016ysw, DiFranzo:2016uzc, Bhattacharya:2017fid, Ahmed:2017dbb, Bhattacharya:2018cqu, Aoki:2018gjf, Barman:2018esi, YaserAyazi:2018lrv, Poulin:2018kap, Chakraborti:2018aae, Bhattacharya:2018cgx, Bernal:2018aon, Elahi:2019jeo, Borah:2019epq, Bhattacharyya:2022trp, Belanger:2021lwd, Chakrabarty:2021kmr, DuttaBanik:2020jrj, Biswas:2019ygr, Kuncinas:2024zjq, Boto:2024tzp, Qi:2024uiz, YaserAyazi:2024dxk, Coleppa:2023vfh, Belanger:2022esk, BasiBeneito:2022qxd, Chakrabarty:2024pvf, Borah:2024emz, Qi:2025jpm, Hernandez-Sanchez:2022dnn, Basu:2023wgo, Das:2022oyx, DiazSaez:2021pmg, DiazSaez:2023wli, Taramati:2024kkn}) can have distinct features in direct and indirect detections (explored in \cite{Profumo:2009tb, Aoki:2013gzs,  Geng:2013nda, Gu:2013iy, Herrero-Garcia:2017vrl, Bhattacharya:2025eym}). These constraints can be mitigated in multicomponent DM frameworks if the relative densities of different DM components are appropriate \cite{Cao:2007fy,Bhattacharya:2016ysw,Bhattacharya:2019fgs}. The approach to identifying two-component DM at colliders has been studied in \cite{Bhattacharya:2022wtr, Bhattacharya:2022qck}, showing that it can produce two peaks in the missing energy distribution.

The connection between the origin of neutrino mass and DM is of interest, particularly in scotogenic scenarios \cite{Ma:2008ym}, where the $\mathcal{Z}_2$ odd particles are involved in generating light neutrino masses and the lightest $\mathcal{Z}_2$ odd particle is the DM candidate. This common origin allows for constraints from both sectors, enhancing the model's predictability. While there have been several studies on single-component DM scenarios and their connection to neutrino masses, fewer studies on multi-component DM's role in the origin of neutrino mass exist \cite{Bhattacharya:2024ohh, Bhattacharya:2019fgs}. In this work, we will consider a minimal scenario accommodating two-component DM with the correct relic abundance, satisfying constraints from neutrino oscillation data \cite{Esteban:2024eli}, constraints from DD and IDs. We show that consistency with DM experimental constraints throughout the mass spectrum can be connected to neutrino masses, connecting two open questions in the SM.

This work thus aims to address two deficiencies of the SM. We start by studying the single component DM (fermion and scalar doublets). Both scenarios have shortcomings when compared to the experimental constraints. We try to rectify these by either introducing higher order (dim-5) effective operators, or by constructing an UV-complete theory  incorporating a triplet scalar (as in type II seesaw \cite{Mohapatra:1980yp, Wetterich:1981bx, PhysRevD.25.774, Brahmachari:1997cq} models), plus an additional Higgs doublet  and a doublet fermion serving as DM into the particle spectrum of the SM. We examine an extension of the SM that involves a $\mathcal{Z}_2 \otimes \mathcal{Z}_2^\prime$ symmetry where the  triplet scalar field and SM particles are even, while the extra doublet scalar is odd under $\mathcal{Z}_2$ and the doublet fermion is odd under $\mathcal{Z}_2^\prime$. This setup can fulfill two objectives: the neutral components of the extra  scalar and fermion doublets, being odd under the $\mathcal{Z}_2$ and $\mathcal{Z}_2^\prime$ symmetry, respectively, can serve as a potential DM candidates, and the small vacuum expectation value (\textit{vev}) of the triplet scalar field (introduced to ensure agreement with direct detection data), which satisfies the electroweak precision test, can be utilized to explain tiny neutrino masses through the type-II seesaw mechanism without requiring heavy right-handed neutrinos.

The organization of this paper is as follows: In section \ref{sec:singleDMmodel}, we will introduce the single component DM i.e., the doublet fermion in \ref{subsec:fermion} and the scalar doublet in \ref{subsec:scalar} model. We then investigate the two-component DM scenario in \ref{sec:twocomp}. In \ref{subsec:2compDIM5} we introduce the two component DM, consisting of one fermion and one scalar doublet, with the addition of dim-5 operators that insure compliance with direct detection data. In \ref{subsec:UVcomplete} we show that, if an additional triplet scalar is added to the theory, the two-component model is consistent with DM constraints while also incorporating the seesaw mechanism for generating neutrino masses. We analyze some consequences and signatures of the model in \ref{sec:colliders} and conclude in \ref{sec:conclusion}. 

\section{Single Component Dark Matter}
\label{sec:singleDMmodel}

The electroweak multiplets are widely studied as minimal DM candidates and are well-motivated due to their minimal parameters and predictive detectability. Before investigating a two-component DM framework with a scalar and a fermion $SU(2)_L$ doublet in the sub-TeV mass range, we begin with a brief review of the single-component scenarios, which have been extensively studied in the literature \cite{Hambye:2009pw,LopezHonorez:2006gr,Bhattacharya:2018fus,Dey:2022whc}, highlighting their shortcomings, to motivate our further exploration of multiple component DM. We present the fermion doublet in \ref{subsec:fermion} and the scalar doublet in \ref{subsec:scalar}.
\subsection{Single Component Fermion Doublet DM}
\label{subsec:fermion}

The SM is extended with a $SU(2)_L$ vector-like fermion doublet (VLFD), $\Psi=\left( \begin{matrix} \psi^0 ~~ \psi^- \end{matrix} \right)^T$, where the neutral component acts as the DM candidate \cite{Bhattacharya:2018fus}. The SM gauge group is augmented by a discrete symmetry $\mathcal{Z}_2$, under which the $\Psi$ transforms as $\Psi \to -\Psi$ (as shown in Table \ref{tab:VLFD}) while the SM fields remain unchanged. 

\begin{table}[htb!]
\begin{tabular}{|c|c|c|c|}
\hline \multicolumn{2}{|c}{Field}&  \multicolumn{1}{|c|}{ $\underbrace{ SU(3)_C \otimes SU(2)_L \otimes U(1)_Y}$ $\otimes \underbrace{ \mathcal{Z}_2 }$} \\ \hline 
{VLFD} & $\Psi = \left(\begin{matrix}
 \psi^0 \\  \psi^- 
\end{matrix}\right)$ & ~~1 ~~~~~~~~~~~2~~~~~~~~~~-1~~~~~~~~~~~-  \\
\hline
\end{tabular}
\caption{Charge assignment of vector-like fermion doublet under the gauge group $\mathcal{G} \equiv \mathcal{G}_{\rm SM} \otimes \mathcal{G}_{\rm DM}$  where $\mathcal{G}_{\rm SM}\equiv SU(3)_C \otimes SU(2)_L \otimes U(1)_Y$ and $\mathcal{G}_{\rm DM} \equiv \mathcal{Z}_2$. All the SM particles are even under the $\mathcal{Z}_2$ symmetry.}
    \label{tab:VLFD}
\end{table}

The interaction Lagrangian involving the field $\Psi$, relevant for the DM analysis is given by,
\begin{align}
    \mathcal{L}^{\rm VLFD} & = \overline{\Psi} \left[i\gamma^{\mu}(\partial_{\mu} - i g_2 \frac{\sigma^a}{2}W_{\mu}^a - i g_1\frac{Y_\Psi}{2}B_{\mu})- m_\Psi \right] \Psi  ~~~~~~~~~ {\rm with} ~~ Y_\Psi=-1~.
\end{align}
The notation used here follows standard conventions, with $m_\Psi$ representing the bare mass of both the neutral ($\psi^0$) and charged ($\psi^\pm$) components. However, the quantum mass correction at one loop breaks the degeneracy between these two states, which can be expressed as \cite{Thomas:1998wy}:
\begin{eqnarray}
    m_{\psi^0} = m_\Psi, ~~~ m_{\psi^\pm}  = m_\Psi + \delta m ~~~~{\rm with}~~\delta m = \frac{\alpha_{\rm EM}}{2} M_Z~ f\Big(\frac{m^2_{\Psi}}{M^2_Z} \Big),  \nonumber \\ 
   {\rm where}~f(r)=\frac{\sqrt{r}}{\pi}\int^1_0 dx (2-x)~ \ln\Bigg[ 1+ \frac{x}{r {(1-x)^2}}\Bigg].  \nonumber
\end{eqnarray}
The value of $\delta m$ found to be in the range of $\sim \{250-340\}$ MeV for $m_\Psi=\{100-1000\}$ GeV. Therefore, $\psi^0$ is the lightest neutral state and acts as the DM candidate, and the $\mathcal{Z}_2$ symmetry ensures the stability of the DM. This framework has only one free parameter,  $m_{\psi^0}$, and thus it is  very predictive.

The fermion doublet DM has strong gauge interactions with the thermal bath particles (SM particles) at the early time of the universe. As a consequence, the abundance of $\psi^0$ follows the standard freeze-out mechanism, as in the already mentioned WIMP case, where,  when the interaction rate falls below the universe's expansion rate, the DM freezes-out of equilibrium and attains its observed relic density. The abundance of DM is governed by its annihilation ($\psi^0 \psi^0 \to {\rm SM~SM}$) and co-annihilation ($\psi^+ \psi^- , \psi^0 \psi^\pm \to {\rm SM~SM}$) processes to the SM particles, as shown in the Feynman diagrams in Fig. \ref{Feyn-ann-FD} and Fig. \ref{Feyn-coann-FD}, respectively.

\begin{figure}[htb!]
   \begin{center}
\begin{tikzpicture}[line width=0.4 pt, scale=1.0]
\draw[solid] (-5.0,1)--(-4.0,0);
\draw[solid] (-5.0,-1)--(-4.0,0);
\draw[snake] (-4.0,0)--(-2.4,0);
\draw[solid] (-2.4,0)--(-1.4,1);
\draw[solid] (-2.4,0)--(-1.4,-1);
\node  at (-5.3,1.2) {$\psi^0$};
\node at (-5.3,-1.2) {$\overline{\psi^0}$};
\node [above] at (-3.2,0.05) {$Z$};
\node at (-0.5,1.2){f/$W^+$/h};
\node at (-0.5,-1.2) {$ \overline{f}$/$W^-$/Z};
\draw[solid] (1.5,1.0)--(2.5,0.5);
\draw[solid] (1.5,-1.0)--(2.5,-0.5);
\draw[solid] (2.5,0.5)--(2.5,-0.5);
\draw[snake] (2.5,0.5)--(3.5,1.0);
\draw[snake] (2.5,-0.5)--(3.5,-1.0);
\node at (1.3,1.2) {$\psi^0$};
\node at (1.2,-1.2) {$\overline{\psi^0}$};
\node at (2.9,0.04) {$\psi^+$};
\node at (3.9,1.2) {$W^+$};
\node at (3.9,-1.2) {$W^-$};
\draw[solid] (5.5,1.0)--(6.5,0.5);
\draw[solid] (5.5,-1.0)--(6.5,-0.5);
\draw[solid] (6.5,0.5)--(6.5,-0.5);
\draw[snake] (6.5,0.5)--(7.5,1.0);
\draw[snake] (6.5,-0.5)--(7.5,-1.0);
\node at (5.3,1.2) {$\psi^0$};
\node at (5.2,-1.2) {$\overline{\psi^0}$};
\node at (6.9,0.06) {$\overline{\psi^0}$};
\node at (7.8,1.2) {$Z$};
\node at (7.8,-1.2) {$Z$};
\end{tikzpicture}
 \end{center}
\caption{Feynman diagrams for DM annihilation into SM particles for the VLFD DM ($\psi^0$).}
\label{Feyn-ann-FD}
\end{figure}
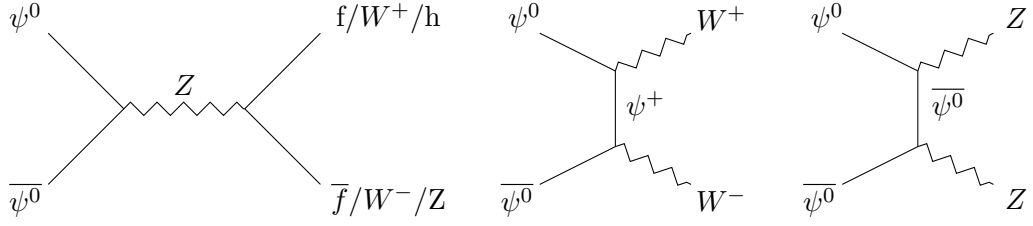
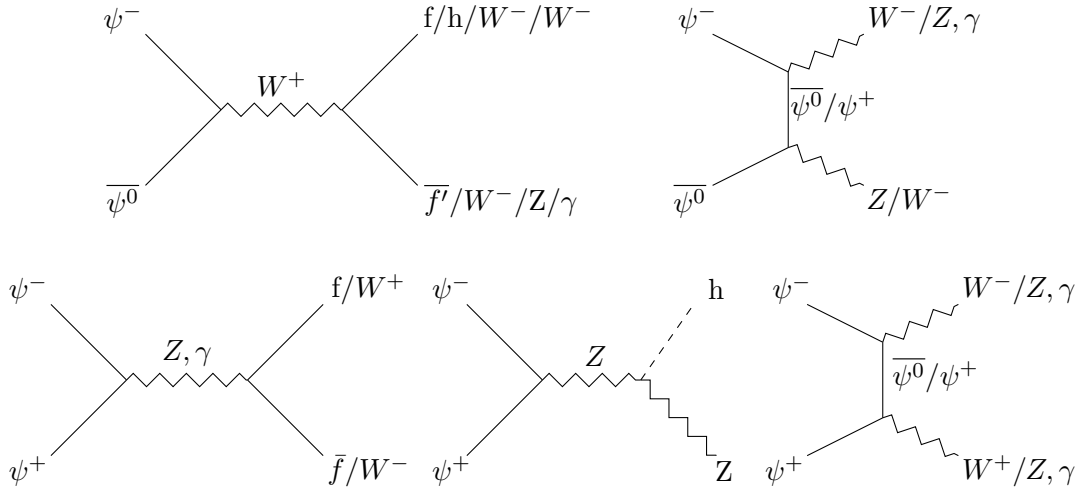
\begin{figure}[htb!]
 \begin{center}
\begin{tikzpicture}[line width=0.4 pt, scale=1.0]
\draw[solid] (-5.0,1)--(-4.0,0);
\draw[solid] (-5.0,-1)--(-4.0,0);
\draw[snake] (-4.0,0)--(-2.4,0);
\draw[solid] (-2.4,0)--(-1.4,1);
\draw[solid] (-2.4,0)--(-1.4,-1);
\node  at (-5.3,1.2) {$\psi^-$};
\node at (-5.3,-1.2) {$\overline{\psi^0}$};
\node [above] at (-3.2,0.05) {$W^+$};
\node at (-0.2,1.2){f/h/$W^-$/$W^-$ };
\node at (-0.3,-1.2) {$\overline{f^\prime}$/$W^-$/Z/$\gamma$};
\draw[solid] (2.5,1.0)--(3.5,0.5);
\draw[solid] (2.5,-1.0)--(3.5,-0.5);
\draw[solid] (3.5,0.5)--(3.5,-0.5);
\draw[snake] (3.5,0.5)--(4.5,1.0);
\draw[snake] (3.5,-0.5)--(4.5,-1.0);
\node at (2.3,1.2) {$\psi^-$};
\node at (2.2,-1.2) {$\overline{\psi^0}$};
\node at (4.1,0.07) {$\overline{\psi^0}/\psi^+$};
\node at (5.3,1.2) {$W^-/Z,\gamma$};
\node at (5.1,-1.2) {$Z/W^-$};
\end{tikzpicture}
  \end{center}    
 \begin{center}
\begin{tikzpicture}[line width=0.4 pt, scale=1.0]
\draw[solid] (-5.0,1)--(-4.0,0);
\draw[solid] (-5.0,-1)--(-4.0,0);
\draw[snake] (-4.0,0)--(-2.4,0);
\draw[solid] (-2.4,0)--(-1.4,1);
\draw[solid] (-2.4,0)--(-1.4,-1);
\node  at (-5.3,1.2) {$\psi^-$};
\node at (-5.3,-1.2) {$\psi^+$};
\node [above] at (-3.2,0.05) {$Z,\gamma$};
\node at (-0.8,1.2){f/$W^+$ };
\node at (-0.8,-1.2) {$\bar{f}$/$W^-$};
\draw[solid] (0.5,1)--(1.5,0);
\draw[solid] (0.5,-1)--(1.5,0);
\draw[snake] (1.5,0)--(2.8,0);
\draw[dashed] (2.8,0)--(3.5,1);
\draw[snake] (2.8,0)--(3.8,-1);
\node  at (0.3,1.2) {$\psi^-$};
\node at (0.3,-1.2) {$\psi^+$};
\node [above] at (2.2,0.05) {$Z$};
\node at (3.8,1.2){h};
\node at (3.9,-1.2) {Z};
\draw[solid] (5,1.0)--(6,0.5);
\draw[snake] (6,0.5)--(7,1.0);
\draw[solid] (6,0.5)--(6,-0.5);
\draw[solid] (6,-0.5)--(5,-1.0);
\draw[snake] (6,-0.5)--(7,-1.0);
\node at (4.8,1.2) {$\psi^-$};
\node at (4.7,-1.2) {$\psi^+$};
\node at (6.7,0.07) {$\overline{\psi^0}/\psi^+$};
\node at (7.8,1.2) {$W^-/Z,\gamma$};
\node at (7.8,-1.2) {$W^+/Z,\gamma$};
\end{tikzpicture}
  \end{center}
%
\caption{Top panel: Feynman diagrams for DM co-annihilation to SM particles for the VLFD DM ($\psi^0$) associated with the heavier states, $\psi^\pm$. Bottom panel: Annihilation of the charged pair of VLFD into SM particles, contributing to co-annihilation channels.}
\label{Feyn-coann-FD}
 \end{figure}
\noindent The evolution of total number density, $n_{\Psi}=n_{\psi^0}+n_{\psi^\pm}$, as a function of time is described by following Boltzmann equation (BEQ) \cite{Kolb:1990vq,Griest:1990kh,Edsjo:1997bg}:
\begin{equation}
    \frac{d n_{\Psi}}{dt} + 3 \mathcal{H} n_{\Psi} = - \langle \sigma v \rangle_{\Psi}^{\rm eff} \left( {n_{\Psi}}^2 - {n_{\Psi}^{\text{eq}}}^2 \right),
    \label{eq:BEQ_VLLD}
\end{equation}
where $n_{\Psi}^{\rm eq}=n_{\psi^0}^{\rm eq}+n_{\psi^\pm}^{\rm eq}$. The equilibrium density defined as: $n_i^{\rm eq}=\frac{g_i}{2\pi^2} m_i^2~ T~ K_2[\frac{m_i}{T}]$ with $g_i$ denotes the internal degree of freedom of the $i^{th}$ species, and $K_2$ is the second-order modified Bessel function. The effective thermal average cross-section, $\left< \sigma v\right>_{\Psi}^{\rm{eff}}$ connected with the number changing processes: annihilation ($\psi^0 \psi^0 \to {\rm SM~SM}$) and co-annihilation ($\psi^+ \psi^- , \psi^0 \psi^\pm \to {\rm SM~SM}$), can be expressed as follows \cite{Griest:1990kh, Edsjo:1997bg}: 
\begin{align}
    \langle \sigma v \rangle_{\Psi}^{\rm{eff}} & = \frac{g_{\psi^0}^2}{g_{\rm eff}^2} \left< \sigma v \right>_{\psi^0\psi^0} +\frac{g_{\psi^\pm}^2}{g_{\rm eff}^2} \left< \sigma v \right>_{\psi^+ \psi^-}(1+\delta_{\psi^\pm})^3e^{-2\zeta\delta_{\psi^\pm}}  \nonumber \\
    & ~~~~~~~~+ \frac{2 g_{\psi^0} g_{\psi^\pm}}{g_{\rm eff}^2} \left< \sigma v \right>_{\psi^0\psi^\pm}(1+\delta_{\psi^\pm})^{\frac{3}{2}}~ e^{-\zeta~\delta_{\psi^\pm}} ,
\end{align}
${\rm where}~ g_{\rm eff}=g_{\psi^0}+g_{\psi^\pm}(1+\delta_{\psi^\pm})^{\bl{\frac{3}{2}}}~ e^{-\zeta\delta_{\psi^\pm}},~\zeta= \frac{m_{\psi^0}}{T}~{\rm and}~ \delta_{\psi^\pm}=\frac{m_{\psi^\pm}-m_{\psi^0}}{m_{\psi^0}}$ with $g_{\psi^0}$ and $g_{\psi^\pm}$ are the internal degrees of freedom associated with the dark lepton states $\psi^0$ and $\psi^\pm$ respectively. 

Before freeze-out, both $\psi^0$ and $\psi^\pm$ were in thermal equilibrium via scattering processes such as $\psi^\pm (\psi^0) +~\rm{SM}~\leftrightarrow \psi^0 +~\rm{SM}~$ and $\psi^\pm \leftrightarrow \psi^0 +~\rm{SM}$. As the temperature drops to $T\sim m_{\psi^0}$, these particles begin to freeze-out. The freeze-out temperature of DM ($T=T_{\rm FO}$) can be obtained by solving the above BEQ, which also allows us to compute the relic abundance in terms of $\Omega h^2$ (with $\Omega = m_{\rm DM} ~n_{\rm DM}/\rho_c$ \cite{Kolb:1990vq}). Since the mass splitting $\delta m = m_{\psi^\pm}-m_{\psi^0}$ ($\sim \{250-340\}$ MeV) is much smaller than $T_{\rm FO}$, both states remain nearly equally populated at freeze-out ($n_{\psi^0} \sim n_{\psi^\pm}$). After decoupling from the thermal bath, the combined comoving number density $n_{\Psi} = n_{\psi^0} + n_{\psi^\pm}$ remains constant. And over time, the heavier charged component $\psi^\pm$ becomes Boltzmann suppressed and decays into DM $\psi^0$, $\psi^\pm \to \psi^0 +~\mathrm{SM}$, resulting in only $\psi^0$ surviving as DM today ($n_{\Psi}= n_{\psi^0} + n_{\psi^\pm} (\to n_{\psi^0}) )$. The decay of $\psi^\pm$ must occur before Big Bang Nucleosynthesis ($\tau_{\psi^\pm \to \psi^0 ~\pi^\pm} < \tau_{\rm BBN}$) to remain consistent with cosmological constraints. We used {\tt MicrOmegas} \cite{Alguero:2023zol}, which takes into account the dynamics described above, to compute the relic density ($\Omega_{\psi^0} h^2$).
 The model files relevant for {\tt MicrOmegas} are generated using {\tt LanHEP} \cite{Semenov:2014rea}. The abundance of DM after freeze-out approximately follows $\Omega_{\psi^0} h^2 \propto 1/\langle \sigma v \rangle_{\Psi}^{\rm eff}$ \cite{Kolb:1990vq}, reflecting its dependence on the interaction rates.
\begin{figure}[htb!]
    \centering
    \includegraphics[width=0.5\linewidth]{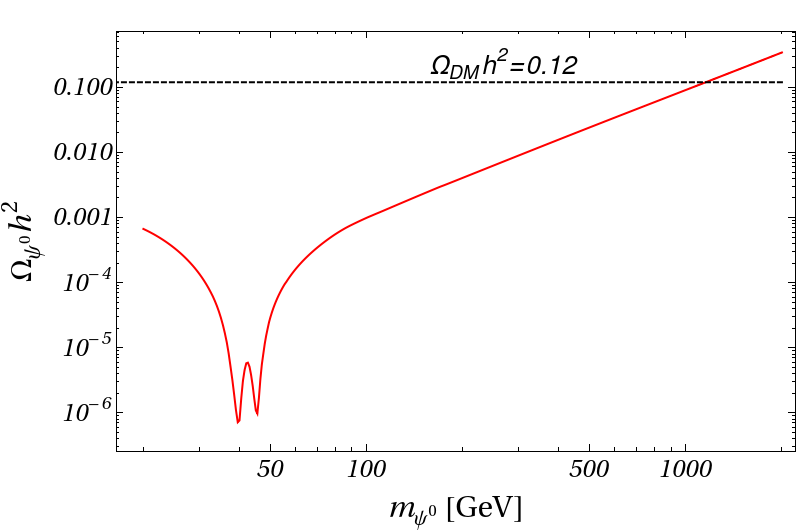}
    \caption{Variation of relic density as a function of the mass of the VLFD DM. The black dashed line corresponds to the observed DM abundance as measured by PLANCK \cite{Planck:2018vyg}. }
    \label{fig:VLL_relic}
\end{figure}

In Fig. \ref{fig:VLL_relic}, we show the variation of the relic density as a function of the only free parameter, the DM mass $m_{\psi^0}$ as shown by the solid red line. The observed relic density is achieved for a DM mass of $m_{\psi^0} \sim 1.2$ TeV. The gauge boson mediated interactions and a small mass splitting ($\delta m \sim \{250-340\}$ MeV) enhance the annihilation and co-annihilation processes, resulting in a large $\left< \sigma v\right>_{\Psi}^{\rm{eff}}$, which, in turn, leads to DM under-abundance for masses below $\sim 1.2$ TeV. With increasing DM mass, $\left< \sigma v\right>_{\Psi}^{\rm{eff}}$ decreases due to its mass suppression ($\propto 1/m_{\psi^0}^2$), resulting in an increase in relic density, as depicted in the Fig. \ref{fig:VLL_relic}. The two sharp drops around $m_{\psi^0} \sim M_W/2$ and $M_Z/2$ are due to W and Z resonances, respectively.

\begin{figure}[htb!]
    \centering
\includegraphics[width=0.47\linewidth]{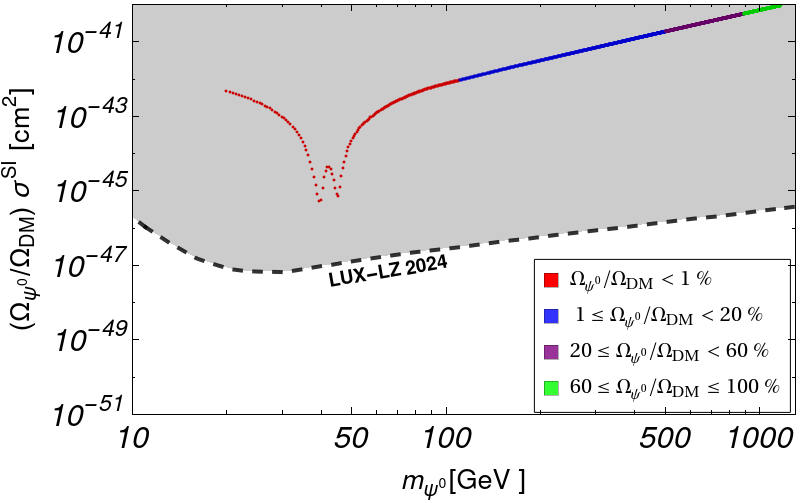} \qquad
\includegraphics[width=0.47\linewidth]{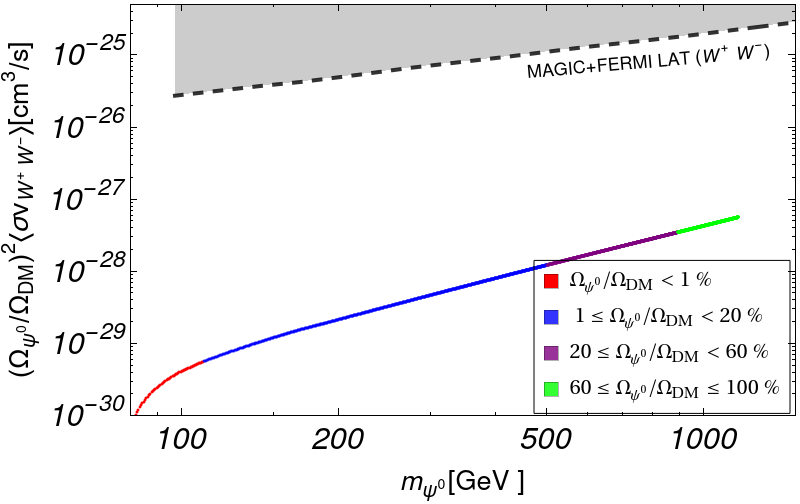}
\caption{Left panel: The effective spin-independent direct detection $\psi^0-N$ scattering cross section, $ \sigma_{\rm{eff}}^{\rm{SI}} (\psi^0) = f_{\psi^0} \sigma^{\rm SI}$ ($f_{\psi^0}={\Omega_{\psi^0}}/{\Omega_{\rm DM}} \leq 1$) is plotted against the DM mass $m_{\psi^0}$. Different coloured points represent the respective percentage of relic contributions compared to the total observed abundance, as a function of DM masses. The exclusion region from LUX-LZ 2024 DD experiment (grey shaded region) is shown for comparison purposes. Right panel: Plot showing the effective indirect cross section $f_{\psi^0}^2 \langle \sigma v\rangle_{W^+W^-}$ vs. the mass of the VLFD DM. }
    \label{fig:VLL_dd}
\end{figure}

The VLFD DM ($\psi^0$) faces constraints from the non-observation of direct searches via the Z-mediated DM-nucleon spin-independent (SI) scattering process, where SI $\psi^0 - N$ elastic scattering happens via the $t$-channel Z-boson mediation. We obtain the SI DM-nucleon scattering cross-section, $\sigma^{\rm SI}$ by using {\tt MicrOmegas} \cite{Alguero:2023zol} along with the DM relic density. In the left panel of Fig. \ref{fig:VLL_dd}, we show the DM-nucleon scattering cross-section scaled with the fractional DM density, defined as $\sigma_{\rm{eff}}^{\rm{SI}} = f_{\psi^0} \times \sigma^{\rm SI}$ with $f_{\psi^0}={\Omega_{\psi^0}}/{\Omega_{\rm DM}}$, as a function of DM mass $m_{\psi^0}$. This plot includes both observed relic points and under-abundant relic points ($ f_{\psi^0} \leq 1$). The different colored points shown in the figure represent the percentage of relic contributions $f_{\psi^0}$, which is also a function of $m_{\psi^0}$. 
We compare the parameter space of the VLFD DM in the same plane with the experimental upper bounds from the most recent LUX-LZ 2024 \cite{LZ:2024zvo} data, shown by the black dashed line. The grey shaded region, which is just above the dashed line, is excluded from the non observation of any DM signal from the corresponding DD experiment LUX-LZ 2024. The figure indicates that the entire  mass region for $m_{\psi^0}$ is excluded from the direct detection constraints. 

For completeness, we plot the constraints from indirect detection experiments. Indirect detection constraints on thermal DM arise from searches for excess gamma-ray flux, which is produced via DM annihilation into SM-charged particle pairs ($X^+X^-$ where $X=\{\mu, \tau, b, W\}$), followed by their subsequent decays. Non-observation of gamma-ray flux in indirect search experiments such as Fermi-LAT \cite{Fermi-LAT:2015att} and MAGIC \cite{MAGIC:2016xys} puts an upper bound on the model parameters in terms of $f_{\psi^0}^2 \times \langle \sigma v \rangle_{X^+ X^-}$ with $f_{\psi^0} \leq 1$, where $\langle \sigma v \rangle_{X^+ X^-}$ represents the thermally averaged annihilation cross-section of $\psi^0 \overline{\psi^0} \to X^+ X^-$ channel(s). To compare with the experimental upper bound, we show the variation of the effective thermal averaged cross-section for a given DM annihilation process, $\psi^0 \overline{\psi^0} \to W^+W^-$ as a function of $m_{\psi^0}$ for all points with $f_{\psi^0}(m_{\psi^0})\leq 1$ in the right panel of Fig. \ref{fig:VLL_dd}. We show the $W^+W^-$ decays as the other modes are less restrictive. In the same plane, we present the combined Fermi-LAT and MAGIC exclusion region for the $W^+ W^-$ channel, shown in the grey region. It shows the theoretical prediction is well below the experimental upper bound and does not face any severe constraints from indirect search experiments.

In summary, while the minimal VLFD DM achieves the observed DM density around a DM mass of $1.2$ TeV, and below this mass, DM is under-abundant, the entire mass region $m_{\psi^0} \lesssim 1.2$ TeV, and is safe from the observational upper limit from the indirect detection searches, is excluded by direct detection constraints. The large direct detection cross section is directly linked to interactions mediated by the Z boson. As the VLFD satisfies all other constraints, it is natural to explore if we can suppress or avoid all-together the Z-mediated interactions.

\subsubsection*{How to evade Z-mediated DD constraints}
\label{subsubsec:DDZconstraints}
The elastic direct detection cross-section bound for DM ($\psi^0$) can be decreased if the DM turns out to be a pseudo-Dirac state, in which case the $Z$ mediated neutral current vanishes \cite{Ghosh:2023dhj}. Two possible solutions are : $(i)$ introduce dim-5 effective "Weinberg-like" interaction: ${{\frac{1}{\Lambda} \overline{\Psi^c} \tilde{\Phi}~\tilde{\Phi}^\dagger \Psi}}$ with $\Lambda$ being the new physics scale and $\Phi$ is the SM Higgs doublet and $(ii)$ introduce Yukawa interactions with a $SU(2)_L$ triplet scalar $\Delta$ with $Y=2$: $ y_{\psi} \overline{\Psi^c}i \sigma^2\Delta \Psi$. To work with the minimal parameter space, here we focus on the first possibility, while in \ref{subsec:UVcomplete} of this paper, we will discuss the second possibility as a UV-complete framework of the first one. The new Lagrangian for the fermion doublet DM involving the dim-5 operator is given by,
\begin{equation}
    \mathcal{L}^{\rm pD-VLFD} = \mathcal{L}^{\rm VLFD}  -  {{\frac{1}{\Lambda} \overline{\Psi^c} \tilde{\Phi}~\tilde{\Phi}^\dagger \Psi}} + {h.c.}
\end{equation}
Note that the symmetry of the current framework also allows the dim-5 Weinberg operator $ {\cal O}_W \simeq \frac{C_{W}}{\Lambda_W} \overline{L^c} \tilde{\Phi} ~\tilde{\Phi}^\dagger L$, which is also responsible for neutrino mass generation. The effective scales $\Lambda $ and $\Lambda_W $ with $\Lambda < \Lambda_W$ are associated with different dynamics in the respective complete UV models.

\noindent The mass matrix for the neutral fermions in the basis $(\psi^0 ~\psi^{0^c})^T$ can be written as
\begin{eqnarray}
\mathcal{L}^{\rm pD-VLFD}_{\rm mass}&=&  
\frac{1}{2} \overline{\left( \begin{matrix} 
                  \psi^0 && {\psi^0}^c   \end{matrix} \right)}
\left( \begin{matrix} 
                  m_\Psi  && \frac{ v^2}{\Lambda}    \\
                    \frac{ v^2}{\Lambda} && m_\Psi   
                        \end{matrix} \right)
\left( \begin{matrix} 
                  \psi^0 \\ {\psi^0}^c   \end{matrix} \right) \nonumber \\
&=&\frac{1}{2} \overline{\left( \begin{matrix} 
                  \psi_1 && \psi_2   \end{matrix} \right)}
\left( \begin{matrix} 
                  m_{\psi_1} && 0   \\
                   0 && m_{\psi_2}  
                        \end{matrix} \right)
\left( \begin{matrix} 
                  \psi_1 \\ \psi_2   \end{matrix} \right),
\label{eq:pstate0}
\end{eqnarray}
with $\langle \Phi \rangle = \frac{v}{\sqrt{2}}$. Here, $\psi_1$ and $\psi_2$ are two pseudo-Dirac states with mass $m_{\psi_1}$ and $m_{\psi_2}$ respectively and can be expressed as
\bea
\psi_1 & = &\frac{i}{\sqrt2}({\psi^0}-{\psi^0}^c)~{\rm with}~m_{\psi_1}= \left( m_\Psi -  \delta m_{12} \right)\nonumber \\
\psi_2&=&\frac{1}{\sqrt2}({\psi^0}+{\psi^0}^c)~{\rm with}~m_{\psi_2}= \left( m_\Psi + \delta m_{12} \right) \nonumber \\
\psi^\pm && ~~~~~~~~~~~~~~~~~~~{\rm with}~m_{\psi^\pm}=m_\Psi+\delta m
\label{eq:pstate}
\eea

\begin{figure}[htb!]
\includegraphics[scale=0.28]{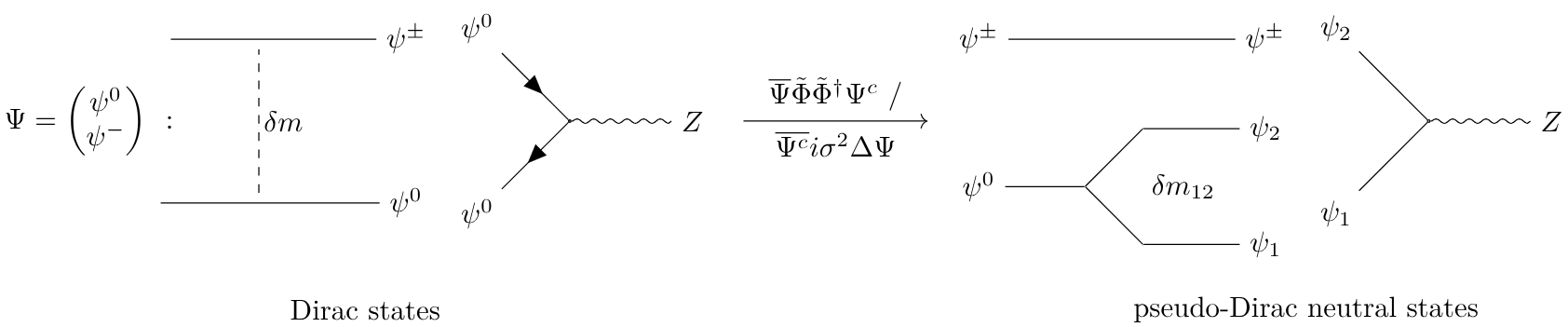}
\caption{Pictorial representations of different mass eigenstates (Dirac states at the left, pseudo-Dirac states at the right), along with the corresponding mass splittings.}
\label{fig:mass_spltng}
\end{figure}

\noindent The mass splitting between $\psi_1$ and $\psi_2$ is defined as $\delta m_{12} \equiv (m_{\psi_2}-m_{\psi_1}) = \frac{2 v^2}{\Lambda}$. The different mass states of the $\Psi$ multiplet along with their corresponding mass eigenvalues are shown pictorially in Fig. \ref{fig:mass_spltng} where the two different mass splittings $\delta m$ and $\delta m_{12}$ were induced through the one-loop quantum correction as well as the introduction of the dim-5 operator, respectively. Here the lightest neutral $\mathcal{Z}_2$-odd particle $\psi_1$ corresponds to the DM. The neutral and charged current interactions involving $\psi^0$ can be modified in terms of the pseudo-Dirac states ($\psi_{1,2}$) as: 
\begin{eqnarray}
    \overline{\psi^0} \gamma^{\mu} Z_{\mu} \psi^0 &\rightarrow& \overline{\psi_2} \gamma^{\mu} Z_{\mu} \psi_1  ~,\nonumber \\
    \overline{\psi^0} \gamma^{\mu} W^+_{\mu} \psi^- &\rightarrow& \frac{1}{\sqrt{2}}\overline{(\psi_1+i \psi_2)} \gamma^{\mu} W_{\mu}^+ \psi^- ~. \nonumber 
\end{eqnarray}
\noindent Note that the terms that involve diagonal neutral current interactions, $\overline{\psi_i} \gamma^{\mu} Z_{\mu} \psi_i$, are absent due to the Majorana nature of the state $\psi_i$ for which $\psi_i^c=\psi_i$. This can be understood from the following relation.
\begin{eqnarray}
    \overline{\psi_i} \gamma^{\mu}  \psi_j = \overline{\psi_i^c} C^{-1}(\gamma^{\mu})^T C \psi_j^c 
= - \overline{\psi_i} \gamma^{\mu}  \psi_j\,~~\text{using}~~ C^{-1}(\gamma^{\mu})^T C= - \gamma^{\mu} . \nonumber
\end{eqnarray}
\noindent Thus, for $i=j$, the diagonal terms mentioned above vanish identically.

\begin{figure}[htb!]
 \includegraphics[scale=0.3]{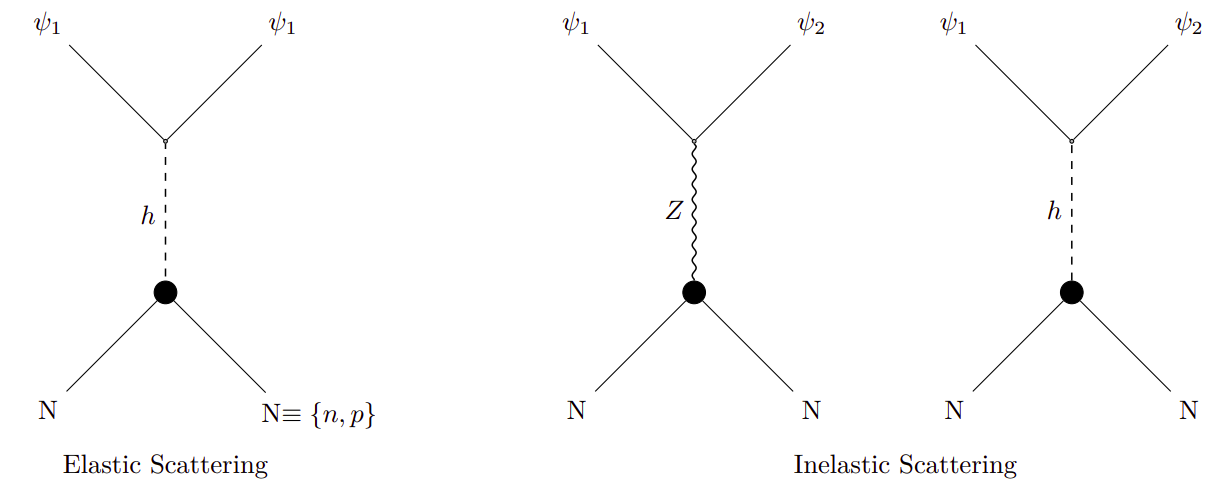}
  \caption{Left panel: Elastic scattering responsible for DD detection in presence of pseudo-Dirac splittings between two neutral states, now via Higgs mediation. Right panel: Inelastic scattering of $\psi_1 N \to \psi_2 N$ via Z and Higgs boson mediations.}
  \label{fig:VLFV_MS}
 \end{figure}

Now, with this setup, there will be two different diagrams that can contribute to the direct detection of the corresponding DM, that is, the elastic scattering of lighter pseudo-Dirac state off the nucleus via SM Higgs mediation (left panel of Fig. \ref{fig:VLFV_MS}) $\psi_1 N \to \psi_1 N$, as well as an inelastic scattering of the DM particle via Z and h-bosons mediation (right panel of Fig. \ref{fig:VLFV_MS}) $\psi_1 N \to \psi_2 N$. However, this type of inelastic scattering process is allowed kinematically only if DM ($\psi_1$) has sufficient kinetic energy to overcome mass splitting $\delta m_{12}$ \cite{Tucker-Smith:2001myb}. This sets an upper limit on $\delta m_{12}$:
\bea
(\delta{m}_{12})^{\rm max} < \frac{1}{2} \big( \frac{ m_{\psi_1} m_{N} }{m_{\psi_1}+ m_{N}}\big) v_{\rm DM}^2 .
\eea
\noindent For $v_{\rm DM} = 650$ km/s and the target nucleus mass of the XENON1T experiment $m_{\mathcal{N}}=130$ amu, the upper limit on mass splitting $\delta{m}_{12}$ lies between 130 keV and 250 keV for $m_{\psi_1}$ ranging from 100 to 1000 GeV. For $\delta m_{12} \gtrsim 250$ kev, the inelastic processes ($\psi_1 ~N \to \psi_2~N$) mediated by $Z$ and $h$ no longer contribute to the DD constraints if the mass of DM is below $\sim 1$ TeV. On the other hand, for $\delta m_{12} \lesssim 250$ keV, the pseudo-elastic Z-mediated diagram contributes significantly to the DM DD cross-section, and eventually is excluded by the DD constraints (see the left panel of Fig.\ref{fig:VLL_dd}).

\begin{figure}[htb!] 
\includegraphics[scale=0.43]{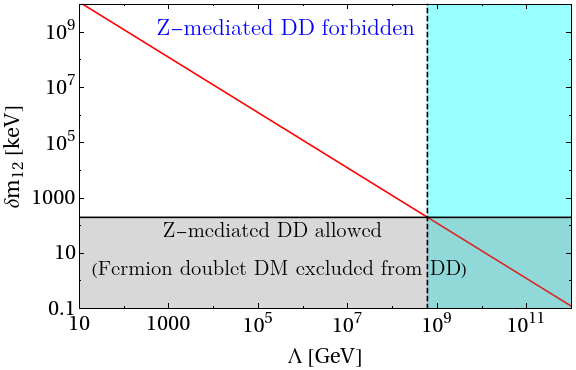}
\includegraphics[scale=0.3]{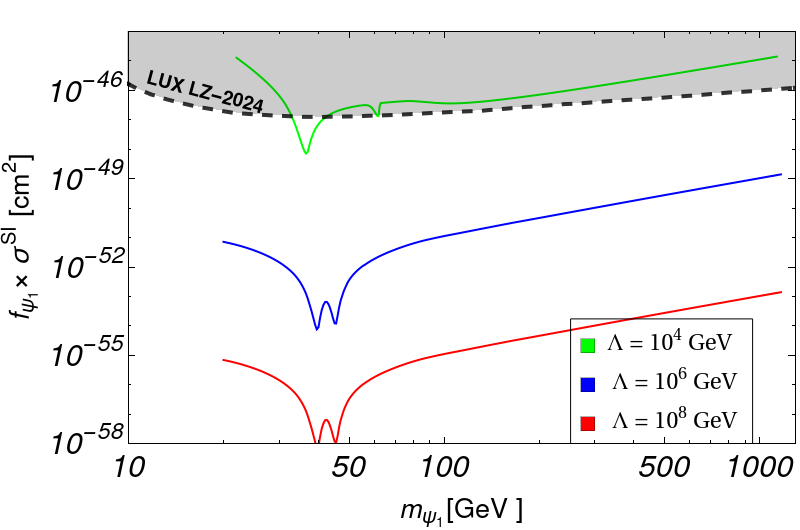}
  \caption{Left panel: Parameter space between pseudo-Dirac splitting $\delta m_{12}$ vs new physics scale $\Lambda$, the grey shaded region corresponds to the disallowed parameter space from DD constraints. The red curve describes our model prediction in presence of effective coupling. Right panel: Dependence of effective DM-nucleon scattering cross-section scaled with the fractional DM density for $\psi_1$ with respect to DM mass. Here, different colour schemes denote different $\Lambda$ values. Here, we have considered the mass splitting $\delta m_{12} > 250$ keV to evade the $Z$ mediated direction detection constraint.}
  \label{fig:VLFD_inelas}
 \end{figure}

 In the left panel of Fig. \ref{fig:VLFD_inelas}, we plot the dependence of the new physics scale $\Lambda$ on the generated pseudo-Dirac splittings $\delta m_{12}$. Here, the tiny splitting $\lesssim 250$ keV for $m_{\psi_1}\sim ~\mathcal{O}$ (1 TeV) \cite{Ghosh:2023dhj} between the pseudo-Dirac states $\psi_1$ and $\psi_2$ will allow the Z-mediated DD scattering which is disfavoured by the present upper bound from LUX-LZ 2024 observation (grey shaded region), while for higher mass splittings we can evade the Z-mediated DD constraints. The red curve corresponds to our model prediction in the presence of the effective interaction. From this parameter space, we can set an upper limit on the scale of new physics,  $\Lambda \lesssim 6 \times 10^{8}$ GeV (vertical dotted lined). Similarly, in the right panel of Fig. \ref{fig:VLFD_inelas}, we plotted the effective DM-nucleon scattering cross-section scaled with the fractional DM density $f_{\psi_1} \times \sigma^{\rm{SI}}$, where $f_{\psi_1} \equiv \Omega_{\psi_1}/\Omega_{\rm{DM}} \leq 1$ with respect to the DM mass $m_{\psi_1}$. Different colour curves denote different $\Lambda$ values {\it i.e.}, $10^4$ GeV (green), $10^6$ GeV (blue) and $10^8$ GeV 
 (red). We have also shown the upper bound from LUX-LZ 2024 along with the disfavoured region (shown in grey) in context of DD signature. For $\Lambda = 10^4$ GeV and $m_{\psi_1} \geq 50$ GeV,  our framework prediction for the DD fractional DM density $f_{\psi_1} \times 
 \sigma^{\rm{SI}}$  (green curve) almost matches the DD constraint from LUX-LZ 2024 limit. Correspondingly, we have set a lower limit, 
 $\Lambda \gtrsim 10^4$ GeV to evade the DD constraint. Furthermore, the two dips again arise from the W and Z resonances in the relic density ($f_{\psi_1}$), as discussed earlier. So, using 
 these two arguments to satisfy the direct detection constraints in our framework, varying the corresponding new physics scale within the range, $10^4$ GeV $\lesssim \Lambda \lesssim 6 \times 10^{8}$ GeV will yield results consistent with all DM constraints.  

 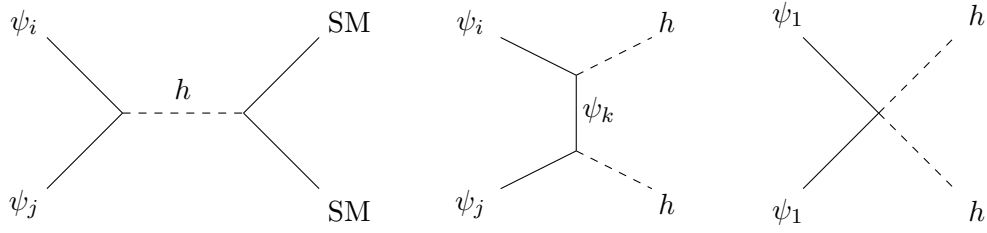
\begin{figure}[htb!]
   \begin{center}
     \begin{tikzpicture}[line width=0.4 pt, scale=1.0]
    \draw[solid] (-5.0,1)--(-4.0,0);
\draw[solid] (-5.0,-1)--(-4.0,0);
\draw[dashed] (-4.0,0)--(-2.4,0);
\draw[solid] (-2.4,0)--(-1.4,1);
\draw[solid] (-2.4,0)--(-1.4,-1);
\node  at (-5.3,1.2) {$\psi_i$};
\node at (-5.3,-1.2) {$\psi_j$};
\node [above] at (-3.2,0.05) {$h$};
\node at (-1,1.2){SM};
\node at (-1,-1.3) {SM};
       \draw[solid] (1.0,1.0)--(2.0,0.5);
\draw[solid] (1.0,-1.0)--(2.0,-0.5);
\draw[solid] (2.0,0.5)--(2.0,-0.5);
\draw[dashed] (2.0,0.5)--(3.0,1.0);
\draw[dashed] (2.0,-0.5)--(3.0,-1.0);
\node at (0.6,1.2) {$\psi_i$};
\node at (0.6,-1.2) {$\psi_j$};
\node at (2.3,0.04) {$\psi_k$};
\node at (3.2,1.2) {$h$};
\node at (3.2,-1.2) {$h$};
 \draw[solid] (5,1)--(6.0,0);
\draw[solid] (5,-1)--(6.0,0);
\draw[dashed] (6.0,0)--(7.0,1);
\draw[dashed] (6.0,0)--(7.0,-1);
\node at (7.3,1.3) {$h$};
\node at (7.3,-1.3) {$h$};
\node at (4.8,1.3) {$\psi_1$};
\node at (4.8,-1.3) {$\psi_1$};
\end{tikzpicture}
  \end{center}    
\caption{Feynman diagrams for DM annihilation and co-annihilation into SM particles ($\psi_1~\psi_1 \to $ SM~SM and $\psi_1~\psi_2 \to $ SM~SM, respectively), enabled by the effective dim-5 operator. Here $\{i,j,k\}=\{1,2\}$.}
\label{fig:Feyn-ann-psi1}
 \end{figure}
  \begin{figure}[htb!]
   \begin{center}
     \begin{tikzpicture}[line width=0.4 pt, scale=1.0]
    \draw[solid] (-5.0,1)--(-4.0,0);
\draw[solid] (-5.0,-1)--(-4.0,0);
\draw[snake] (-4.0,0)--(-2.4,0);
\draw[solid] (-2.4,0)--(-1.4,1);
\draw[solid] (-2.4,0)--(-1.4,-1);
\node  at (-5.3,1.2) {$\psi_1$};
\node at (-5.3,-1.2) {$\psi_2$};
\node [above] at (-3.2,0.05) {$Z$};
\node at (-1,1.2){$h/f/W^+$};
\node at (-1,-1.3) {$Z/\bar{f}/W^-$};
       \draw[solid] (0.8,1.0)--(1.8,0.5);
\draw[solid] (0.8,-1.0)--(1.8,-0.5);
\draw[solid] (1.8,0.5)--(1.8,-0.5);
\draw[snake] (1.8,0.5)--(2.8,1.0);
\draw[snake] (1.8,-0.5)--(2.8,-1.0);
\node at (0.6,1.2) {$\psi_1$};
\node at (0.6,-1.2) {$\psi_2$};
\node at (2.5,0.04) {$\psi_1/\psi^\pm$};
\node at (3.2,1.3) {$Z/W^\mp$};
\node at (3.2,-1.3) {$Z,/W^\pm$};
        \draw[solid] (5.5,1.0)--(6.5,0.5);
\draw[solid] (5.5,-1.0)--(6.5,-0.5);
\draw[solid] (6.5,0.5)--(6.5,-0.5);
\draw[snake] (6.5,0.5)--(7.5,1.0);
\draw[snake] (6.5,-0.5)--(7.5,-1.0);
\node at (5.3,1.2) {$\psi_i$};
\node at (5.2,-1.2) {$\psi^\pm$};
\node at (7.6,0.06) {$\psi_j/\psi^\pm$};
\node at (7.8,1.3) {$Z/W^\pm$};
\node at (7.8,-1.3) {$W^\pm/Z,\gamma$};

\end{tikzpicture}
  \end{center}    
 \begin{center}
  \begin{tikzpicture}[line width=0.4 pt, scale=1.0]

  \draw[solid] (-5.0,1)--(-4.0,0);
\draw[solid] (-5.0,-1)--(-4.0,0);
\draw[snake] (-4.0,0)--(-2.4,0);
\draw[solid] (-2.4,0)--(-1.4,1);
\draw[solid] (-2.4,0)--(-1.4,-1);
\node  at (-5.3,1.2) {$\psi_i$};
\node at (-5.3,-1.2) {$\psi^\pm$};
\node [above] at (-3.2,0.05) {$W$};
\node at (-1,1.2){$h/f/W^\pm$};
\node at (-1,-1.3) {$W^\pm/\bar{f^\prime}/Z,\gamma$};

\draw[solid] (1.0,1)--(2.0,0);
\draw[solid] (1.0,-1)--(2.0,0);
\draw[snake] (2.0,0)--(3.5,0);
\draw[solid] (3.5,0)--(4.5,1);
\draw[solid] (3.5,0)--(4.5,-1);
\node  at (1,1.2) {$\psi^+$};
\node at (1,-1.2) {$\psi^-$};
\node [above] at (2.7,0.05) {$Z/\gamma$};
\node at (4.5,1.2){$h/f/W^+$};
\node at (4.5,-1.3) {$Z/\bar{f}/W^-$};

\draw[solid] (6,1.0)--(7,0.5);
\draw[solid] (6,-1.0)--(7,-0.5);
\draw[solid] (7,0.5)--(7,-0.5);
\draw[snake] (7,0.5)--(8,1.0);
\draw[snake] (7,-0.5)--(8,-1.0);
\node at (5.8,1.2) {$\psi^+$};
\node at (5.8,-1.2) {$\psi^-$};
\node at (8.1,0.06) {$\psi^+/\psi_i$};
\node at (8.3,1.3) {$Z,\gamma/W^+$};
\node at (8.7,-1.3) {$Z, \gamma /W^-$};
     \end{tikzpicture}
 \end{center}
\caption{Feynman diagrams for DM co-annihilation into SM particles. Here $\{ i\neq j \}=\{1,2\}$. This set of diagrams is present in the Dirac case ($\psi^0$). Due to the pseudo-Dirac splitting ($\psi_{1,2}$), additional diagrams appear.}
\label{fig:Feyn-coann-psi1}
 \end{figure}
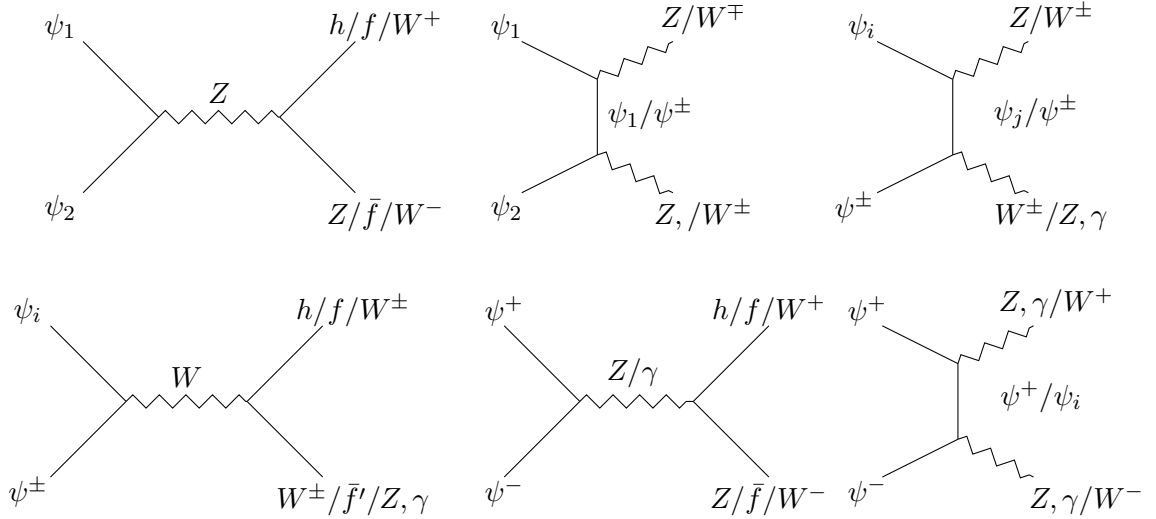

The dim-5 operator ${{\frac{1}{\Lambda} \overline{\Psi^c} \tilde{\Phi}~\tilde{\Phi}^\dagger \Psi}}$ opens up new Higgs-mediated annihilation and co-annihilation processes, contributing to both the dark matter relic abundance and direct detection cross-section. Relevant Feynman diagrams for the annihilation and co-annihilation of the DM particle have been presented in Fig. \ref{fig:Feyn-ann-psi1} and \ref{fig:Feyn-coann-psi1}. The splitting of the $\psi^0$ state into pseudo-Dirac states $\psi_{1,2}$, required for evading direct detection bounds, also opens up new co-annihilation processes. The effective thermal average cross-section, $\left< \sigma v\right>_\Psi^{\rm{eff}}$ in Eqn.\ref{eq:BEQ_VLLD} is modified as \cite{Griest:1990kh,Edsjo:1997bg}:
\begin{align}
    \langle \sigma v \rangle_\Psi^{\rm{eff}} & = \frac{g_{\psi_1}^2}{g_{\rm eff}^2} \left< \sigma v \right>_{\psi_1\psi_1} + \frac{2 g_{\psi_1} g_{\psi_2}}{g_{\rm eff}^2} \left< \sigma v \right>_{\psi_1\psi_2}(1+\delta_{\psi_2})^{\frac{3}{2}}~ e^{-\xi \delta_{\psi_2}} +  \frac{2 g_{\psi_1} g_{\psi^\pm}}{g_{\rm eff}^2} \left< \sigma v \right>_{\psi_1\psi^\pm}(1+\delta_{\psi^\pm})^{\frac{3}{2}}~ e^{-\xi \delta_{\psi^\pm}}\nonumber \\
    & +  \frac{2 g_{\psi_2} g_{\psi^\pm}}{g_{\rm eff}^2} \left< \sigma v \right>_{\psi_2\psi^\pm}(1+\delta_{\psi_2})^{\frac{3}{2}}~(1+\delta_{\psi^\pm})^{\frac{3}{2}}~ e^{-\xi \left( \delta_{\psi_2}+\delta_{\psi^\pm} \right)}\nonumber \\
   & +  \frac{g_{\psi_2}^2}{g_{\rm eff}^2} \left< \sigma v \right>_{\psi_2\psi_2}(1+\delta_{\psi_2})^3~ e^{-2 \xi \delta_{\psi_2}}  +  \frac{g_{\psi^\pm}^2}{g_{\rm eff}^2} \left< \sigma v \right>_{\psi^+ \psi^-}(1+\delta_{\psi^\pm})^3~ e^{-2\xi \delta_{\psi^\pm}}~,
\end{align}
with $g_{\rm eff} = g_{\psi_1}+g_{\psi_2}(1+\delta_{\psi_2})^{\frac{3}{2}}~ e^{-\xi \delta_{\psi_2}}+g_{\psi^\pm}(1+\delta_{\psi^\pm})^{\frac{3}{2}}~ e^{-\xi \delta_{\psi^\pm}}$ and $\xi=\frac{m_{\psi_1}}{T}$. Here $\delta_{\psi_2}=\frac{m_{\psi_2}-m_{\psi_1}}{m_{\psi_1}} = \frac{2\delta m_{12}}{m_{\psi_1}}$ and $\delta_{\psi^\pm}=\frac{m_{\psi^\pm}-m_{\psi_1}}{m_{\psi_1}} = \frac{\delta m~+~\delta m_{12}}{m_{\psi_1}}$. The internal degrees of freedom  $g_{\psi_{1,2}}$ and $g_{\psi^\pm}$ are associated with the dark fermion states, $\psi_{1,2}$ and $\psi^\pm$ respectively. In the BEQ (in Eqn.\ref{eq:BEQ_VLLD}), $n_{\Psi}$ denotes the combined number density of all relevant states as: $n_{\Psi}=n_{\rm \psi_1}+n_{\psi_2}+n_{\psi^\pm}$.
 
\begin{figure}[htb!]
\centering
\includegraphics[width=0.45\linewidth]{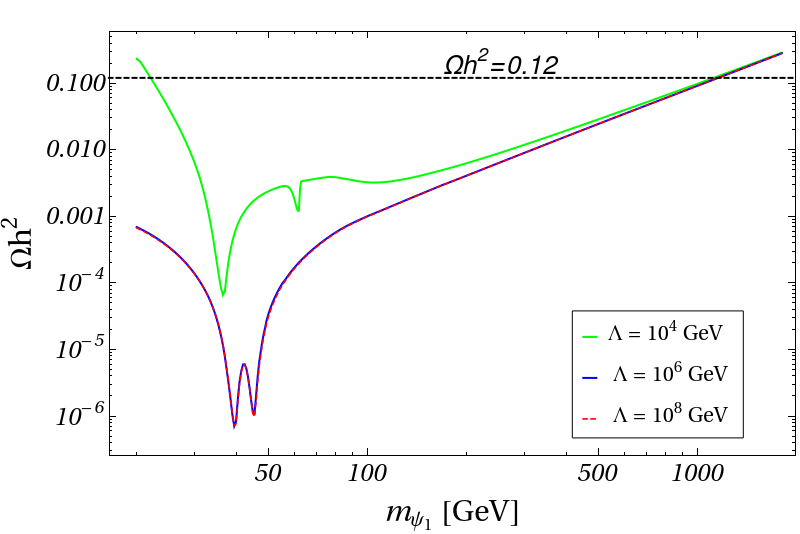}
\includegraphics[width=0.45\linewidth]{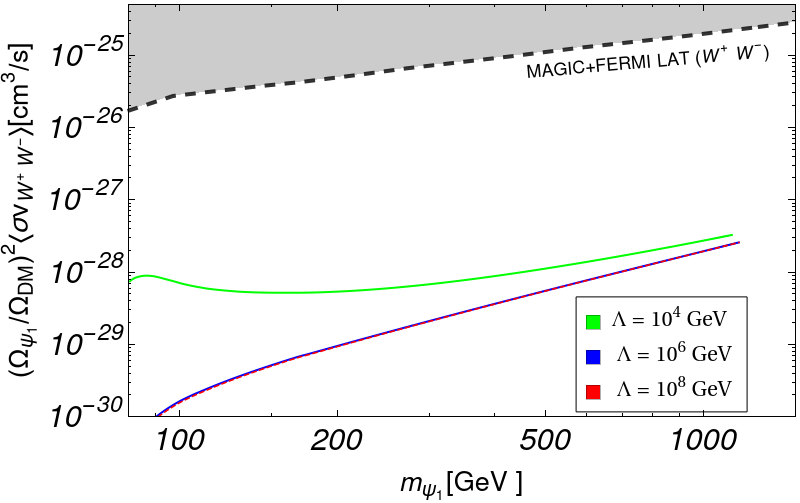}
\caption{Left panel: Plot showing the relic density vs. the mass of the DM for three different values of $\Lambda$. Right panel: The scaled indirect detection annihilation cross-section into the $W^+ W^-$ channel.}
\label{fig:VLF_relic}
\end{figure}

We have also analyzed the relic density as a function of the DM mass in the presence of the effective coupling. The corresponding 
dependence over the entire mass region is shown in left panel of Fig. \ref{fig:VLF_relic} for different $\Lambda$ values i.e., $10^4$ GeV (green), $10^6$ GeV (blue) and $10^8$ GeV (red dashed). Here, also the entire parameter space for $m_{\psi_1} \lesssim 1.2$ TeV is under-abundant (except lower mass region $m_{\psi_1} \leq 20$ GeV). For comparatively smaller value of $\Lambda = 10^4$ GeV, the only effective
coupling $\frac{1}{\Lambda}$ between DM and SM Higgs becomes quite tiny, therefore, the contribution of annihilation process $\psi_1 \psi_1 \to hh$ to total annihilation cross-section becomes negligible. However, the pseudo-Dirac mass splittings between $\psi_1$ and 
$\psi_2$ becomes $\sim \mathcal{O}$ (GeV), correspondingly the contribution of co-annihilation channels via $\psi_1-\psi_2$ and $\psi_1-\psi^\pm$ interactions is negligible. Now, by increasing the $\Lambda$ value, we effectively decrease the pseudo-Dirac splitting between the fermionic states, which eventually increases the contribution of co-annihilation to the total annihilation cross-section and 
decreases the relic density contribution even further for higher $\Lambda$ values, as shown in the left panel of the figure. Therefore, through the higher values of $\Lambda \gtrsim 10^4$ GeV and the sub-TeV DM mass can evade the the DD constraint,  in this scenario the pseudo-Dirac fermion DM still it fails to saturate the correct DM relic density measured by PLANCK. In the right panel of the same 
figure, we have presented the scaled indirect detection cross-section to the most important contribution due to $W^+ W^-$ annihilation, $f_{\psi_1}^2 \times \langle \sigma v_{W^+ W^-} \rangle$, for three different values of $\Lambda$'s. The similar behaviour between 
different $\Lambda$ values is also visible here via the scaling factor $(\Omega_{\psi_1}/\Omega_{\rm{DM}})^2$, as we have noticed in the left panel. The corresponding indirect detection combined upper limit from MAGIC and FERMI-LAT data on $W^+ W^-$ annihilation channel is shown , along with the disfavoured region shown in grey. Therefore, again the indirect detection constraint allows the entire parameter 
space of this scenario considered, and while DD constraints are satisfied (though tightly), the relic density is under-abundant for almost all parameter space.

\subsection{Single Component Scalar Doublet DM}
\label{subsec:scalar} 
We now consider the SM extended only by an inert scalar $SU(2)_L$ doublet (ISD), $H$, the lightest neutral component of which will serve as a stable DM candidate. The stability of the DM is ensured by the discrete symmetry $\mathcal{Z}_2$ under which the new scalar doublet transforms as $H \to -H$, while all SM particles are $\mathcal{Z}_2$-even. The new scalar $H$ transforms similarly to the SM Higgs doublet under the SM gauge group, as indicated in Tab. \ref{tab:ISD}. The corresponding kinetic as well as interaction terms  of $H$ with SM gauge bosons can be written as,
\begin{align}
    \mathcal{L}^{\rm ISD} & = \Big|\Big( \partial_\mu - i g_2 \frac{\sigma^a}{2} W_{\mu}^a-i g_1 \frac{Y_{H}}{2} B_{\mu}\Big) H \Big|^2 - V(H,\Phi) ~~~~~~~~~ {\rm with} ~~ Y_H=1~.
\end{align}
\begin{table}[htb!]
 \begin{tabular}{|c|c|c|}
\hline 
\multicolumn{2}{|c|}{Fields} & \multicolumn{1}{|c|}{$\underbrace{ SU(3)_C \otimes SU(2)_L \otimes U(1)_Y } \otimes \underbrace{ \mathcal{Z}_2 }$} \\ 
\hline
\textbf{ISD} & $\mathbf{H} = \left(\begin{matrix}
 H^+ \\  \frac{1}{\sqrt{2}}( H^0 +i A^0) 
\end{matrix}\right)$ & ~~1 ~~~~~~~~~~~2~~~~~~~~~~1~~~~~~~~~~~-  \\
\hline
\textbf{Higgs doublet} & $\mathbf{\Phi} = \left(\begin{matrix} \phi^+ \\ \frac{1}{\sqrt{2}} (v + h + i z) \end{matrix}\right)$ & ~~1 ~~~~~~~~~~~2~~~~~~~~~~1~~~~~~~~~~+\\
\hline
\end{tabular}
\caption{Charge assignments of inert scalar doublet and SM Higgs doublet under the gauge group $\mathcal{G} \equiv \mathcal{G}_{\rm SM} \otimes \mathcal{G}_{\rm DM}$  where $\mathcal{G}_{\rm SM}\equiv SU(3)_C \otimes SU(2)_L \otimes U(1)_Y$ and $\mathcal{G}_{\rm DM} \equiv \mathcal{Z}_2$.  }
    \label{tab:ISD}
\end{table}

\noindent The scalar potential involving two scalar doublets $\Phi$ and $H$ in this scenario can be written as,
\begin{align}
   V(H, \Phi) & = - \mu_H^2(H^\dagger H)-\lambda_H (H^\dagger H)^2  -\lambda_{1} (\Phi^\dagger \Phi)(H^\dagger H) -\lambda_{2} (\Phi^\dagger H)(H^\dagger \Phi) -\frac{\lambda_{3}}{2}[(\Phi^\dagger H)^2+h.c.] .
\end{align}
and the relations between the physical masses and the couplings are given by :
\begin{align}
  & \mu_{H}^2 = m_{H^0}^2 - v^2 \lambda_L, ~~~~ \lambda_{1} = 2 \lambda_L - \frac{2}{v^2} (m_{H^0}^2 - m_{H^\pm}^2), \nonumber \\
  & \lambda_{2} = \frac{1}{v^2} (m_{H^0}^2 + m_{A^0}^2 - 2m_{H^\pm}^2), ~~~~ \lambda_{3} = \frac{1}{v^2} (m_{H^0}^2 - m_{A^0}^2),
\end{align}
where $\lambda_L = \frac{1}{2} (\lambda_{1} + \lambda_{2} + \lambda_{3})$. 
Correspondingly, one can invert these relations and write down the mass eigenvalues of different new states in terms of above mentioned couplings as,
\begin{align}
    & m_{H^0}^2 = \mu_H^2 + \frac{1}{2}v^2 (\lambda_1 + \lambda_2 + \lambda_3), \nonumber \\
    & m_{A^0}^2 = \mu_H^2 + \frac{1}{2}v^2 (\lambda_1 + \lambda_2 - \lambda_3), \nonumber \\
    & m_{H^\pm}^2 = \mu_H^2 + \lambda_1 v^2.
    \label{eq:Phieigenstates}
\end{align}

Considering $\lambda_3 < 0$, it is easily inferred from these relations that the CP-even scalar $H^0$ is the lightest neutral state and acts as the single component scalar DM candidate for our discussion. The hierarchy between CP-odd state $A^0$ and charged scalar $H^\pm$ depends on the relative magnitudes of  $\lambda_1\, , \lambda_2$ and $\lambda_3$. Throughout this work, we have considered the hierarchy between the scalars as, $m_{H^0} < m_{A^0} \leq  m_{H^\pm}$. Due to strong gauge-mediated interactions with the visible sector, the DM $H^0$ was in thermal equilibrium in the early universe. At low temperatures ($T < m_{H^0}$), when the interaction rate falls below the universe's expansion rate, the DM freezes out of thermal equilibrium and attains its observed relic density via the usual thermal freeze-out mechanism. The abundance of DM is governed by its annihilation ($H^0 H^0 \to {\rm SM~SM}$) as well as co-annihilation ($H^0 A^0 , H^0 H^\pm \to {\rm SM~SM}$) processes to the SM particles as shown in the Feynman diagrams in Fig. \ref{Feyn-ann-SD} and Fig. \ref{Feyn-coann-SD}, respectively.
\begin{figure}[htb!]
 \begin{center}
    \begin{tikzpicture}[line width=0.4 pt, scale=1.0]
        \draw[dashed] (-6,1)--(-5,0);
\draw[dashed] (-6,-1)--(-5,0);
\draw[dashed] (-5,0)--(-4,1);
\draw[dashed] (-5,0)--(-4,-1);
\node at (-6.2,1.1) {$H^0$};
\node at (-6.2,-1.1) {$H^0$};
\node at (-3.8,1.1) {$h$};
\node at (-3.8,-1.1) {$h$};
\draw[dashed] (-1.8,1.0)--(-0.8,0.5);
\draw[dashed] (-1.8,-1.0)--(-0.8,-0.5);
\draw[dashed] (-0.8,0.5)--(-0.8,-0.5);
\draw[dashed] (-0.8,0.5)--(0.2,1.0);
\draw[dashed] (-0.8,-0.5)--(0.2,-1.0);
\node at (-2.1,1.1) {$H^0$};
\node at (-2.1,-1.1) {$H^0$};
\node at (-1.2,0.07) {$H^0$};
\node at (0.4,1.1) {$h$};
\node at (0.4,-1.1) {$h$};
        \draw[dashed] (2.4,1)--(3.4,0);
\draw[dashed] (2.4,-1)--(3.4,0);
\draw[dashed] (3.4,0)--(4.4,0);
\draw[solid] (4.4,0)--(5.4,1);
\draw[solid] (4.4,0)--(5.4,-1);
\node  at (2.1,-1.1) {$H^0$};
\node at (2.1,1.1) {$H^0$};
\node [above] at (3.9,0.05) {$h$};
\node at (5.8,1.0){{\rm SM}};
\node at (5.8,-1.0) {{\rm SM}};
     \end{tikzpicture}
\end{center}
\begin{center}
    \begin{tikzpicture}[line width=0.4 pt, scale=1.0]
        \draw[dashed] (-6,1)--(-5,0);
\draw[dashed] (-6,-1)--(-5,0);
\draw[snake] (-5,0)--(-4,1);
\draw[snake] (-5,0)--(-4,-1);
\node at (-6.2,1.1) {$H^0$};
\node at (-6.2,-1.1) {$H^0$};
\node at (-3.4,1.1) {$W^+(Z)$};
\node at (-3.4,-1.1) {$W^-(Z)$};
\draw[dashed] (-1.8,1.0)--(-0.8,0.5);
\draw[dashed] (-1.8,-1.0)--(-0.8,-0.5);
\draw[dashed] (-0.8,0.5)--(-0.8,-0.5);
\draw[snake] (-0.8,0.5)--(0.2,1.0);
\draw[snake] (-0.8,-0.5)--(0.2,-1.0);
\node at (-2.1,1.1) {$H^0$};
\node at (-2.1,-1.1) {$H^0$};
\node at (-0.1,0.07) {$H^{\pm}(A^0)$};
\node at (0.8,1.2) {$W^{\pm}(Z)$};
\node at (0.8,-1.2) {$W^{\mp}(Z)$};
\draw[dashed] (2.4,1)--(3.4,0);
\draw[dashed] (2.4,-1)--(3.4,0);
\draw[dashed] (3.4,0)--(4.4,0);
\draw[snake] (4.4,0)--(5.4,1);
\draw[snake] (4.4,0)--(5.4,-1);
\node  at (2.1,-1.1) {$H^0$};
\node at (2.1,1.1) {$H^0$};
\node [above] at (3.9,0.05) {$h$};
\node at (6.0,1.2){$W^+(Z)$};
\node at (6.0,-1.2){$W^-(Z)$};
     \end{tikzpicture}
\end{center}
\caption{Feynman diagrams for DM annihilation into SM particles for the scalar DM candidate ($H^0$). }
\label{Feyn-ann-SD}
 \end{figure}
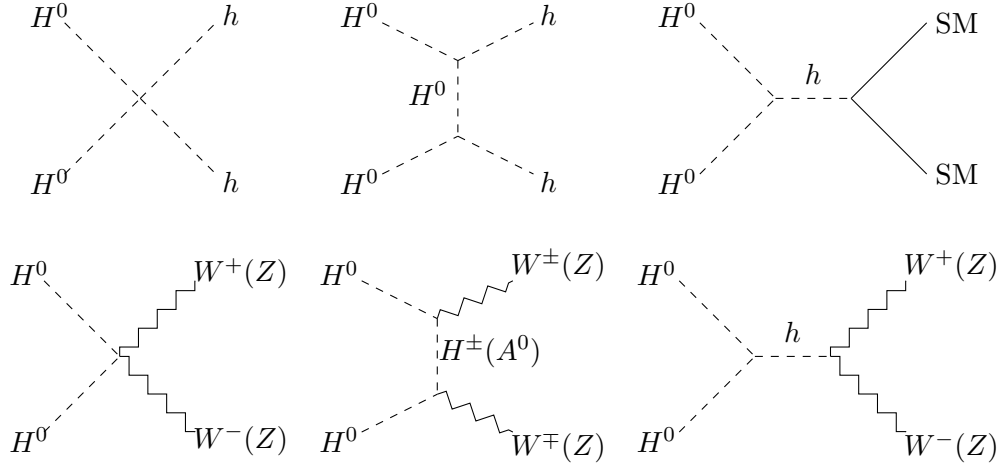
\begin{figure}[htb!]
 \begin{center}
    \begin{tikzpicture}[line width=0.4 pt, scale=1.0]
        \draw[dashed] (-5.0,1)--(-4.0,0);
\draw[dashed] (-5.0,-1)--(-4.0,0);
\draw[snake] (-4.0,0)--(-2.4,0);
\draw[solid] (-2.4,0)--(-1.4,1);
\draw[solid] (-2.4,0)--(-1.4,-1);
\node  at (-5.3,1.2) {$H^0$};
\node at (-5.7,-1.2) {$A^0 (H^\pm)$};
\node [above] at (-3.2,0.05) {$Z(W^\pm)$};
\node at (-1.0,1.2){{\rm SM}};
\node at (-1.0,-1.2) {{\rm SM}};
\draw[dashed] (0.8,1)--(1.8,0);
\draw[dashed] (0.8,-1)--(1.8,0);
\draw[snake] (1.8,0)--(2.8,1);
\draw[snake] (1.8,0)--(2.8,-1);
\node at (0.6,1.2) {$H^0$};
\node at (0.5,-1.2) {$H^\pm$};
\node at (3.4,1.2) {$Z,\gamma$};
\node at (3.4,-1.2) {$W^{\pm}$};
     \end{tikzpicture}
\end{center}
\begin{center}
    ~~~~~~~~~~ ~~ \begin{tikzpicture}[line width=0.4 pt, scale=1.0]
       \draw[dashed] (-1.8,1.0)--(-0.8,0.5);
\draw[dashed] (-1.8,-1.0)--(-0.8,-0.5);
\draw[dashed] (-0.8,0.5)--(-0.8,-0.5);
\draw[snake] (-0.8,0.5)--(0.2,1.0);
\draw[snake] (-0.8,-0.5)--(0.2,-1.0);
\node at (-2.1,1.1) {$H^0$};
\node at (-2.1,-1.1) {$A^0$};
\node at (0.4,0.07) {$H^0,A^0 (H^\pm)$};
\node at (1.0,1.2) {$h,Z (W^\pm)$};
\node at (1.0,-1.2) {$Z,h(W^\pm)$};
\draw[dashed] (4.0,1.0)--(5.0,0.5);
\draw[dashed] (4.0,-1.0)--(5.0,-0.5);
\draw[dashed] (5.0,0.5)--(5,-0.5);
\draw[snake] (5,0.5)--(6,1.0);
\draw[snake] (5,-0.5)--(6,-1.0);
\node at (3.8,1.2) {$H^0$};
\node at (3.8,-1.2) {$H^\pm$};
\node at (5.8,0.07) {$H^{\pm}(A^0)$};
\node at (7.1,1.2) {$W^{\pm}(Z,h,\gamma)$};
\node at (7.1,-1.2) {$h,Z,\gamma(W^\pm)$};
     \end{tikzpicture}
\end{center}
\caption{Feynman diagrams for DM co-annihilation into SM particles for the scalar DM ($H^0$) associated with heavier states ($A^0,H^\pm$). }
\label{Feyn-coann-SD}
\end{figure}
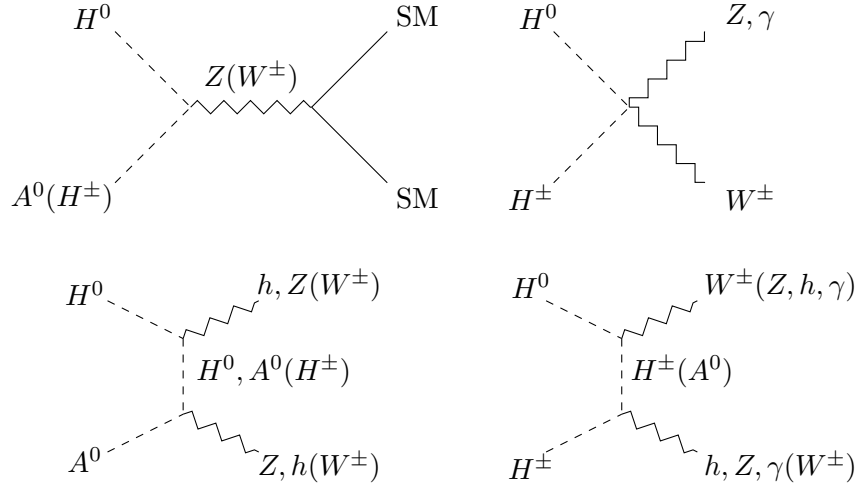

The evolution of the total number density, $n_{H}=n_{H^0}+n_{A^0}+n_{H^\pm}$, involving both annihilation and co-annihilation processes in the early universe, as a function of time, is described by the BEQ as \cite{Kolb:1990vq,Griest:1990kh,Edsjo:1997bg}:
\begin{equation}
    \frac{d n_{H}}{dt} + 3 \mathcal{H} n_{H}  = - \langle \sigma v \rangle_{H}^{\rm eff}  \left( n_{H}^2 - {n^{\text{eq}}_{H}}^2 \right) .
\end{equation}
The effective thermally averaged cross-section, $\langle \sigma v \rangle_{H}^{\rm eff}$, can be expressed as \cite{Griest:1990kh,Edsjo:1997bg}:
\begin{align}
    \langle \sigma v \rangle_{H}^{\rm{eff}} & = \frac{g_{H^0}^2}{g_{\rm eff}^2} \left< \sigma v \right>_{H^0 H^0} + \frac{ 2 g_{H^0} g_{A^0}}{g_{\rm eff}^2} \left< \sigma v \right>_{H^0 A^0}(1+ \delta_{A^0})^{\frac{3}{2}}~ e^{-\xi \delta_{A^0}} +  \frac{2 g_{H^0} g_{H^\pm}}{g_{\rm eff}^2} \left< \sigma v \right>_{H^0 H^\pm}(1+ \delta_{H^\pm})^{\frac{3}{2}}~ e^{-\xi \delta_{H^\pm}}\nonumber \\
    & +  \frac{2 g_{A^0} g_{H^\pm}}{g_{\rm eff}^2} \left< \sigma v \right>_{A^0 H^\pm}(1+\delta_{A^0})^{\frac{3}{2}}~(1+\delta_{H^\pm})^{\frac{3}{2}}~ e^{-\xi \left( \delta_{A^0}+\delta_{H^\pm} \right)}\nonumber \\
   & +  \frac{g_{A^0}^2}{g_{\rm eff}^2} \left< \sigma v \right>_{A^0 A^0}(1+\delta_{A^0})^3~ e^{-2 \xi \delta_{A^0}}  +  \frac{g_{H^\pm}^2}{g_{\rm eff}^2} \left< \sigma v \right>_{H^+ H^-}(1+\delta_{H^\pm})^3~ e^{-2\xi \delta_{H^\pm}}~,
\end{align}
with $g_{\rm eff} = g_{H^0}+g_{A^0}(1+\delta_{A^0})^{\frac{3}{2}}~ e^{-\xi \delta_{A^0}}+g_{H^\pm}(1+\delta_{H^\pm})^{\frac{3}{2}}~ e^{-\xi \delta_{H^\pm}}$ and $\xi=\frac{m_{H^0}}{T}$. Here $\delta_{A^0}=\frac{m_{A^0}-m_{H^0}}{m_{H^0}}$ and $\delta_{H^\pm}=\frac{m_{H^\pm}-m_{H^0}}{m_{H^0}}$. The internal degrees of freedom  $g_{H^0}$, $g_{A^0}$ and $g_{H^\pm}$ are associated with the dark scalar states, $H^0$, $A^0$ and $H^\pm$ respectively.

\begin{figure}[htb!]
    \centering
\includegraphics[width=0.4\linewidth]{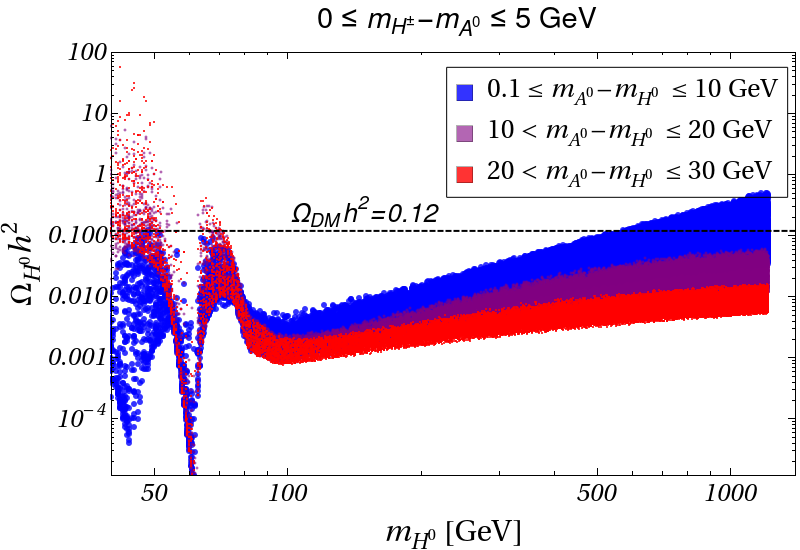}
\includegraphics[width=0.4\linewidth]{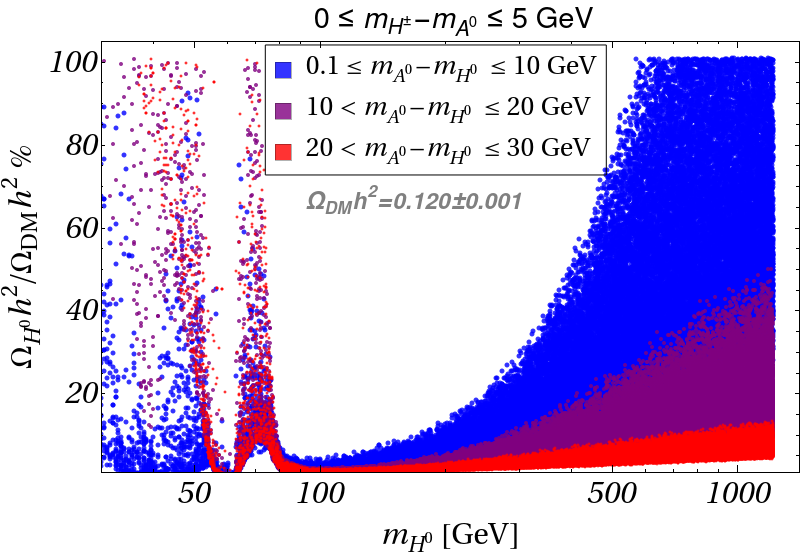}
\caption{Left panel: The variation of the relic density as a function of scalar doublet DM mass. Different colour schemes represent different mass splittings between CP-even and CP-odd new scalar states, as indicated in the enclosed panels. Right panel: Percentage contribution of scalar relic density to the required one is shown considering same mass splittings. For these plots, we have varied the mass splittings between the charged scalar and the pseudoscalar states ($m_{H^\pm} - m_{A^0}$) within [0, 5] GeV.}
\label{fig:IDM_relic_mass}
\end{figure}

The variation in relic density ($\Omega_{H^0}h^2$) as a function of scalar doublet DM mass ($m_{H^0}$) is shown in the left panel of Fig. \ref{fig:IDM_relic_mass}. Here, we have varied the mass splitting between $H^\pm$ and $A^0$ within the range [0, 5] GeV, while the different mass splittings between the CP-odd and CP-even states are differentiated by using different colors: [0.1, 10] GeV (blue), [10, 20] GeV (purple) and [20, 30] GeV (red). In this scenario, the co-annihilation effects along with the annihilation channels are quite significant for smaller mass splittings between the scalar states, mostly denoted by blue points. Therefore in this regime it can be inferred that, for $m_{H^0} \gtrsim 525$ GeV as well as some scattered mass values below 80 GeV or so, the scalar DM alone can saturate the DM relic density criteria, denoted as black dashed line. Strong gauge-mediated (co-)annihilation processes are responsible for this.
However, if we increase the mass splitting between $H^0$ and $A^0$, we need $m_{H^0} \gtrsim 1$ TeV to satisfy this bound\footnote{We note that the DM mass region that satisfied relic constraints is larger for the scalar doublet than for the fermion doublet due in part to the larger number of the parameters in the former compared to the latter case.}. In the right panel of the same figure, we have shown the percentage contribution of the DM relic density normalized to the current required DM relic density ($f_{H^0} \equiv \Omega_{H^0}h^2 / \Omega_{\rm{DM}}h^2$) for the same mass difference definitions as in the left panel.

\begin{figure}[htb!]
    \centering
\includegraphics[width=0.25\linewidth]{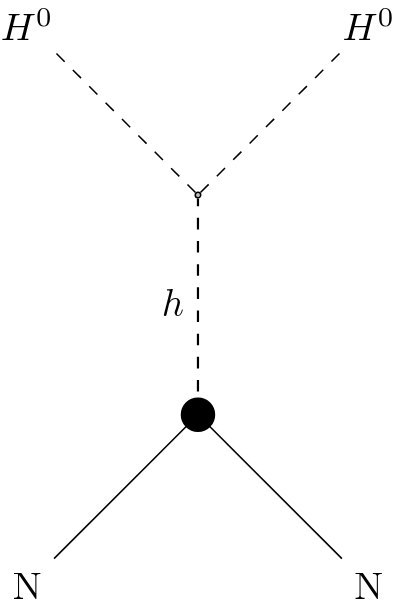}\qquad
\includegraphics[width=0.57\linewidth]{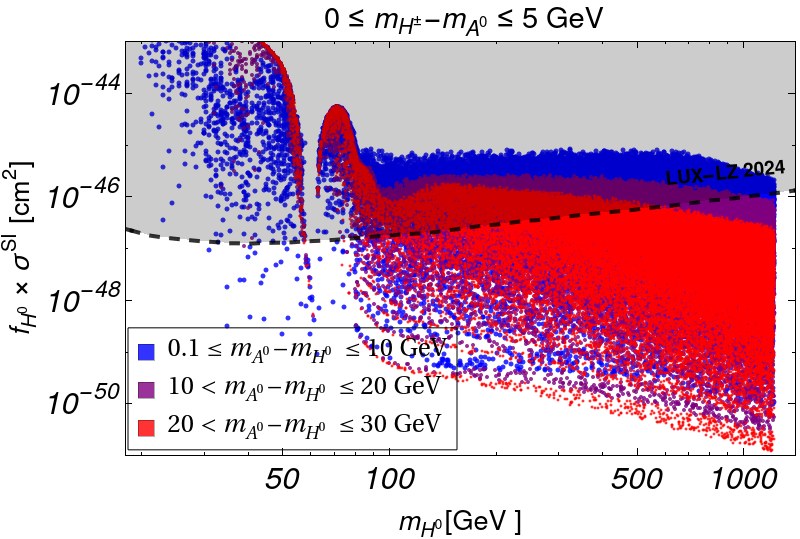} 
\caption{Left panel: Feynman diagram for Higgs mediated DM-nucleon scattering. Right panel: Dependence of DM-nucleon elastic scattering cross-section scaled by fractional DM density with respect to DM mass. Different colour schemes consider different mass splittings between CP-even and CP-odd states.}
    \label{fig:IDM_DD}
\end{figure}

The trilinear Higgs portal interaction allows for $t$-channel Higgs-mediated DM-nucleon elastic scattering, is responsible for DM direct detection, and is shown in the left panel of Fig. \ref{fig:IDM_DD}.
The relevant spin-independent DM-nucleon scattering cross section for scalar or inert doublet DM candidate scaled with the fractional DM density ($f_{H^0} \leq 1$) can be expressed as: $\sigma_{\rm{eff}}^{\rm{SI}} (H^0) = f_{H^0} \times \sigma^{\rm{SI}}$.
This dependence as a function of ISD mass $m_{H^0}$ is shown in the right panel of Fig. \ref{fig:IDM_DD}. Here, we also have varied the mass difference ($m_{H^\pm} - m_{A^0}$) within [0, 5] GeV. From the upper bound from LUX-LZ 2024 experiment \cite{LZ:2024zvo}, denoted by the black dashed curve, one can see that all three different mass splittings between CP-even and CP-odd scalars can evade the DD constraints for $m_{H^0} \gtrsim 35$ GeV (smaller mass splitting) as well as for $m_{H^0} \gtrsim 60$ GeV (larger mass splittings).

\begin{figure}[htb!]
    \centering
\includegraphics[width=0.47\linewidth]{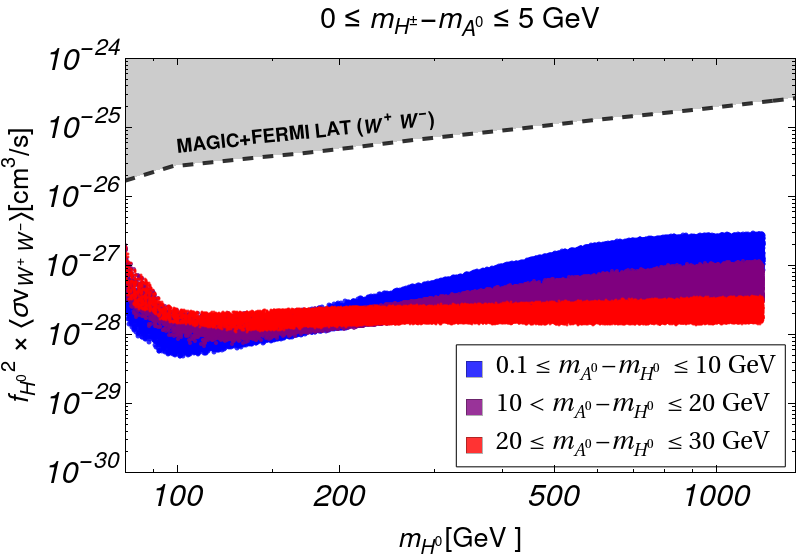} \qquad
\includegraphics[width=0.47\linewidth]{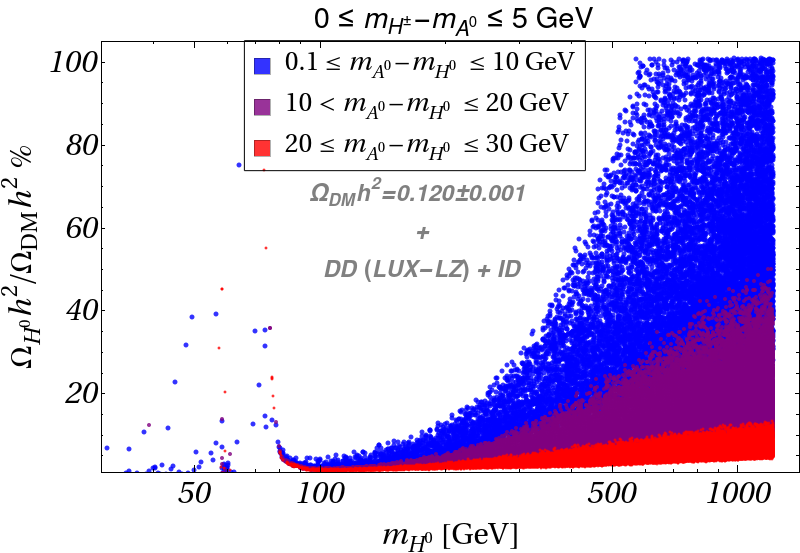}
    \caption{Left panel: Indirect detection cross-section scaled by fractional DM density for $W^+ W^-$ annihilation channel. Right panel: Allowed parameter space from direct detection and indirect detection bounds in the parameter space spanned by fractional DM density in percentage contribution versus ISD mass. 
    }
    \label{fig:IDM_DMNC_mass}
\end{figure}

Furthermore, we investigate the parameter space spanned by the indirect detection cross-section of ISD annihilating into $W^+ W^-$ scaled by the fractional DM density squared \textit{vs} the ISD mass. All different mass splittings can clearly saturate the upper bound coming from the combined FERMI-LAT + MAGIC data, shown in the left panel of Fig. \ref{fig:IDM_DMNC_mass}. We also investigated the parameter space between fractional DM density as a percentage contribution \textit{vs} ISD mass in the right panel of the same figure. Considering both the DD (LUX-LZ 2024) as well as ID (MAGIC + FERMI-LAT) constraints, we  found the allowed parameter points in this plane for different mass splitting scenario. It is evident that, in the mass range 80 GeV $\lesssim m_{H^0} \lesssim 525$ GeV, the ISD alone cannot saturate three current bounds from relic density + DD + ID observations. 

Therefore, in summary, we see from these two previous subsections that a sub-TeV single component VLFD DM  cannot saturate the current relic density criteria and cannot evade the DD upper bound, while introducing the effective dim-5 "Weinberg-like" operator can generate the pseudo-Dirac mass splittings between neutral $\mathcal{Z}_2$-odd states, and in turn evade the Z-mediated DD constraints. However, this scenario still cannot saturate the relic density abundance, indicating that in itself it does not form a complete DM spectrum. In comparison, the single component ISD can easily saturate the DD and ID constraints from LUX-LZ 2024 and MAGIC + FERMI-LAT collaborations, respectively, but the scenario fails to satisfy the correct required relic density criteria within the mass range 80 GeV $\lesssim m_{H^0} \lesssim 525$ GeV. 

\section{Two component dark matter: scalar plus fermion doublet}
\label{sec:twocomp}

Thus within this sub-TeV region, the single component DM candidates individually cannot be a viable DM candidate. In what follows, we aim to find whether a viable two-component DM scenario with the combination of an ISD mass between $\sim \{ 80 - 525 \}$ GeV and a VLFD DM mass below 1 TeV  can saturate all the three bounds from relic density + DD + ID observations.  This motivates us to study the two component DM scenario with  the combined relic density, $\Omega_{H^0} h^2 + \Omega_{\psi_1} h^2  = 0.120 \pm 0.001$ \cite{Planck:2018vyg}. We build upon the previous section to investigate a two component (one scalar plus one fermion doublet), with the aim to satisfy all DM experimental constraints coming from relic abundance, and direct and indirect detection cross sections. Based on our previous considerations, we must include additional operators or fields to avoid or suppress Z boson mediated decays, which yield direct detection cross sections in conflict with experimental bounds. As discussed previously, there are two remedies for this: one is to introduce dim-5 operators, the other to construct an UV-complete model which includes a triplet scalar with hypercharge $Y=2$. We analyze both scenarios in turn.

\subsection{Two component dark matter with Effective operators (dim-5)}
\label{subsec:2compDIM5}
Combining the two single component scenarios discussed earlier, we introduce two $ \mathcal{Z}_2$ symmetries to stabilize the two-component DM. The charge assignments of BSM fields along with the SM Higgs doublet under the gauge group $\mathcal{G} \equiv \mathcal{G}_{\rm SM} \otimes \mathcal{G}_{\rm DM}$ (where $\mathcal{G}_{\rm SM}\equiv SU(3)_C \otimes SU(2)_L \otimes U(1)_Y$ and $\mathcal{G}_{\rm DM} \equiv \mathcal{Z}_2 \otimes \mathcal{Z}^\prime_2$) are given in Tab. \ref{tab:tab1}.
 \begin{table}[htb!]
 \begin{tabular}{|c|c|c|c|}
\hline \multicolumn{2}{|c}{Fields}&  \multicolumn{1}{|c|}{ $\underbrace{ SU(3)_C \otimes SU(2)_L \otimes U(1)_Y}$ $\otimes \underbrace{ \mathcal{Z}_2 \otimes \mathcal{Z}^\prime_2 }$} \\ \hline
\multirow{2}{*} 
{SD DM} & $H = \left(\begin{matrix}
 H^+ \\  \frac{1}{\sqrt{2}}( H^0+i A^0) 
\end{matrix}\right)$ & ~~1 ~~~~~~~~~~~2~~~~~~~~~~1~~~~~~~~~~~-~~~~~+  \\
\hline
{FD DM} &  $\Psi=\left(\begin{matrix} \psi^0 \\ \psi^- \end{matrix}\right)$&  ~~1 ~~~~~~~~~~~2~~~~~~~~~~-1~~~~~~~~~~+~~~~~-  \\
\hline
\hline
Higgs doublet & $\Phi = \left(\begin{matrix} \phi^+ \\ \frac{v + h + i z}{\sqrt{2}} \end{matrix}\right)$ & ~~1 ~~~~~~~~~~~2~~~~~~~~~~1~~~~~~~~~~+~~~~~+ \\
\hline
\end{tabular}
\caption{Charge assignments of DM doublets and SM Higgs doublet fields under the gauge group $\mathcal{G} \equiv \mathcal{G}_{\rm SM} \otimes \mathcal{G}_{\rm DM}$  where $\mathcal{G}_{\rm SM}\equiv SU(3)_C \otimes SU(2)_L \otimes U(1)_Y$ and $\mathcal{G}_{\rm DM} \equiv \mathcal{Z}_2 \otimes \mathcal{Z}^\prime_2 $.  }
    \label{tab:tab1}
\end{table}

\noindent The corresponding interaction Lagrangian is given by,
\begin{equation}
    \mathcal{L}^{\rm{eff}} \supset  \mathcal{L}^{\rm FD~ DM} +\mathcal{L}^{\rm SD~ DM}+\mathcal{L}^{\rm SD+FD~ DM}.
\end{equation}
where the Lagrangian for Fermion DM involving dim-5 operators is given by,
\begin{equation}
    \mathcal{L}^{\rm FD~DM} = \mathcal{L}^{\rm{VLFD}} -  {\frac{1}{\Lambda} \overline{\Psi^c} \tilde{\Phi}~\tilde{\Phi}^\dagger \Psi} + {\rm h.c.}~.
\end{equation}
 This effective interaction between the SM Higgs doublet and the $\Psi$ multiplet generates the pseudo-Dirac mass splitting between neutral states of $\Psi$, as discussed previously. The Lagrangian for scalar DM involving dim-5 operators is given by
\begin{align}
    \mathcal{L}^{\rm SD~DM} & = \mathcal{L}^{\rm{ISD}} -V(H,\Phi) - {\frac{1}{\Lambda_H} \overline{L^c} \tilde{H}~\tilde{H}^\dagger L} + {\rm h.c.}~.
\end{align}
The interaction Lagrangian between the two DM components involving dim-5 operators is given by
\begin{equation}
    \mathcal{L}^{\rm SD+FD~DM} = - {\frac{1}{\Lambda_1} \overline{\Psi^c} \tilde{H} ~\tilde{H}^\dagger {\Psi}} + {\rm h.c.}~.
\end{equation}
 It should be noted that the symmetry of the current framework allows the inclusion of an additional dim-5 operator,  the known Weinberg operator: $ {\cal O}_W \simeq \frac{C_{W}}{\Lambda_W} \overline{L^c} \tilde{\Phi}~ \tilde{\Phi}^\dagger L$, which is responsible for the generation of neutrino masses. The effective scales $\Lambda, \Lambda_H$, $\Lambda_1$ and $\Lambda_W $ with $\Lambda, \Lambda_H, \Lambda_1 < \Lambda_W$ are associated with different dynamics in the respective UV complete models.

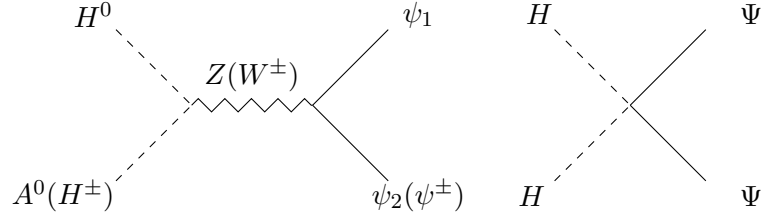
\begin{figure}[htb!]
\begin{center}
\begin{tikzpicture}[line width=0.4 pt, scale=1.0]
\draw[dashed] (-5.0,1)--(-4.0,0);
\draw[dashed] (-5.0,-1)--(-4.0,0);
\draw[snake] (-4.0,0)--(-2.4,0);
\draw[solid] (-2.4,0)--(-1.4,1);
\draw[solid] (-2.4,0)--(-1.4,-1);
\node  at (-5.3,1.2) {$H^0$};
\node at (-5.7,-1.2) {$A^0 (H^\pm)$};
\node [above] at (-3.2,0.05) {$Z(W^\pm)$};
\node at (-1.0,1.2){$\psi_1$};
\node at (-1.0,-1.2) {$\psi_2 (\psi^\pm) $};
\draw[dashed] (0.8,1)--(1.8,0);
\draw[dashed] (0.8,-1)--(1.8,0);
\draw (1.8,0)--(2.8,1);
\draw (1.8,0)--(2.8,-1);
\node at (0.6,1.2) {$H$};
\node at (0.5,-1.2) {$H$};
\node at (3.4,1.2) {$\Psi$};
\node at (3.4,-1.2) {$\Psi$};
\end{tikzpicture}
\end{center}
\caption{Feynman diagrams for DM $\to$ DM conversion. }
\label{Feyn-conv}
\end{figure}

Therefore, in our framework, there are two DM candidates, $H^0$ and $\psi_1$. As both candidates contribute to the DM relic density determined by the PLANCK experiment, they must satisfy \cite{Planck:2018vyg}:
\begin{equation}
    \Omega_{\rm DM} h^2 \equiv \Omega_{H^0} h^2 + \Omega_{\psi_1} h^2  = 0.120 \pm 0.001~.
    \label{eq:OmegaTotal}
\end{equation}
The $\Omega_{H^0} h^2$ and $\Omega_{\psi_1} h^2$ represent the relic densities for scalar DM $H^0$ and fermionic DM $\psi_1$, respectively. Due to the presence of annihilation channels of these DM particles into the SM particles, and  allowing for the exchange of particles between $H^0$ and $\psi_1$ (depending on their respective masses), the BEQs governing the abundance of the two DM candidates are generally coupled. As a result, there is no simple formula that can be used to estimate the abundance of each individual DM component in this scenario, as was possible in the case of single component DM discussed earlier. Before moving into the specifics of the coupled BEQs, it is necessary to identify and categorize the DM conversion processes between $H^0$ and $\psi_1$, as shown in the Fig. \ref{Feyn-conv}.

The process of thermal freeze-out in a framework involving two-component DM is governed by a set of coupled BEQs \cite{Bhattacharya:2016ysw,Ahmed:2017dbb}. These can be written as
\begin{align}
    \frac{dn_{\Psi}}{dt}+3 \mathcal{H} n_{\Psi} & = - \langle \sigma v \rangle_{ \Psi }^{\rm eff} \left( n_{\Psi}^2 - (n_{\Psi}^{\rm{eq}})^2 \right) \nonumber \\
    & - \langle \sigma v \rangle_{\bar{\Psi} \Psi \rightarrow H H} \left( n_{\Psi}^2 - \frac{(n_{\Psi}^{\rm{eq}})^2}{(n_{H}^{\rm{eq}})^2}n_{H}^2 \right) \Theta \left( m_{\Psi} - m_{H} \right) \nonumber \\ 
    & + \langle \sigma v \rangle_{H H \rightarrow \bar{\Psi} \Psi } \left( n_{H}^2 - \frac{(n_{H}^{\rm{eq}})^2}{(n_{\Psi}^{\rm{eq}})^2}n_{\Psi}^2 \right) \Theta \left(  m_{H} - m_{\Psi} \right) , \nonumber \\
    \frac{dn_{H}}{dt} +3 \mathcal{H} n_{H} & = - \langle \sigma v \rangle_{H}^{\rm eff} \left( n_{H}^2 - (n_{H}^{\rm{eq}})^2 \right) \nonumber \\
    & +  \langle \sigma v \rangle_{\bar{\Psi} \Psi \rightarrow H H} \left( n_{\Psi}^2 - \frac{(n_{\Psi}^{\rm{eq}})^2}{(n_{H}^{\rm{eq}})^2}n_{H}^2 \right) \Theta \left( m_{\Psi} - m_{H} \right) \nonumber \\ 
    & - \langle \sigma v \rangle_{H H \rightarrow \bar{\Psi} \Psi } \left( n_{H}^2 - \frac{(n_{H}^{\rm{eq}})^2}{(n_{\Psi}^{\rm{eq}})^2}n_{\Psi}^2 \right) \Theta \left(  m_{H} - m_{\Psi} \right).
\end{align}

\begin{figure}[htb!]
    \centering
\includegraphics[width=0.55\linewidth]{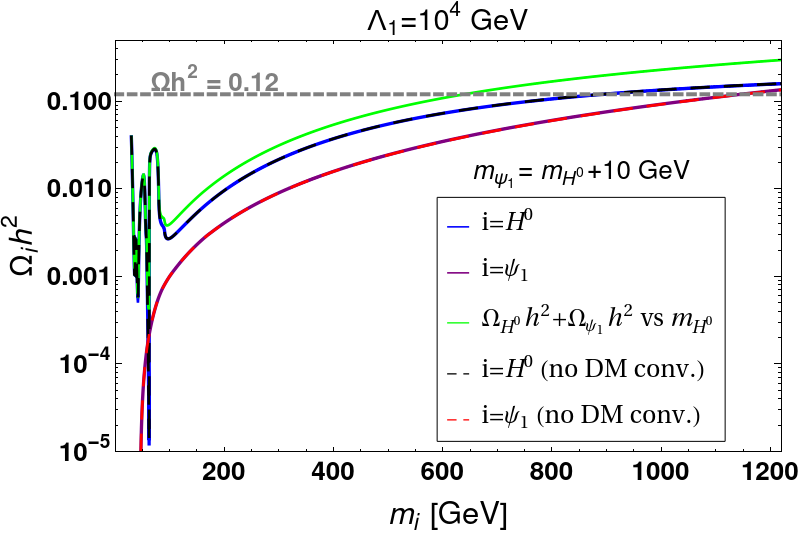}
    \caption{The dependence of the DM relic density on the corresponding DM mass in this two-component framework. The dotted lines indicate the absence of DM conversion. The black dashed line indicates the DM abundance measured by the Planck. For illustration, the other parameters are fixed as: $\{ m_{H^\pm}-m_{A^0}=1~{\rm{GeV}}, m_{A^0}-m_{H^0}=5~ {\rm{GeV}},~\lambda_L=0.01 \}$ (for scalar DM sector), $\Lambda = 10^6$ GeV (for fermion DM sector),  $\Lambda_1 = 10^4$ GeV and ~$ m_{\psi_1}=m_{H^0}+10$ GeV.}
\label{fig:omega_2cdm}
\end{figure}

The abundance of each DM component ($\Omega_i h^2$ with $i=\{H^0, \psi_1\}$) in the two-component framework is determined by solving the above coupled BEQs that govern the freeze-out of the individual components, taking into account annihilation, co-annihilation, and DM conversion processes. The total relic density of DM for the two-component case is $\Omega_{\rm{DM}}h^2$ given in Eq. \ref{eq:OmegaTotal}~ is the sum of the relic densities of the individual component in the interacting framework. In Fig.\ref{fig:omega_2cdm}, we show the variation of the individual and total abundances of DM as a function of the corresponding DM mass. The solid blue and purple lines correspond to the abundances of $H^0$ and $\psi_1$, respectively. The solid green line represents the total abundance of DM as a function of $m_{H^0}$, indicating that DM masses around 600 GeV meet the relic density requirement. The black and red dashed lines show the cases for $H^0$ and $\psi_1$, respectively, where the DM conversion ($\Psi~\Psi \to H~ H$ with $m_{\psi_1} > m_{H^0}$) is turned off. As these lines match those with conversion included, it indicates that the conversion process has a negligible impact on the DM relic abundance.

\begin{figure}[htb!]
    \centering
\includegraphics[width=0.47\linewidth]{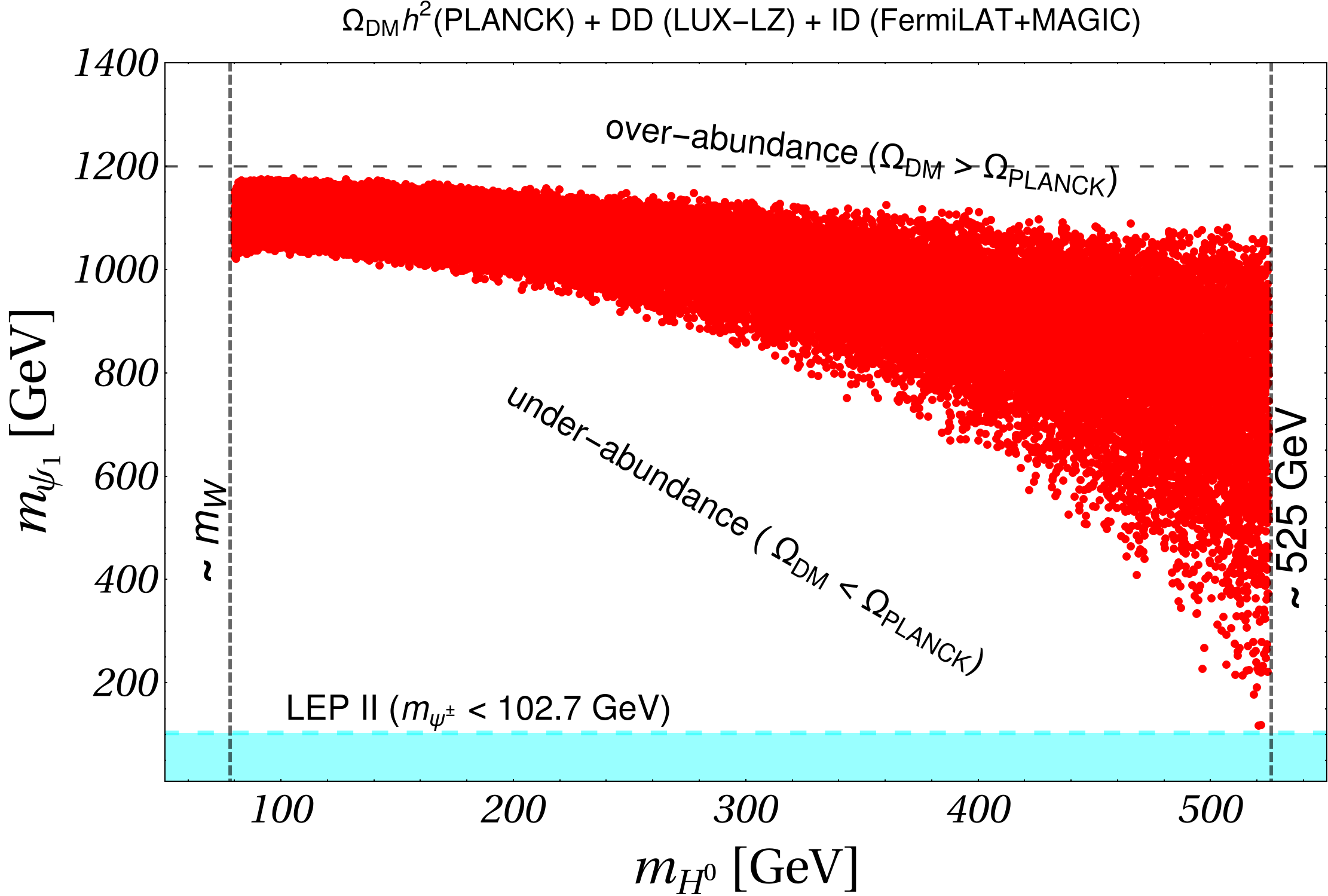} \qquad
\includegraphics[width=0.47\linewidth]{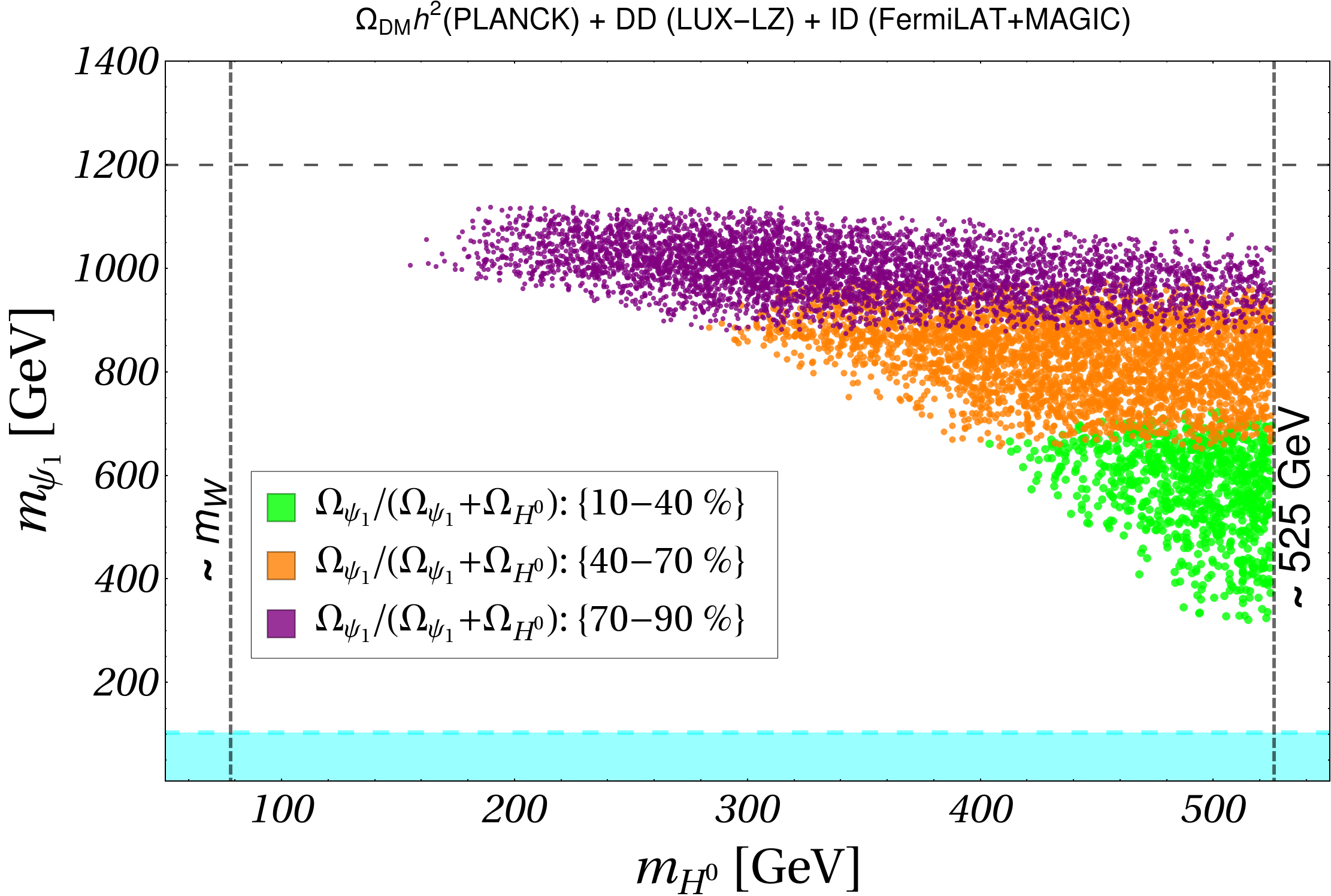}
    \caption{Left  panel: Allowed parameter space  for both DM masses satisfying correct relic density (PLANCK) + DD (LUX LZ 2024)+ ID (FermiLAT+MAGIC) observations. The region below the red points shows that both DM components are not sufficient to meet the observed relic density ($\Omega_{\rm DM} h^2 < \Omega_{\rm PLANCK} h^2$). And the region above the red points overshoots the observed DM abundance ($\Omega_{\rm DM} h^2 > \Omega_{\rm PLANCK} h^2$). The parameter space with $m_{\psi^\pm} \lesssim 102.7$ GeV is excluded by LEP \cite{L3:2001xsz} and is shown as the cyan area. Above the grey dashed line ($m_{\psi_1}=1200$ GeV), fermionic dark matter ($\psi_1$) is overproduced. Right panel: The relative fractional contribution of fermionic sector to the total relic density in the plane of the two DM masses. The region between the two vertical lines with $m_{W} \lesssim m_{H^0} \lesssim 525$ GeV, is our region of interest. To illustrate, we set the other parameters as: $(m_{H^\pm}-m_{A^0})\in [0,~5] ~{\rm GeV}, ~ (m_{A^0}-m_{H^0})\in [0.1,~10] ~ {\rm GeV},~\lambda_L \in [10^{-3}-0.1]$ and the new physics scales are set at $\Lambda =10^6$ GeV and $\Lambda_1 = 10^5$ GeV.}
\label{fig:twocomponent_allowedregion}
\end{figure}

The allowed points satisfying the correct relic density requirement for the combined DM relic density as well as both DD and ID experimental upper bounds in the $m_{\psi_1} - m_{H^0}$ plane are shown in the left panel of Fig. \ref{fig:twocomponent_allowedregion}. Here, we considered $(m_{H^\pm}-m_{A^0})\in [0,~5] ~{\rm GeV}, ~ (m_{A^0}-m_{H^0})\in [0.1,~10] ~ {\rm GeV}$ for the scalar doublet mass splittings, and the coupling $\lambda_L$ varied within $[10^{-3}-0.1]$. The new physics scales are set at $\Lambda = 10^6$ GeV and $\Lambda_1=10^5$ GeV.
In this simulation, we have considered specifically the mass region 80 GeV $\lesssim m_{H^0} \lesssim 525$ GeV and $m_{\psi_1} \lesssim 1.2$ TeV, where these DM candidates are individually under-abundant, but together they cover a significant portion of the allowed parameter space. For smaller mass splitting and smaller $m_{H^0}$, the larger co-annihilation and annihilation cross-section yields a quite suppressed scalar contribution, and this is  why we  needed the larger mass value for $\psi_1$ to saturate the relic density limit. When we go to larger $m_{H^0}$ masses, the scalar contribution to total combined relic density increases, so even very light values of $m_{\psi_1}$ can saturate the relic density requirement. 


In the right panel of Fig. \ref{fig:twocomponent_allowedregion}, we plot the relative
fractional contribution of fermionic sector to the total relic density in the plane of the two DM masses.
Here, three different colours represent relative fermionic relic density contributions with respect to the total relic density in the framework i.e., $\Omega_{\psi_1}/(\Omega_{\psi_1} + \Omega_{H^0}) = 10 - 40 \%$ (green), $40 - 70 \%$ (orange), $70 - 90 \%$ (purple). For heavier $\psi_1 \sim$ TeV, the fractional fermionic contribution becomes quite dominant (purple points), so entire range of $m_{H^0}$, individually under-abundant, can contribute to saturate the relic density criteria, depending on the corresponding mass splittings between $H^0$ and $A^0$. On the other hand, for lighter $\psi_1 \sim 400 - 700$ GeV (green points), the relative contribution from fermionic sector falls below 40 \%, so  heavier $H^0$ are needed to saturate the remaining $> 60\%$ part of the required relic density. The intermediate orange region represents the mid-region between these two extreme scenarios. 

In summary, we have now found a significant parameter space in the $m_{\psi_1} - m_{H^0}$ plane which eventually saturates all the relevant criterion in DM phenomenological study like correct relic density, DD as well as ID observations and this two-component DM scenario can be a viable DM option within the interesting mass range for 80 GeV $\lesssim m_{H^0} \lesssim 525$ GeV and $m_{\psi_1} \lesssim 1.2$ TeV, where these single component DMs individually fail to accommodate the observed relic density.


\subsection{Two Component DM : UV Complete Theory}
\label{subsec:UVcomplete}
To construct a UV complete model of the above-mentioned two-component DM framework, one can introduce an additional $SU(2)_L$ triplet scalar, $\Delta$, with $Y = 2$ where the new physics scale introduced is represented by this new scalar mass and couplings with other particles. This also allows generation of neutrino masses and mixing through the type-II seesaw. The charge assignment of such $SU(2)_L$ triplet scalar under the gauge group $\mathcal{G} \equiv \mathcal{G}_{\rm SM} \otimes \mathcal{G}_{\rm DM}$  where $\mathcal{G}_{\rm SM}\equiv SU(3)_C \otimes SU(2)_L \otimes U(1)_Y$ and $\mathcal{G}_{\rm DM} \equiv \mathcal{Z}_2 \otimes \mathcal{Z}^\prime_2 $ is presented in Tab. \ref{tab:tab2} (where the charge assignment, under this gauge group, of new fermion and scalar doublets as well as SM Higgs doublet is as given in Table \ref{tab:tab1}).
\begin{table}[htb!]
 \begin{tabular}{|c|c|c|c|}
\hline \multicolumn{2}{|c}{Fields}&  \multicolumn{1}{|c|}{ $\underbrace{ SU(3)_C \otimes SU(2)_L \otimes U(1)_Y}$ $\otimes \underbrace{ \mathcal{Z}_2 \otimes \mathcal{Z}^\prime_2 }$} \\ \hline
Scalar Triplet & $\Delta=\left(\begin{matrix} \frac{\delta^+}{\sqrt{2}} & \delta^{++} \\ \frac{1}{\sqrt{2}}(v_\Delta + \delta^0 + i \delta_I) & -\frac{\delta^+}{\sqrt{2}} \end{matrix}\right)$ & ~~1 ~~~~~~~~~~~3~~~~~~~~~~2~~~~~~~~~~+~~~~~~~+ \\
\hline
\end{tabular}
\caption{Charge assignment of $SU(2)_L$ triplet scalar field $\Delta$ under the gauge group $\mathcal{G} \equiv \mathcal{G}_{\rm SM} \otimes \mathcal{G}_{\rm DM}$.  }
    \label{tab:tab2}
\end{table}

The corresponding interaction Lagrangian is given by
\begin{equation}
    \mathcal{L}^{\rm{UV}} \supset  \mathcal{L}^{\rm Scalar} + \mathcal{L}^{\rm Yuk.}_\Delta + \mathcal{L}^{\rm FD~ DM}_{\rm{UV}} + \mathcal{L}^{\rm SD~ DM}_{\rm{UV}}.
\end{equation}
The $\mathcal{L}^{\rm{Scalar}}$ term is made up of kinetic terms for SM Higgs doublet and triplet scalar as well as their corresponding scalar potential term \cite{Arhrib:2011uy},
\begin{equation}
   \mathcal{L}^{\rm scalar}=\left(D^{\mu} \Phi \right)^{\dagger} \left(D_{\mu}  \Phi \right) + {\rm Tr} \left[\left(D^{\mu}\Delta\right)^{\dagger}\left(D_{\mu}\Delta\right)\right] - V(\Delta, \Phi), 
\end{equation}
with the covariant derivatives for both doublet and triplet scalar are defined as,
\begin{align}
    & D_{\mu} \Phi = \Big(\partial_\mu - i g_2 \frac{\sigma^a}{2} W_{\mu}^a-i g_1 \frac{Y_{\Phi}}{2} B_{\mu}\Big) \Phi , \nonumber \\
    & D_{\mu}\Delta = \partial_{\mu}\Delta - i g_2 \left[\frac{\sigma^{a}}{2} W_{\mu}^a ,\Delta\right] - {i g_1} \frac{Y_\Delta}{2} B_{\mu}\Delta ~.
\end{align}
where $\sigma^a$ is the Pauli matrix and $g_2$ and $g_1$ are coupling constants for the gauge groups $SU(2)_L$ and $U(1)_Y$, respectively. The most general renormalizable scalar potential of this model can be written as,
\begin{align}
   V(\Delta,\Phi) & =  -\mu_{\Phi}^2 (\Phi^\dagger \Phi)+ \lambda_{\Phi} (\Phi^\dagger \Phi)^2 + \mu_{\Delta}^2 {\rm Tr}\left[\Delta^{\dagger}\Delta\right]  + \lambda_{1}^\prime \left(\Phi^{\dagger} \Phi \right){\rm Tr}\left[\Delta^{\dagger}\Delta\right] + \lambda_2^\prime \left({\rm Tr}[\Delta^{\dagger}\Delta]\right)^2 \nonumber \\
   & + \lambda_3^\prime {\rm Tr} [\left(\Delta^{\dagger}\Delta\right)^2] + \lambda_4^\prime \left(\Phi^{\dagger}\Delta\Delta^{\dagger} \Phi \right)+ \left[\mu_1\left(\Phi^T i \sigma^2 \Delta^{\dagger} \Phi \right)+ h.c.\right]~.
   \label{eq:Vscalar}
\end{align}

\noindent With $\mu_\Delta^2 > 0$, the triplet scalar $\Delta$ does not acquire any \textit{vev}. However, after EWSB, the cubic term $\Phi^T i \sigma^2 \Delta^{\dagger}\Phi$ generates a small induced \textit{vev} $v_\Delta$. In alignment limit, we have $\sqrt{v^2+2 v_\Delta^2} = 246$ GeV with $v \gg v_\Delta$. Minimization of the scalar potential around the \textit{vev}s ($v$ and $v_\Delta$) leads to the following relations:
\begin{eqnarray}
    \frac{\mu_1 v^2}{\sqrt{2} v_\Delta} \equiv M_\Delta^2 &=& \mu_\Delta^2+v_\Delta^2 (\lambda^\prime_2+\lambda^\prime_3)+\frac{1}{2} v^2 (\lambda^\prime_1+\lambda^\prime_4) ~,\nonumber \\
    \mu_H^2 &=& -\frac{2 M_\Delta^2 v_\Delta^2 }{v^2} +\frac{1}{2} v_\Delta^2 (\lambda^\prime_1+\lambda^\prime_4)+\lambda_H v^2 .\nonumber 
\end{eqnarray}

\noindent The masses for other physical Higgs eigenstates can be written as,
\begin{align}
    & m_{hh}^2 = \left(M_\Delta^2+2 v_\Delta^2 (\lambda^\prime_2+\lambda^\prime_3)\right) \sin ^2\alpha + 2 \lambda_\Phi v^2 \cos ^2\alpha-\frac{v_\Delta \sin2\alpha \left(2 M_\Delta^2 - v^2 (\lambda^\prime_1+\lambda^\prime_4)\right)}{v}   \nonumber \\
    & m_{HH}^2 = \left(M_\Delta^2+2 v_\Delta^2 (\lambda^\prime_2+\lambda^\prime_3)\right) \cos ^2\alpha + 2 \lambda_\Phi v^2 \sin^2\alpha+\frac{v_\Delta \sin2\alpha \left(2 M_\Delta^2 - v^2 (\lambda^\prime_1+\lambda^\prime_4)\right)}{v} \nonumber\\
    & m_{AA}^2 = \frac{M_\Delta^2}{v^2} \left(4 v_\Delta^2+v^2\right) \nonumber\\
    & m_{\Delta^\pm}^2 = \frac{1}{4 v^2} \left(2 v_\Delta^2+v^2\right) \left(4 M_\Delta^2 - \lambda^\prime_4 v^2\right) \nonumber \\
    & m_{\Delta^{\pm\pm}}^2 = M_\Delta^2-\lambda^\prime_3 v_\Delta^2-\frac{\lambda^\prime_4 v^2}{2}.
\end{align}
Here, $hh, HH$ are the SM Higgs and heavier CP-even Higgs, respectively, while $AA$ is the CP-odd physical Higgs. Consequently, the singly and doubly charged Higgs are $\Delta^\pm$ and $\Delta^{\pm\pm} \equiv \delta^{\pm\pm}$. These different physical Higgs have been originated via the mixing between SM Higgs doublet $\Phi$ and triplet scalar $\Delta$, where this mixing is characterized by the mixing angle $\alpha$ \cite{Arhrib:2011uy}. For a detailed discussion of the scalar potential with the SM Higgs doublet and triplet scalar, see \cite{Ghosh:2022fzp}.
Precision measurements of the electroweak $\rho$ parameter ($\rho=1.00031 \pm 0.00019 $ \cite{ParticleDataGroup:2024cfk}) impose a upper limit on $v_\Delta \lesssim 2.8$ GeV with $3\sigma$ C.L.. The triplet scalar can generate the Majorana mass for neutrinos. The Yukawa Lagrangian $\mathcal{L}^{\rm Yuk.}_\Delta = -{{ y_{L} \overline{L^c}i \sigma^2\Delta L }} +h.c.$, is responsible for generating neutrino masses in this framework, commonly referred to as the type-II Seesaw mechanism. Neutrino masses are generated via a non-zero \textit{vev} of the neutral component of $\Delta$. In this scenario, the neutrino mass term turns out to be
\begin{equation}
    m_{\alpha \beta}^\nu =   \frac{v_\Delta}{\sqrt{2}} ~(y_L)_{\alpha \beta} ~,
    \label{eq:numass}
\end{equation} 
and this mass matrix $ m_{\alpha \beta}^\nu$ in the flavor basis must be diagonalised in order to obtain the physical neutrino masses through the Pontecorvo-Maki-Nakagawa-Sakata (PMNS) matrix $U_{\text{PMNS}}$. For $v_\Delta$ to be in the range of few MeV $- ~\mathcal{O}(1)$ GeV, generating light neutrino masses, as in Eq. \ref{eq:numass}, in the correct ballpark, $y_L$ needs to be highly fine-tuned to be $\lesssim 10^{-8}$.

Furthermore, the $\mathcal{L}^{\rm{FD~DM}}_{\rm{UV}}$ term consists of the kinetic term for fermionic doublet $\Psi$ and its interaction with triplet scalar, it can be written as \cite{Ghosh:2023dhj,Barman:2019tuo},
\begin{equation}
    \mathcal{L}^{\rm FD~DM}_{\rm{UV}} = \mathcal{L}^{\rm{VLFD}} -  {{ \frac{y_{\psi}}{\sqrt{2}} \overline{\Psi^c}i \sigma^2\Delta \Psi+h.c. }},
\end{equation}

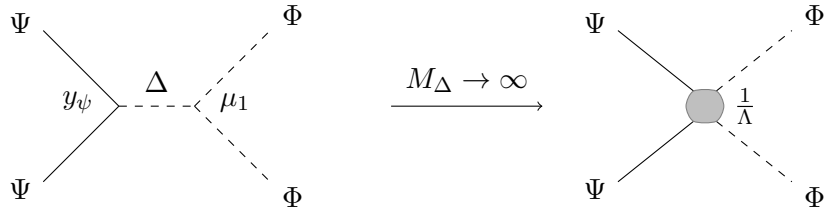
\begin{figure}[htb!]
 \begin{center}
\begin{tikzpicture}[line width=0.4 pt, scale=1.0]
\draw[solid] (2.4,1)--(3.4,0);
\draw[solid] (2.4,-1)--(3.4,0);
\draw[dashed] (3.4,0)--(4.4,0);
\draw[dashed] (4.4,0)--(5.4,1);
\draw[dashed] (4.4,0)--(5.4,-1);
\node  at (2.1,-1.1) {$\Psi$};
\node at (2.1,1.1) {$\Psi$};
\node [left] at (3.2,0.05) {$y_\psi$};
\node [above] at (3.9,0.05) {$\Delta$};
\node at (5.7,1.2){$\Phi$};
\node at (5.7,-1.2){$\Phi$};
\node [right] at (4.6,0.05) {$\mu_1$};

\draw[->] (7,0) -- (9,0);
\node [above] at (8,0.05) {$M_\Delta \rightarrow \infty$};

\draw[solid] (10,1)--(11,0.2);
\draw[solid] (10,-1)--(11,-0.2);
\draw[dashed] (11.3,0.2)--(12.3,1);
\draw[dashed] (11.3,-0.2)--(12.3,-1);
\node  at (9.7,-1.1) {$\Psi$};
\node at (9.7,1.1) {$\Psi$};
\node at (12.6,1.2){$\Phi$};
\node at (12.6,-1.2){$\Phi$};
\draw[fill=gray!50, draw=gray] plot [smooth cycle] coordinates 
    {(11,0.2) (11.3,0.2) (11.4, 0)  (11.3,-0.2) (11,-0.2) (10.9, 0)};
\node [right] at (11.4,0.0) {$\frac{1}{\Lambda}$};
\end{tikzpicture}
 \end{center}
\caption{Left panel: Tree-level Feynman diagram for $\Psi - \Phi$ interaction mediated by triplet scalar $\Delta$ in the UV complete model, characterized by Yukawa-type coupling $y_\psi$ and trilinear Higgs coupling $\mu_1$. Right panel: for $M_\Delta \to \infty$, we can integrate out the triplet scalar and recover the corresponding dim-5 EFT operator  $\frac{1}{\Lambda} \overline{\Psi^c} \tilde{\Phi}~\tilde{\Phi}^\dagger \Psi$, where the blob corresponds to the EFT framework, characterized by the new physics scale $\Lambda$.}
\label{fig:EFT}
 \end{figure}

\noindent As discussed earlier, the presence of the triplet scalar plays a crucial role in evading gauge-mediated DM–nucleon elastic scattering by generating a pseudo-Dirac mass splitting of $\psi^0$. The mass splitting between the two pseudo-Dirac states, $\psi_{1,2}$ turns out to be: $\delta m_{12}=m_{\psi_2}-m_{\psi_1}=2 y_\psi v_\Delta$, followed by the Eqs. \ref{eq:pstate0} and \ref{eq:pstate}.
The corresponding effective dim-5 interaction Lagrangian between $\Psi$ and SM Higgs can be generated from the combination of $\Psi-\Delta$ as well as the trilinear Higgs interaction with dimensionful coupling constant $\mu_1$ (as shown in Fig. \ref{fig:EFT}) i.e., $|\Lambda| = \frac{M_{\Delta}^2}{y_\psi \mu_1}$.

\begin{figure}[htb!]
 \centering
    \includegraphics[scale=0.5]{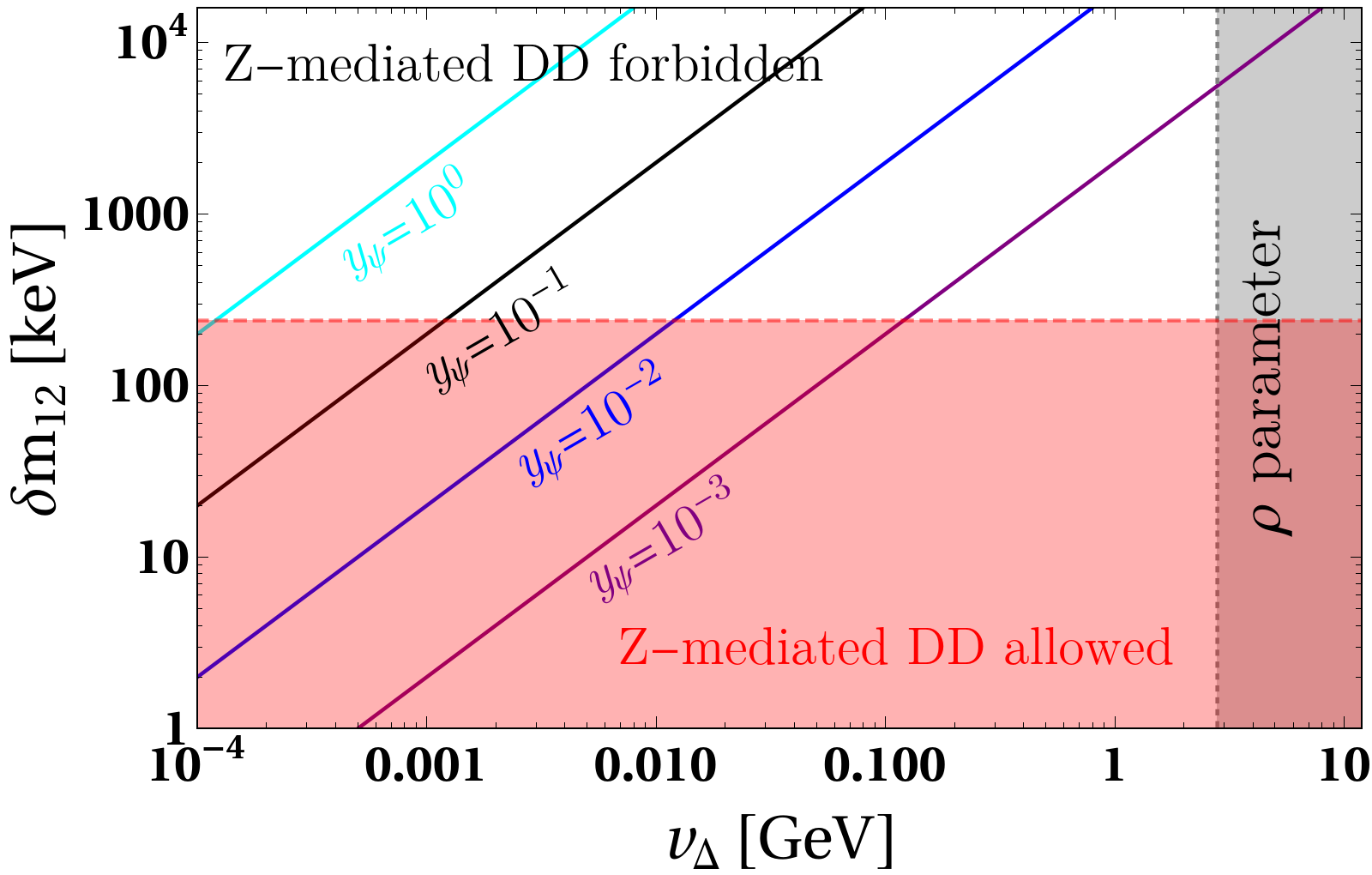}  
\caption{$\delta m_{12}=m_{\psi_2}-m_{\psi_1}$ as a function of $v_\Delta$ with different values of $y_\psi~:$ $10^0$ (cyan), $10^{-1}$ (black), $10^{-2}$ (blue) and $10^{-3}$ (purple). The red-shaded region is excluded by the Z-mediated direct search constraint. The gray-shaded vertical region, $v_\Delta \gtrsim 2.8$ GeV, is forbidden due to constraints from the EW $\rho$ parameter \cite{ParticleDataGroup:2024cfk}. The white region which is our region of interest, provides an opportunity to evade the Z-mediated direct detection limits on fermion doublet DM.}
\label{fig:DD}
 \end{figure}

We show the splitting between two pseudo-Dirac neutral $\mathcal{Z}_2$-odd fermion state masses as $\delta m_{12} = m_{\psi_2} - m_{\psi_1}$ with respect to $v_\Delta$ in  Fig. \ref{fig:DD}. As shown previously, in  this  approach, introducing the heavy triplet scalar $\Delta$ eventually forbids the elastic DM-nucleon scattering diagram mediated by the Z boson, which, in turn, evades the Z-mediated DD constraints. However, in this framework,  elastic scattering of DM-nucleon via the SM Higgs occurs. In this figure, the red shaded region represents the parameter space where the Z-mediation DD process is allowed,  while the white region shows the region where Z-mediation DD is forbidden, which is the favourable region  considered in our  scenario. The vertical grey shaded region is forbidden by $\rho$ parameter measurement in this multi-Higgs scenario i.e., $v_\Delta$ should be $< 2.8$ GeV. We have considered four different values for the Yukawa coupling, $y_\psi = 10^{-3}, 10^{-2}, 10^{-1}, 10^{0}$,  represented by purple, blue, black and cyan solid lines in the figure. Notice that, to evade the Z-mediated DD constraints, we can put lower bounds on the corresponding Yukawa couplings : (i) when $y_\psi = 10^{0}$,  $v_\Delta$ should be $> 10^{-4}$ GeV, (ii) when $y_\psi = 10^{-1}$,  $v_\Delta$ should be $> 0.001$ GeV, (iii) when $y_\psi = 10^{-2}$,  $v_\Delta$ should be $> 0.01$ GeV, (iv) when $y_\psi = 10^{-3}$,  $v_\Delta$ should be $> 0.1$ GeV. This behaviour is  evident as, the mass splittings between these neutral pseudo-Dirac states is mostly governed by $y_\psi v_\Delta$, so eventually, for a fixed mass splittings, we need lower (higher) values of $y_\psi$ for higher (lower) values of $v_\Delta$, which  justifies the behaviour for different curves, and imposes a lower bound on $v_\Delta$ to evade the DD constraints. This allows a choice for $y_\psi$ as well as for $v_\Delta$. 

The Lagrangian term $\mathcal{L}^{\rm{SD~DM}}_{\rm{UV}}$ for scalar ISD and its interaction with other scalar particles in the theory can be written as,
\begin{align}
    \mathcal{L}^{\rm SD~DM}_{\rm{UV}} & = \mathcal{L}^{\rm{ISD}} - V(H,\Phi) - \Big(\lambda_{\Delta H}{\left(H^{\dagger}H \right){\rm Tr}\left[\Delta^{\dagger}\Delta\right]
    + \mu_D (H^T i \sigma^2 \Delta^\dagger H) + \lambda_{4D} H^\dagger \Delta \Delta^\dagger H } \Big)
\end{align}

\noindent For simplicity, we have considered the couplings $\lambda_{4D}, \lambda_{\Delta H}$ to be zero throughout the work. This choice does not further reduce the abundance of scalar doublet DM ($H^0$). However, the non-zero value of $\mu_D$ plays an important role in contributing to the signature of the doubly charged triplet scalar $\Delta^{\pm\pm}$. In the presence of the $\mu_D$ term, the mass eigenvalues (Eq. \ref{eq:Phieigenstates}) will be slightly modified as,
\begin{align}
    & m_{H^0}^2 = \mu_H^2 + \frac{1}{2}v^2 (\lambda_1 + \lambda_2 + \lambda_3) - \sqrt{2}v_\Delta \mu_D, \nonumber \\
    & m_{A^0}^2 = \mu_H^2 + \frac{1}{2}v^2 (\lambda_1 + \lambda_2 - \lambda_3) + \sqrt{2}v_\Delta \mu_D, \nonumber \\
    & m_{H^\pm}^2 = \mu_H^2 + \lambda_1 v^2.
    \label{eq:Phineweigenstates}
\end{align}

Similar as Fig. \ref{fig:EFT}, the other new effective coupling $\frac{1}{\Lambda_1} \overline{\Psi^c} \tilde{H} ~\tilde{H}^\dagger \Psi$ can be generated via the interactions between $\Psi-\Delta$ as well as between $H-\Delta$, considering $\Lambda_1 = M_{\Delta}^2 / y_\psi \mu_D$. Also, this new triplet scalar can generate Majorana mass for neutrinos considering the type-II seesaw interaction term $y_L \overline{L^c} i\sigma^2 \Delta L$, which can also give rise to the effective operators like $\frac{1}{\Lambda_H} \overline{L^c} \tilde{H} ~\tilde{H}^\dagger L$ as well as $\frac{C_W}{\Lambda_W} \overline{L^c}\tilde{\Phi}~\tilde{\Phi}^\dagger L$ where $\Lambda_H = M_\Delta^2 / y_L \mu_D$ and $\Lambda_W = \frac{C_W M_\Delta^2}{y_L \mu_1}$, respectively.

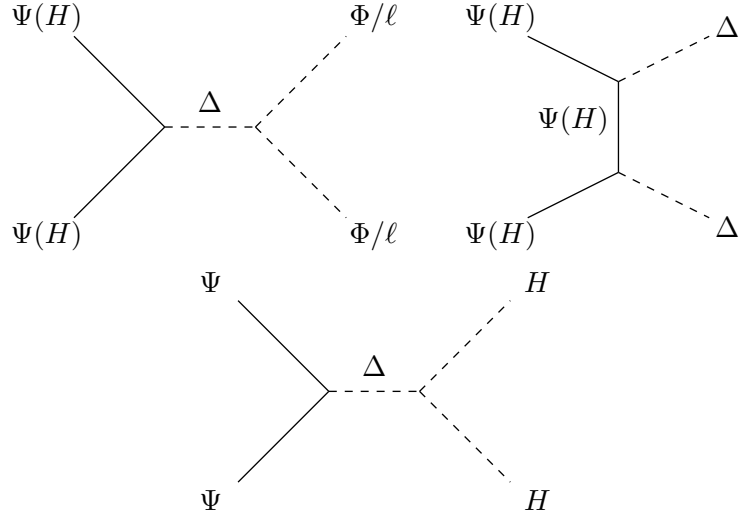
\begin{figure}[htb!]
  \centering
    \begin{tikzpicture}[line width=0.5 pt, scale=1.2]
      \draw[solid] (-3.6,1)--(-2.6,0);
      \draw[solid] (-3.6,-1)--(-2.6,0);
      \draw[dashed] (-2.6,0)--(-1.6,0);
      \draw[dashed] (-1.6,0)--(-0.6,1);
      \draw[dashed] (-1.6,0)--(-0.6,-1);
      \node  at (-3.9,-1.2) {$\Psi (H) $};
      \node at (-3.9,1.2) {$\Psi (H)$};
      \node [above] at (-2.1,0.05) {$\Delta$};
      \node at (-0.3,1.2){$\Phi / \ell$};
      \node at (-0.3,-1.2){$\Phi  / \ell$};

      \draw[solid] (1.4,1.0)--(2.4,0.5);
      \draw[solid] (1.4,-1.0)--(2.4,-0.5);
      \draw[solid] (2.4,0.5)--(2.4,-0.5);
      \draw[dashed] (2.4,0.5)--(3.4,1.0);
      \draw[dashed] (2.4,-0.5)--(3.4,-1.0);
      \node at (1.1,1.2) {$\Psi (H)$};
      \node at (1.1,-1.2) {$\Psi (H)$};
      \node at (1.9,0.07) {$\Psi (H)$};
      \node at (3.6,1.1) {$\Delta$};
      \node at (3.6,-1.1) {$\Delta$};
    \end{tikzpicture}

    \begin{tikzpicture}[line width=0.5 pt, scale=1.2]
      \draw[solid] (-3.6,1)--(-2.6,0);
      \draw[solid] (-3.6,-1)--(-2.6,0);
      \draw[dashed] (-2.6,0)--(-1.6,0);
      \draw[dashed] (-1.6,0)--(-0.6,1);
      \draw[dashed] (-1.6,0)--(-0.6,-1);
      \node  at (-3.9,-1.2) {$\Psi $};
      \node at (-3.9,1.2) {$\Psi$};
      \node [above] at (-2.1,0.05) {$\Delta$};
      \node at (-0.3,1.2){$H$};
      \node at (-0.3,-1.2){$H$};
    \end{tikzpicture}
\caption{Upper panel:  (left)  the Feynman diagram for DM DM $\to \Phi \Phi$ or $\ell \ell$ via $s$-channel $\Delta$ mediation where $\Phi$ is the SM Higgs doublet;  (right) the DM DM $\to \Delta \Delta$ annihilation via $t$-channel process. Lower panel: Feynman diagram for the conversion process between fermionic ($\Psi$) and scalar ($H$) DM in presence of $\Delta$.}
\label{fig:DMannwithDelta}
\end{figure}

First, we study the single-component scenario (DM$=\psi_1$ or $H^0$) with the triplet scalar $\Delta$, followed by the complete two-component scenario. The DM annihilation channels mediated by the scalar triplet $\Delta$ are shown in Fig.~\ref{fig:DMannwithDelta}. While the left diagram in the upper panel corresponds to the $s$-channel process into the SM Higgs doublet $\Phi$ or leptons, the right one shows the $t$-channel annihilation into a pair of triplet scalars. The conversion diagram between fermionic ($\Psi$) and scalar ($H$) DM in the presence of the triplet $\Delta$ is shown in the lower panel of Fig.~\ref{fig:DMannwithDelta}. The presence of the triplet scalar and its interactions with DM particles lead to suppression of relic density. This is because when $m_{\rm DM} > M_\Delta$, the additional process ${\rm DM~DM} \rightarrow \Delta \Delta$, and when $m_{\rm DM} \sim M_\Delta/2$, the additional $\Delta$ mediated  process ${\rm DM~DM} \rightarrow {\rm SM~SM}$ will contribute to the relic density and the corresponding BEQ  will be modified as follows :
\begin{equation}
    \frac{d n_{\text{DM}}}{dt} + 3 \mathcal{H} n_{\text{DM}}  = - [ \langle \sigma v \rangle^{\text{eff}}_{{\rm DM~DM} \rightarrow \text{SM~SM}} + \langle \sigma v \rangle^{\text{eff}}_{{\rm DM~DM} \rightarrow \Delta \Delta} ] \left( n_{\text{DM}}^2 - n_{\text{DM}}^{\text{eq}^2} \right)~.
\end{equation}
The corresponding relic density of the DM is approximately expressed as
\begin{equation}
    \Omega_{\rm DM} h^2 \propto [\langle \sigma v \rangle^{\text{eff}}_{{\rm DM~DM} \rightarrow \text{SM~SM}} + \langle \sigma v \rangle^{\text{eff}}_{{\rm DM~DM} \rightarrow \Delta \Delta}]^{-1} .
\end{equation}
\begin{figure}[htb!]
\includegraphics[scale=0.30]{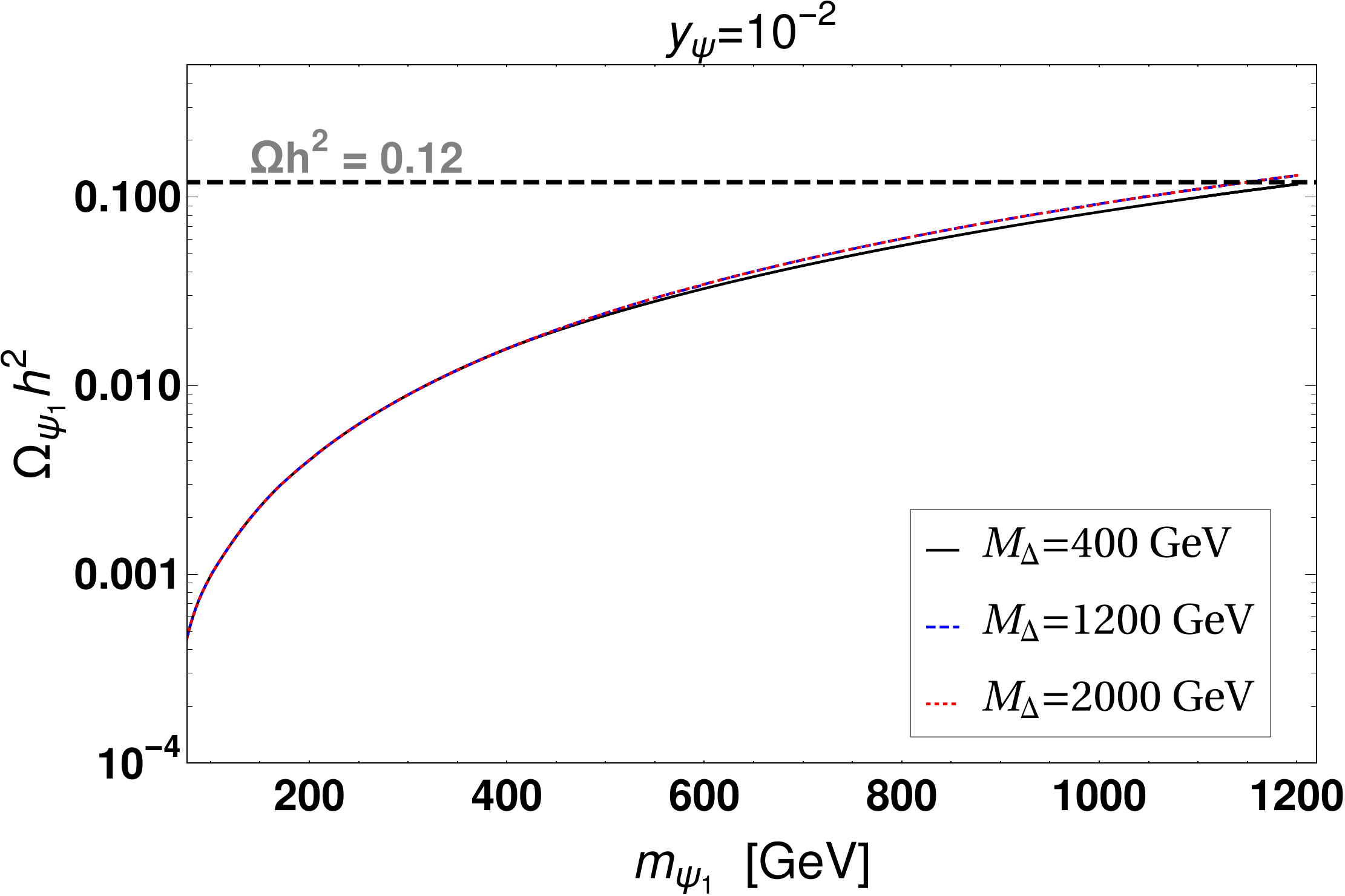}\qquad
\includegraphics[scale=0.30]{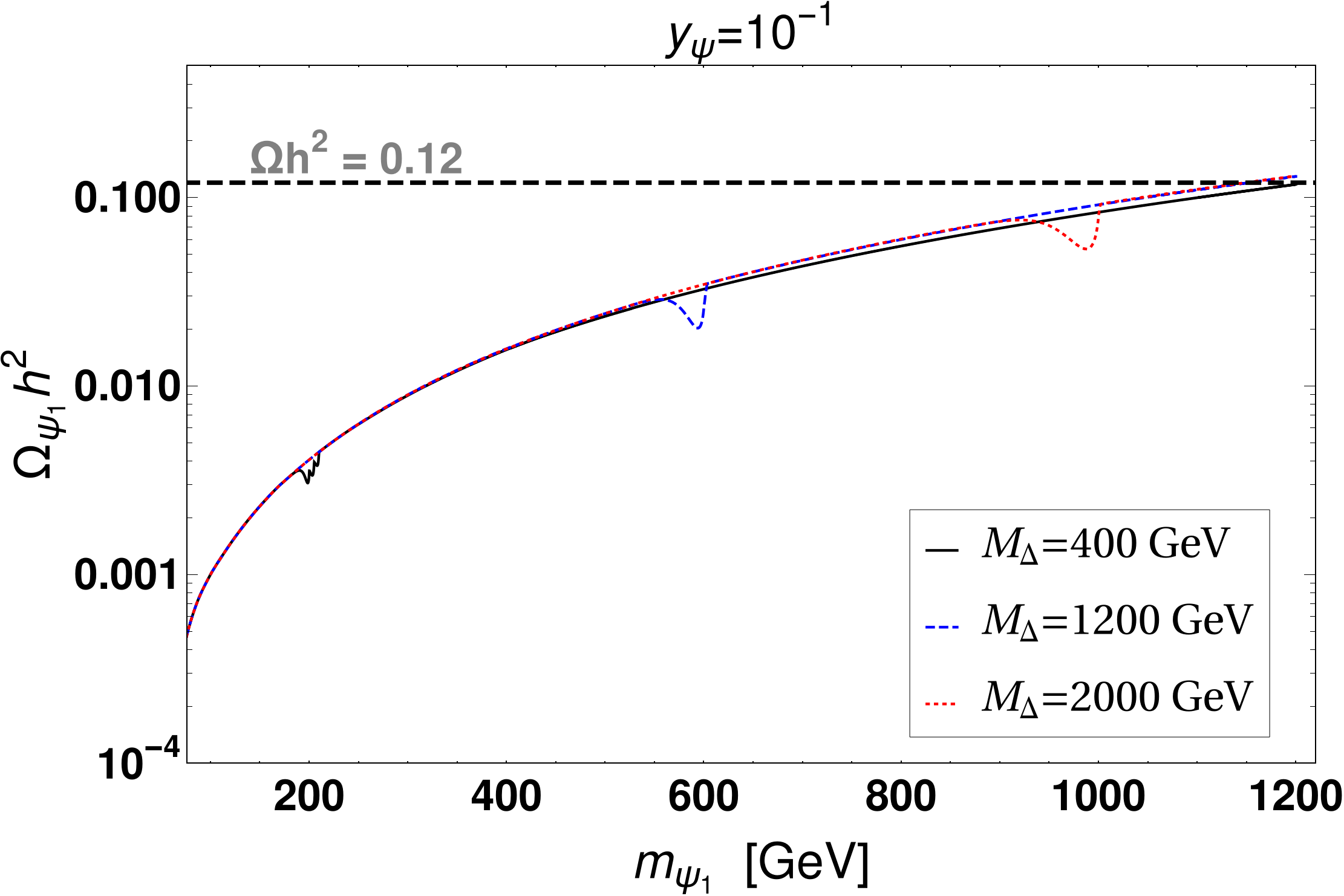}\qquad
\includegraphics[scale=0.35]{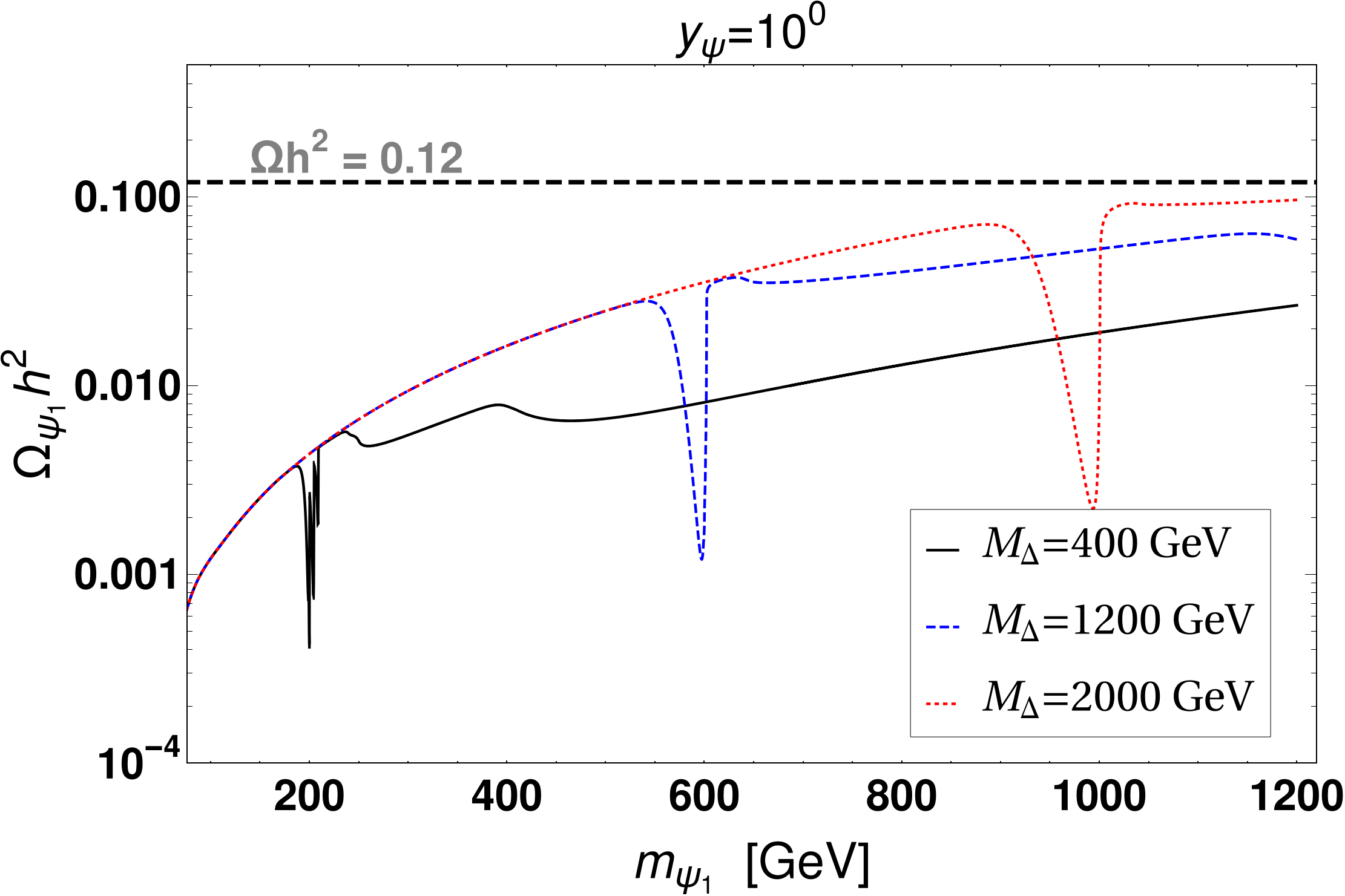}
\caption{The dependence of fermionic relic density with respect of fermionic DM mass while varying the Yukawa couplings $y_\psi = 10^{-2}$ (left), $10^{-1}$ (right), $10^{0}$ (bottom) as well as the triplet sclar mass parameter $M_{\Delta} = 400$ GeV (black solid), 1200 GeV (blue dashed) and 2000 GeV (red dashed). For illustration, we fix the benchmark point (BP) for the triplet scalar as : $\{\lambda_\Phi=0.129,~\lambda_1^\prime=0.2,~\lambda_2^\prime=0.0,~\lambda_3^\prime=0.95,~\lambda_4^\prime=-0.55,~v_\Delta=1.0 ~{\rm GeV}~\}$.}
\label{fig:SC_fermion}
\end{figure}

As the presence of the triplet scalar $\Delta$ and its mass parameter eventually affects the relic density from the fermion sector, in Fig. \ref{fig:SC_fermion} we show the variation of the relic density with respect to the fermion DM masses for three different values of the Yukawa couplings, $y_\psi = 10^{-2}, 10^{-1} ~\text{and}~ 10^0$ in the top left, top right and bottom panels, respectively. In each plot, we  varied the $M_\Delta$ and shown the dependence of the relic abundance on triplet scalar mass for three representative values of $M_\Delta = 400$ GeV (black solid), 1200 GeV (blue dashed) and 2000 GeV (red dashed). We took $v_\Delta$ to remain $\mathcal{O}(1)$ GeV for plotting these contours. Note that the annihilation channels into the Higgs bosons ($\Psi ~\Psi \to \Phi ~\Phi$) are mediated by the scalar triplet $\Delta$ through the $s$-channel diagrams. This diagram nominally depends on both the Yukawa coupling $y_\psi$ and the trilinear scalar coupling $\mu_1$. However, due to the $1/M_\Delta^2$ dependence in the $s$-channel $\Delta$ propagator and the definition of $\mu_1$ (followed by the scalar potential minimization conditions, $\frac{\mu_1 v^2}{\sqrt{2} v_\Delta} = M_\Delta^2 $), the effective strength of the annihilation amplitude becomes mainly controlled by $y_\psi$ and the scalar triplet VEV $v_\Delta$. For smaller values of $y_\psi \lesssim 10^{-2}$, the triplet scalar is only lightly coupled to the fermionic sector, and thus it does not affect the fermionic relic density much. The contribution of $\Delta$ to fermionic relic density can be safely ignored irrespective of $M_\Delta$ values. But if we move towards the higher values of $y_\psi \geq 10^{-1}$, one can  see that dips in the contours  arise due to the opening of new channels in the annihilation cross-section mediated by $\Delta$ around $m_{\psi_1} \sim M_\Delta/2$. This effect becomes more bizarre if we go to $y_\psi \sim 1$ or so. Therefore, to evade these new unwanted contributions on the fermionic relic density, one can choose two different scenarios : either (i) a smaller value of $y_\psi \lesssim 10^{-2}$ for $v_\Delta \sim \mathcal{O}(1)$ GeV (or vice-versa, as the production cross-section of $\Delta$ from $\psi_1$ self-annihilation is proportional to $(y_\psi v_\Delta)^2$), and be able to consider any value of $M_\Delta$, or (ii) larger values of $y_\psi$ and $v_\Delta$ but allowing only higher values of $M_\Delta \sim$ far above the TeV scale. Furthermore, it is important to emphasize that the observed reduction in the relic abundance for dark matter masses satisfying $m_{\psi_1} > M_\Delta$ arises from the kinematic opening of the annihilation channel 
$\Psi ~ \Psi \to \Delta \Delta$. Once this threshold is crossed, the pair production of scalar triplets becomes accessible, significantly enhancing the total annihilation cross section. This results in a more efficient depletion of dark matter during freeze-out, thereby lowering the predicted relic density in this mass regime.

\begin{figure}[htb!]
\includegraphics[scale=0.30]{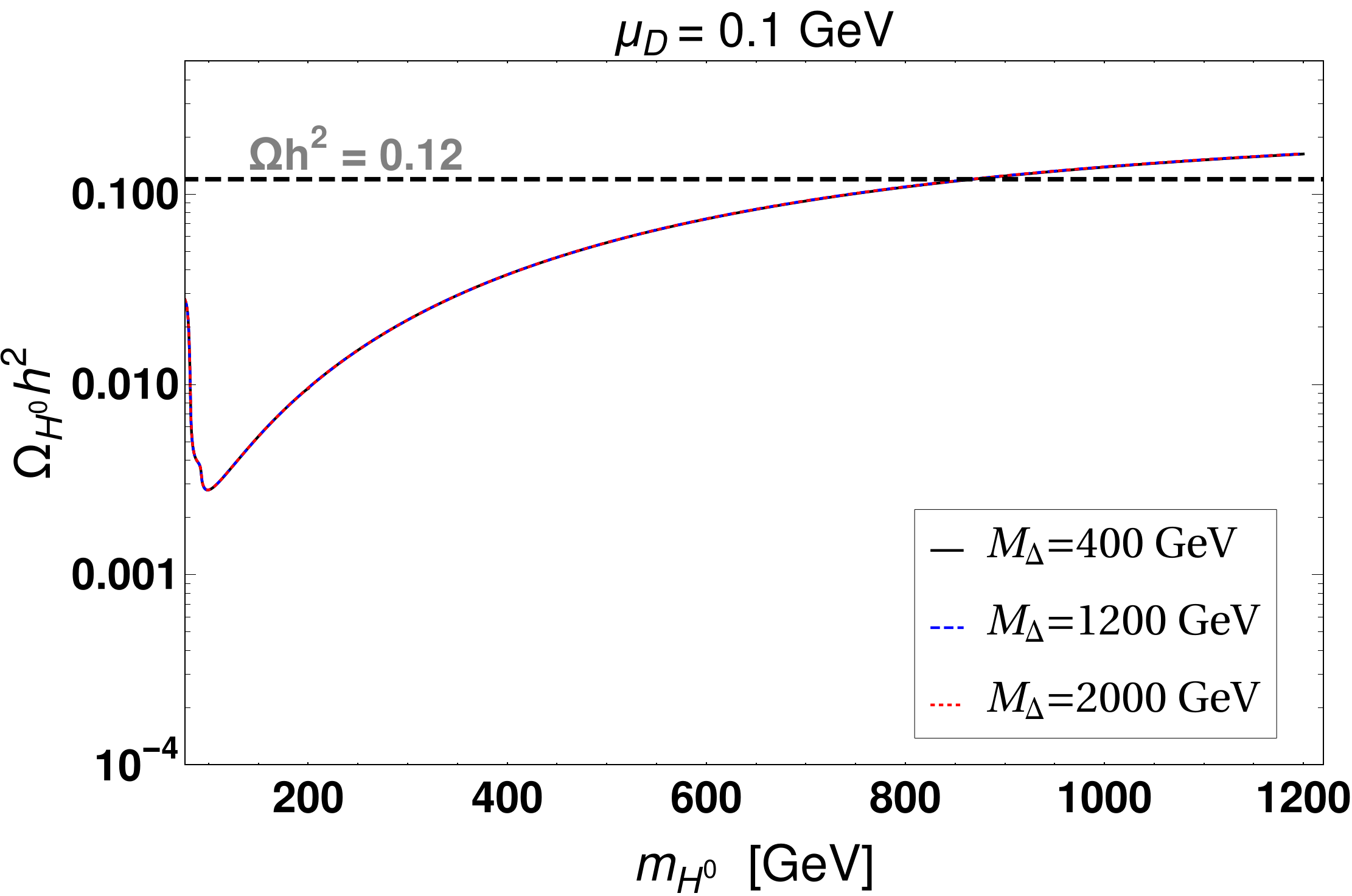}\qquad
\includegraphics[scale=0.30]{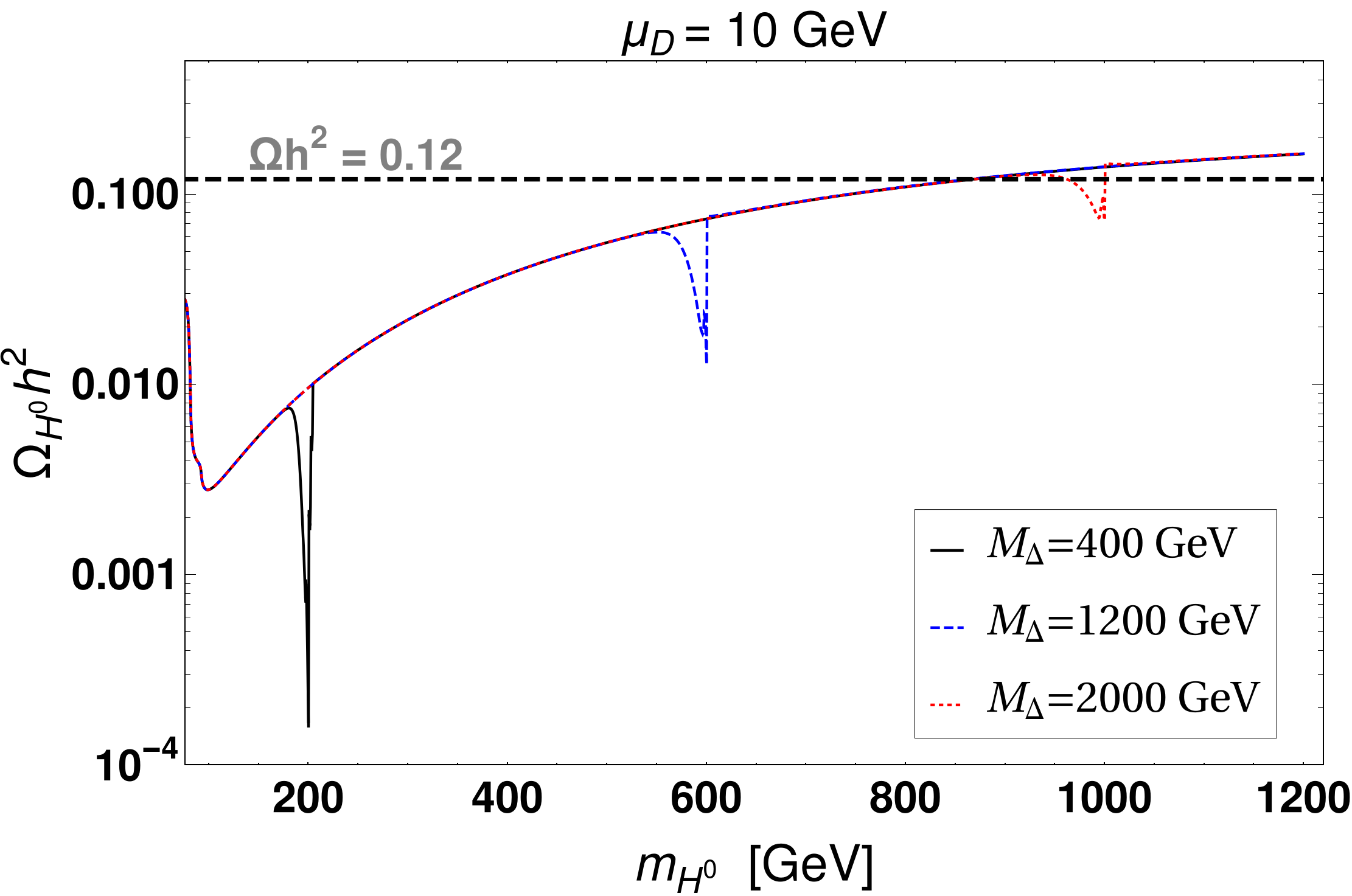}
\caption{The dependence of scalar relic density with respect of scalar DM mass while varying the dimensionful couplings $\mu_D = 0.1$ GeV (left) and $10$ GeV (right) as well as the triplet mass parameter $M_{\Delta} = 400$ GeV (black solid), 1200 GeV (blue dashed) and 2000 GeV (red dashed). Triplet sector parameters are fixed as in Fig. \ref{fig:SC_fermion}.}
\label{fig:SC_scalar}
\end{figure}

Similarly, the triplet scalar also impacts the scalar DM relic density through the dimensionful coupling $\mu_D$. In Fig.\ref{fig:SC_scalar}, we show the variation of the relic density as a function of the scalar DM mass for $\mu_D=0.1$ GeV (left) and $\mu_D=10$ GeV (right). In both cases, we varied $M_\Delta$, showing its effect using three values: 400 GeV (black solid), 1200 GeV (blue dashed), and 2000 GeV (red dashed). When $\mu_D \lesssim 0.1$ GeV, the triplet scalar has a minimal impact on the relic density of the scalar, shown in the left panel figure. However, for larger couplings, $\mu_D \sim 10$ GeV (right panel), the relic density drops significantly near $m_{H^0} \sim M_{\Delta}/2$. To minimize these extra contributions to the scalar relic density, one can choose $\mu_D \lesssim 0.1$ GeV for any value of $M_\Delta$, or larger $\mu_D ~(\gtrsim 10 {\rm GeV})$ if $M_\Delta$ is sufficiently high ($\sim$ TeV scale).

Next, to illustrate the phenomenological viability of our two-component DM framework, we adopt a representative BP for the triplet sector:
$\{M_\Delta=400 ~{\rm GeV},~\lambda_\Phi=0.129,~\lambda_1^\prime=0.2,~\lambda_2^\prime=0.0,~\lambda_3^\prime=0.95,~\lambda_4^\prime=-0.55,~v_\Delta=1.0 ~{\rm GeV}~\}$. 
We choose the value of Yukawa parameter $y_\psi$ as well as the triplet \textit{vev} $v_\Delta$ in a suitable region so that the mass splittings between the two physical pseudo-Dirac states originated from fermionic doublet DM lie around the MeV scale, which is needed to ensure that the direct detection bound is satisfied. 
In accordance with the earlier discussion, we fix $y_\psi=10^{-3}$ and $\mu_D= 1$ GeV, so that the relic density of the individual DM components does not get affected much.
%
%
\begin{figure}[htb!]
\centering
\includegraphics[scale=0.4]{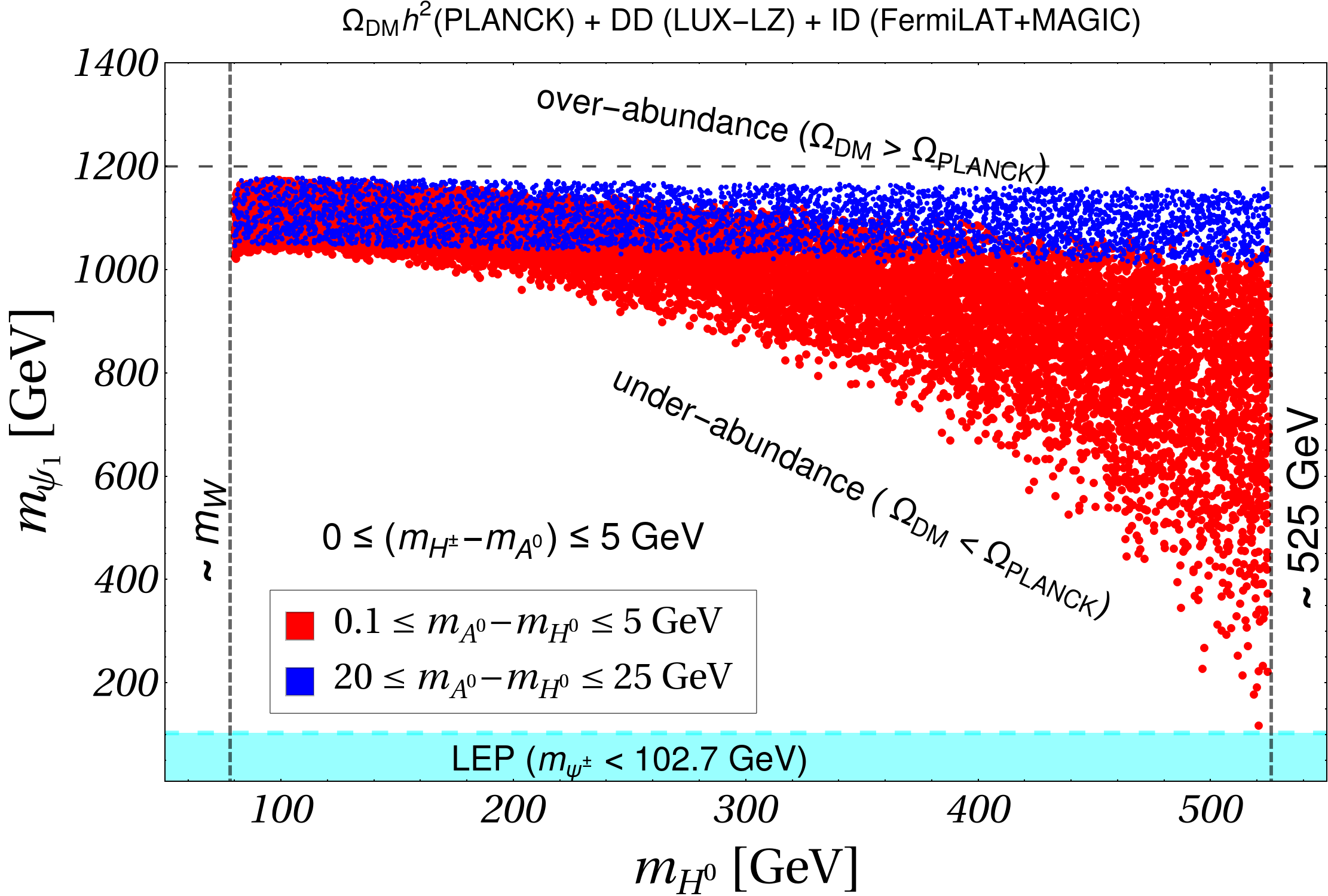}
\caption{Allowed parameter space  for both DM masses satisfying correct relic density (PLANCK) + DD (LUX LZ 2024)+ ID (FermiLAT+MAGIC) observations in the $(m_{\psi_1}, m_{H^0})$ plane. The red and blue points correspond to the mass splittings of the scalar doublet $(m_{A^0} - m_{H^0})$ in the ranges $[0.1,5]$ GeV and $[20,25]$ GeV, respectively. The other parameters of the scalar doublet fixed as: $(m_{H^\pm} - m_{A^0}) \in [0,5]$ GeV and $\lambda_L \in [10^{-3}-0.1]$. The parameters involving the triplet scalar are provided in the text. The exclusion and allowed regions shown here are identical to those in Figure \ref{fig:twocomponent_allowedregion}.}
\label{fig:two_comp}
\end{figure}

In Fig.~\ref{fig:two_comp}, we present the allowed regions that satisfy the correct relic density requirement for the combined DM relic
density as well as the experimental upper bounds of DD (LUX LZ-2024) and ID (FermiLAT + MAGIC) in the $m_{\psi_1}-m_{H^0}$ plane.
The red and blue points represent allowed parameter regions for mass splittings of the scalar doublet $(m_{A^0} - m_{H^0})$ in the ranges $[0.1,5]$ GeV and $[20,25]$ GeV, respectively, while the other mass splitting $(m_{H^\pm} - m_{A^0})$ is fixed within $[0,5]$ GeV. The other parameters are fixed as illustrated in the earlier paragraph. 
From this figure, it is evident that the mass ranges $80 < m_{H^0} < 525$ GeV and $105 \lesssim  m_{H^0} < 1200$ GeV together meet the observed dark matter relic density and satisfy all constraints.
For larger mass splitting scenario (blue points), the allowed parameter space is much more constrained. It is also apparent from Fig. \ref{fig:IDM_relic_mass} that, for larger mass splittings, the scalar contribution is even smaller, so correspondingly we always need heavier $\psi_1$ fermions to saturate the relic density bound.


%


\section{Collider signatures}
\label{sec:colliders}
 
We now proceed to explore the consequences of our scenarios in collider signals. Both DM components in the model have gauge interactions, making the framework attractive for collider detection. The single component scalar and fermion doublet DM can contribute to similar signal processes, for example: $\ell^+ \ell^- + \slashed{E_T}$ (LHC)/ $\slashed{E}$ (ILC), due to
having similar mass spectra: $\{H^0 ({\rm DM1}), A^0, H^{\pm}\}$ for the scalar DM and $\{\psi_1 ({\rm DM2}), \psi_2, \psi^{\pm}\}$ for the fermion DM. The small mass splitting between the charged component and DM for fermion makes it difficult to 
distinguish the signal from the background, as the distribution lies within the large SM background \cite{Bhattacharya:2018fus}. Therefore, the minimal framework including both fermion and scalar doublets fails to produce a distinguishable signature when compared to the single-component scenario at colliders, as discussed in Ref. \cite{Bhattacharya:2022wtr}. 
Moreover, the mass splitting between the scalar component enables significant separation, thanks to fermionic DM, which helps to separate the SM background \cite{Bhattacharya:2019fgs}. 

We concentrate here on the more promising scenario of the complete ultraviolet model, focusing in particular on how the presence of DM affects the decays of the doubly charged component of the triplet scalar boson, $\Delta^{\pm\pm}$. We start by reviewing the experimental constraints on the $\Delta^{\pm\pm}$ masses.

\begin{figure}[htb!]
\centering
\includegraphics[scale=0.40]{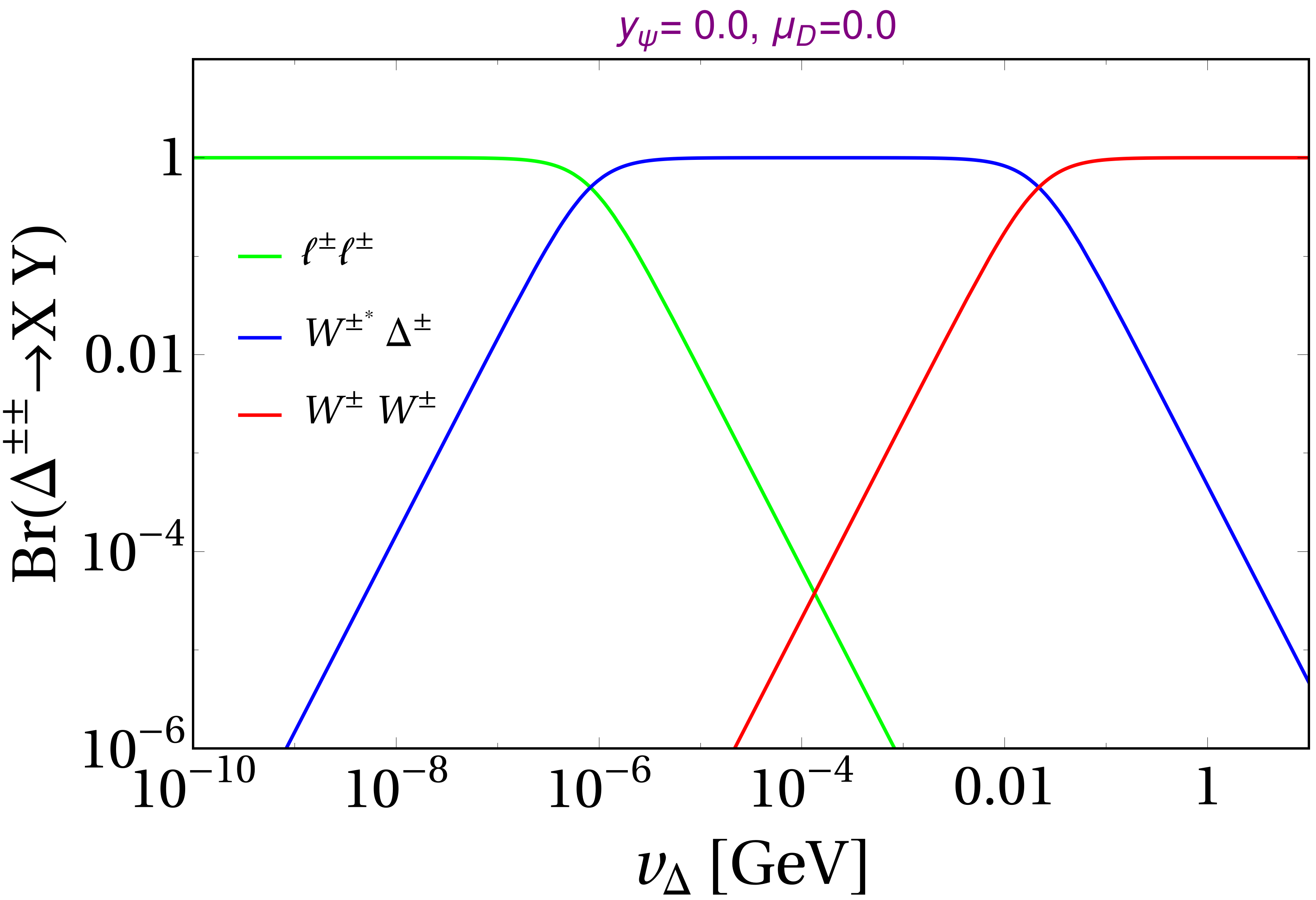} \qquad
    \includegraphics[scale=0.40]{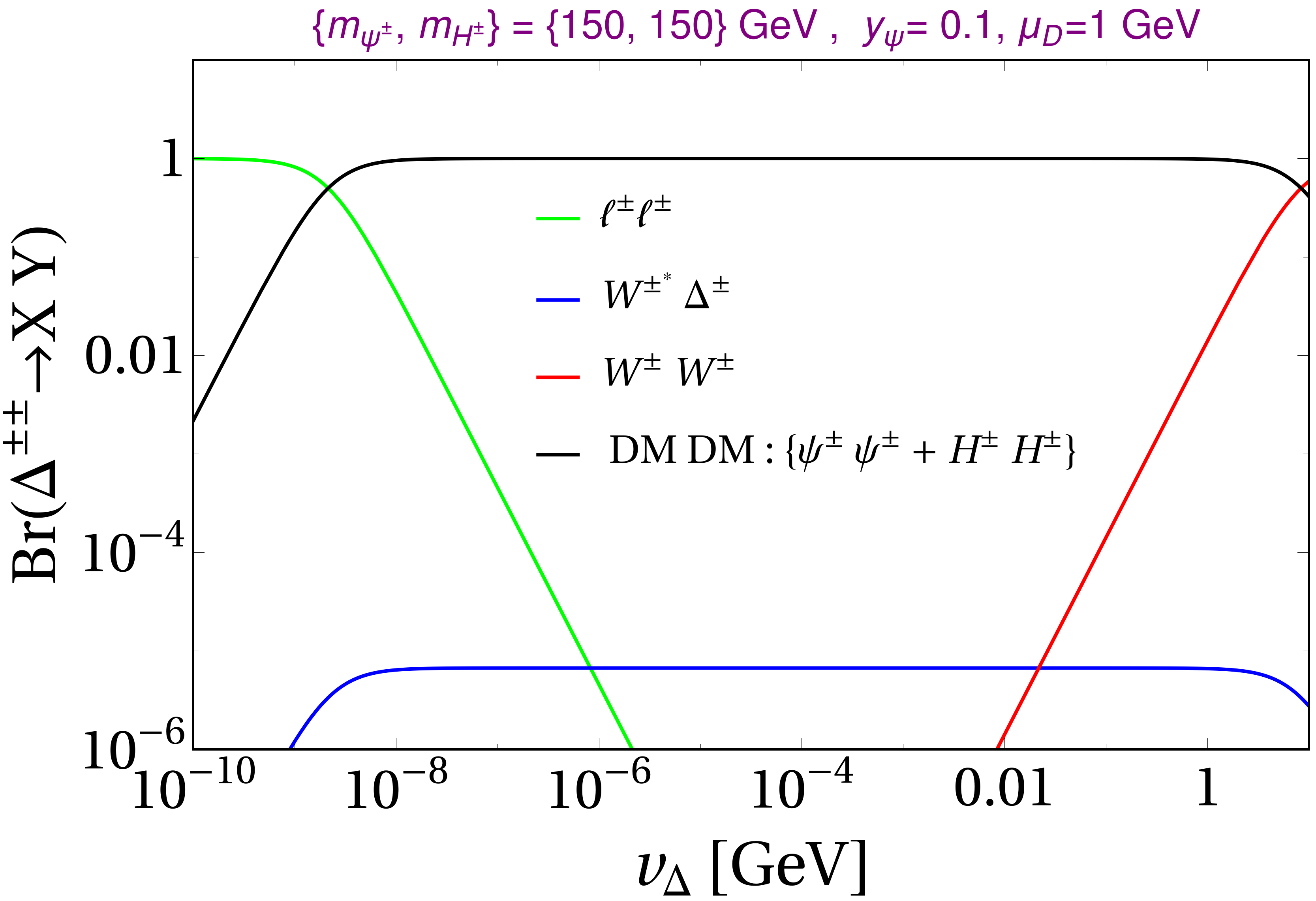}  
\caption{Left panel: Decay branching ratios of $\Delta^{\pm\pm}$ without the dark sector. Right panel: Decay branching ratios of $\Delta^{\pm\pm}$ including contributions from decays into both dark sectors. We take $m_{\Delta^{\pm\pm}}=400$ GeV and $\Delta m \equiv m_{\Delta^{\pm\pm}}-m_{\Delta^{\pm}}= 10$ GeV, as a benchmark value for these plots.}
\label{fig:CC1}
 \end{figure}

The ATLAS collaboration placed the most stringent bound on $m_{\Delta^{\pm\pm}} \gtrsim 870$ GeV, assuming Br($\Delta^{\pm\pm} \rightarrow \mu^{\pm} \mu^{\pm}$) to be 100$\%$~\cite{ATLAS:2017xqs}, and using the data with $36.1$ fb$^{-1}$ and $\sqrt{s}=13$ TeV for $\Delta m = m_{\Delta^{\pm\pm}} - m_{\Delta^\pm}=0$. Furthermore, ATLAS again excludes $m_{\Delta^{\pm\pm}} \gtrsim 350$ GeV in the $W^{\pm} {W^\pm}$ final state with 139 fb$^{-1}$ of the data \cite{ATLAS:2021jol}. For the cascade mode ($\Delta^{\pm \pm} \to \Delta^{\pm}  ~{W^{\pm}}^*$), the lower bound on the doubly charged scalar is loosely constrained from the collider data. As discussed in \cite{Ashanujjaman:2021txz},  for $\Delta m =$ 10 GeV (30 GeV), with $v_\Delta$ around $10^{-3}$ GeV to $10^{-5}$ GeV ($10^{-3}$ GeV to $10^{-6}$ GeV), the high luminosity LHC (3 ab$^{-1}$) will be unable to constrain $m_{\Delta^{\pm\pm}} \gtrsim$ 200 GeV.    
However, in the presence of charged DM states from both DM sectors with mass $m_{\psi^\pm}, m_{H^\pm} < m_{\Delta^{\pm\pm}}/2$, the additional decay mode, $\Delta^{\pm\pm} \to \psi^\pm \psi^\pm ,~ H^\pm H^\pm$, may relax the constraint on $m_{\Delta^{\pm\pm}}$. The decay widths for both processes are given by:
\begin{align}
    & \Gamma_{\Delta^{\pm\pm} \to \psi^\pm ~\psi^\pm} = \frac{y_{\psi}^2~ m_{\Delta^{\pm\pm}}}{8\pi} \Big(1- \frac{2 ~m_{\psi^\pm}^2}{m_{\Delta^{\pm\pm}}^2} \Big) \sqrt{1- \frac{4 ~m_{\psi^\pm}^2}{m_{\Delta^{\pm\pm}}^2} } ~\Theta[m_{\Delta^{\pm\pm}}-2m_{\psi^\pm}] , \nonumber \\
    & \Gamma_{\Delta^{\pm\pm} \to H^\pm ~H^\pm} = \frac{\mu_{D}^2}{8 \pi m_{\Delta^{\pm\pm}}}  \sqrt{1- \frac{4 ~m_{H^\pm}^2}{m_{\Delta^{\pm\pm}}^2} } ~\Theta[m_{\Delta^{\pm\pm}}-2m_{H^\pm}].
\end{align}

\begin{figure}[htb!]
\centering
\includegraphics[scale=0.40]{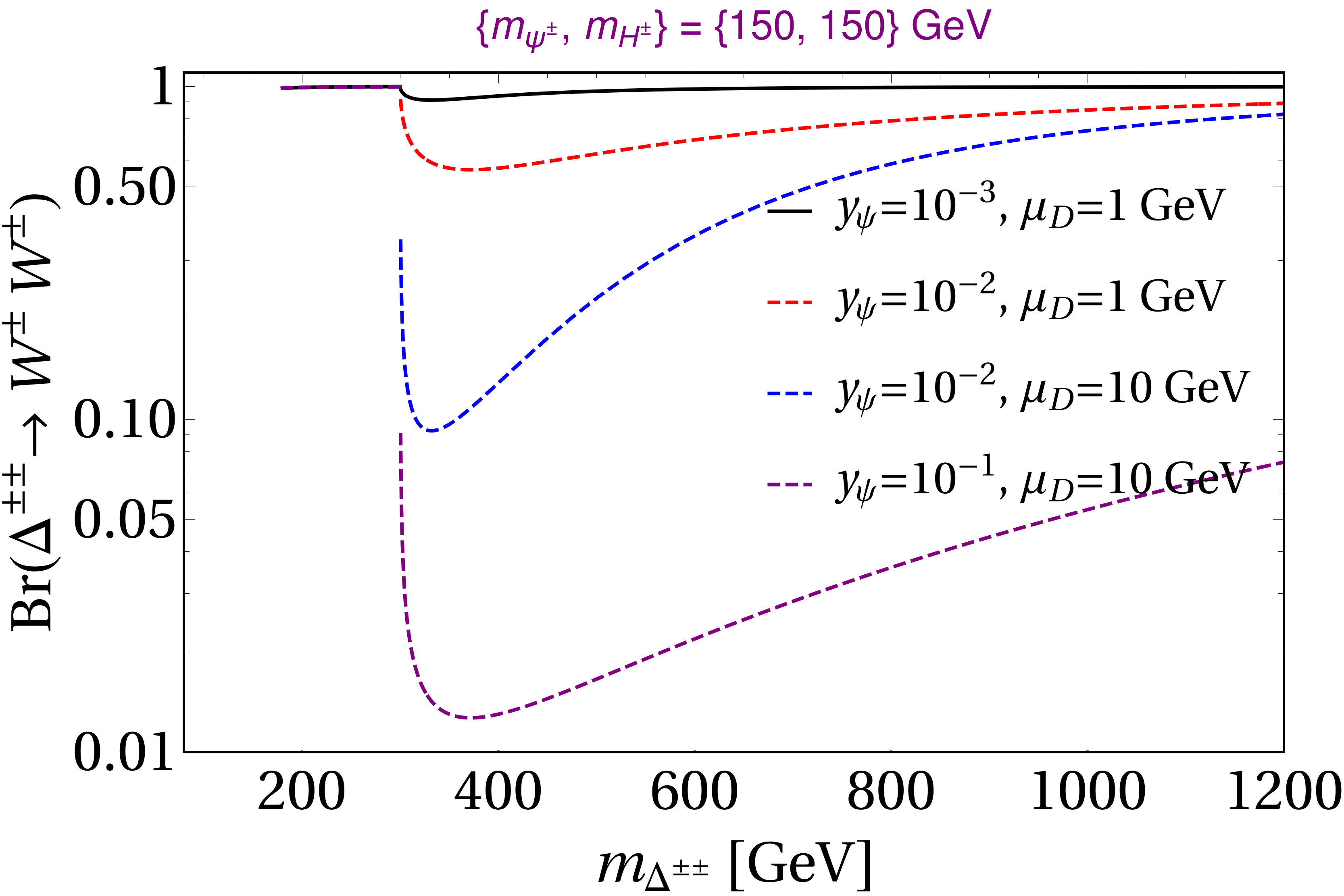}\qquad
\includegraphics[scale=0.40]{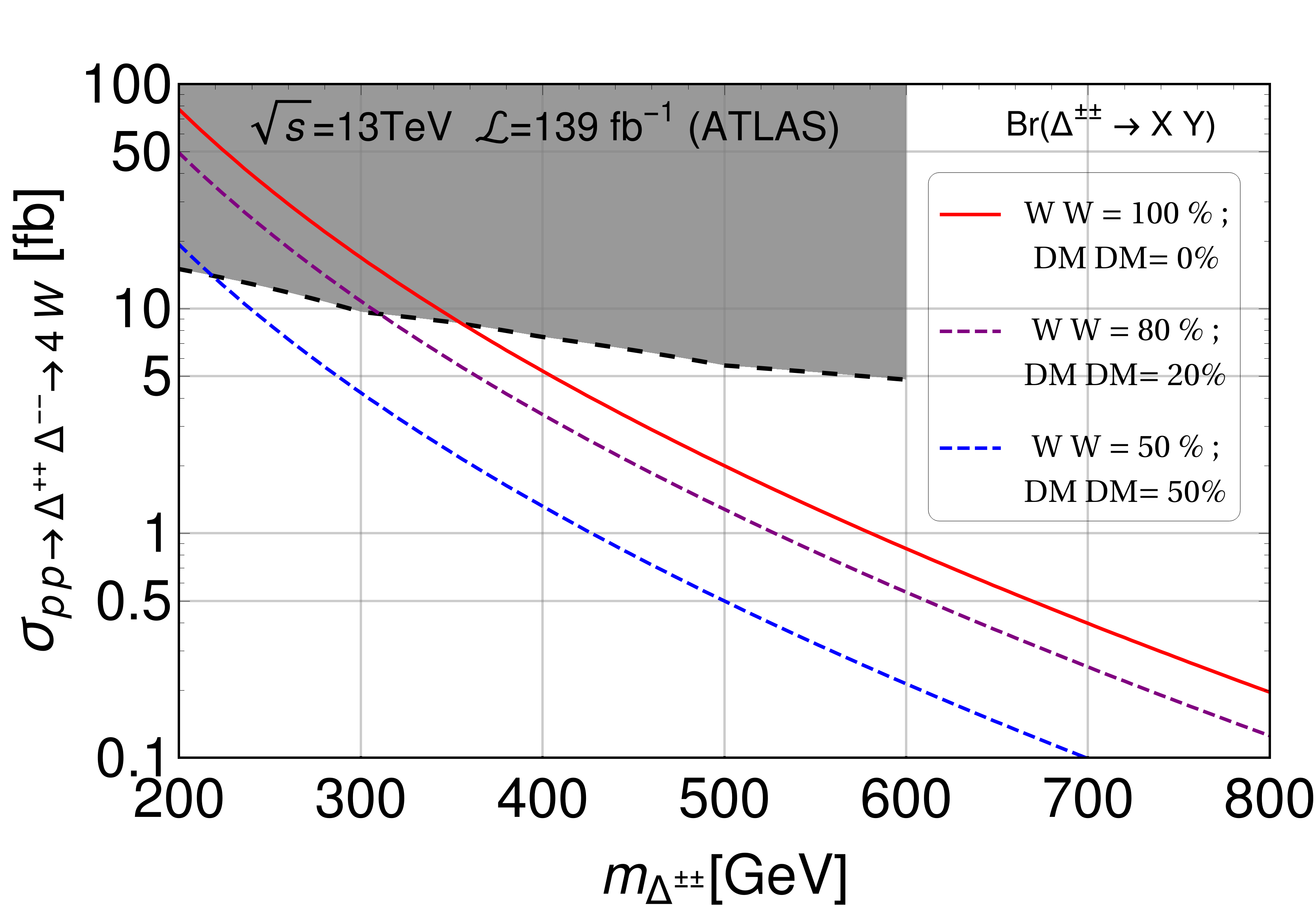} 
\caption{Left panel: Branching ratios of the doubly charged triplet scalar $\Delta^{\pm\pm}$ to $W~W$ as a function of $m_{\Delta^{\pm\pm}}$ with presence of two component DM. For demonstration, we take $\Delta m = 10$ GeV and $v_\Delta = 1$ GeV, while the other parameters are mentioned in the inset of the figure. Right panel: Exclusion bound on the doubly charged scalar triplet mass from $W^+ W^+ W^- W^-$ final states searches at colliders using ATLAS data with $36.1$ fb$^{-1}$ and $\sqrt{s}=13$ TeV in the plane of $m_{\Delta^{\pm\pm}} {~\rm vs~} \sigma_{pp\to \Delta^{\pm\pm} \Delta^{\mp\mp} \to 4W}$. The red line represents the scenario where ${\rm Br}(\Delta^{\pm\pm} \to W^{\pm} W^{\pm}) \sim 100\%$, indicating that masses below $\sim 350$ GeV are excluded by the data. The predictions for our model, including the decay $\Delta^{\pm\pm} \to {\rm DM}~{\rm DM}$ with non zero branching ratios, are shown by the purple and blue dashed lines, indicating a possible relaxation of the lower mass bound.}
\label{fig:CC2}
 \end{figure}

For our discussion, we considered the triplet \textit{vev}, $v_\Delta$, with $v_\Delta \gtrsim 0.01$ MeV, where the decay of $\Delta^{\pm\pm}$ 
to $W~W$ dominates ($\sim 100 \%$) in the absence of the dark sector (as shown in the left panel of Fig. \ref{fig:CC1}). However, the presence of charged fermion ($\psi^\pm$) and scalar ($H^\pm$) states in DM sector modifies the branching ratios, as shown in the right panel of Fig. \ref{fig:CC1}. As a consequence, the collider exclusion bound on doubly-charged particles could be relaxed, as shown in the right panel of Fig. \ref{fig:CC2}. This will constitute a unique feature of our model, signifying a shift in branching ratios for the doubly charged  triplet scalar due to decay into charged particles in the dark sector.
 
\section{Conclusion}
\label{sec:conclusion}

In this work we investigated a two-component DM augmented SM that can satisfy DM constraints from relic density, direct and indirect detection experiments over a large region of DM parameter space. 

We first analyzed a single component DM sector involving a fermion doublet, stabilized by the ${\mathcal Z}_2$ symmetry. This scenario is very simple, as it depends only on one parameter, the fermion mass. For this case,   DM densities consistent with the PLANCK measurements are obtained for a DM mass of 1.2 TeV only, and below this mass the fermion DM is under-abundant. The entire mass region is safe from indirect bounds but unfortunately is nowhere consistent with the spin-independent direct detection bounds, due to the fermion interaction with the Z boson. One can evade this constraint if the DM turns out to be a pseudo-Dirac state, in which case the Z mediated neutral current vanishes. We highlighted one possibility, that is, introducing a dim-5 effective operator, which splits the neutral fermion state in the doublet, yielding two pseudo-Dirac states, the lightest of which is the DM candidate. Same-state interactions with the Z boson are forbidden due to the Majorana nature of these fermions. While for small mass splittings Z-mediated interactions still present a problem, for higher mass splittings between these pseudo-Dirac states (${\cal O}(100$ keV)), only the elastic scattering contributes to direct detection and correspondingly the cross section lies below the direct detection upper bound, while the fermion masses are still required to be in the TeV region to saturate the correct relic density.

The single component scalar doublet fares better, in large part due to the additional parameters (various couplings) in the model. Here, depending on parameters, the CP even or odd neutral component can serve a DM candidate. Due to large annihilation and co-annihilation of the scalar DM particle into SM particles through gauge interactions,  the relic density is under-abundant in the region 80 GeV $ \lesssim m_{H^0} \lesssim$ 525 GeV for small mass splitting between the charged-scalar and pseudo-scalar components, while, if this mass splitting is increased, we need the DM scalar mass to be in the TeV region to be consistent with the relic density. The direct detection bound can be satisfied for both small and large mass regions, and indirect detection constraints are satisfied over a large region of the parameter space. 

These issues associated with the single components motivate the analysis of the two-component scenario (combining the scalar and fermion doublets, each stabilized by two ${\mathcal Z}_2$ symmetries). As this scenario inherits problems with direct detection emerging from the fermion sector, we investigate two solutions: introducing dim-5 effective operators, or constructing an UV-complete model which includes a triplet scalar with hypercharge Y = 2. 

In the first scenario, where effective operators are used, the relic density is now a sum of both fermionic and scalar relic densities, yielding a significant region of parameter space for the pseudo-Dirac fermion and scalar consistent with relic density measurements and bounds on direct and indirect detection cross sections. The precise region depends on whether the scalar mass splitting between CP-even and odd states is smaller or larger. For smaller mass splitting and small scalar masses, the larger co-annihilation and annihilation cross-section result in the scalar contribution to be quite suppressed, and this is why we needed larger fermion mass values. However, if we increase the scalar mass, the contribution to relic density from scalar sector increases and eventually lighter fermionic DM can suffice. For larger mass splittings, the scalar contribution is even smaller, so correspondingly we always need heavier fermions in this scenario to saturate the relic density value.

Finally, we construct an UV-complete theory including the triplet scalar boson added to the SM particle content, whose mass parameter and coupling to the DM fermionic doublet affects the relic density via the fermion sector. Choosing the value of Yukawa parameter as well as the triplet \textit{vev} in a suitable region, so that the mass splittings between the two physical pseudo-Dirac states in the fermionic doublet DM are of ${\cal O}$(MeV), ensures that the direct detection bound is satisfied. With the triplet \textit{vev} set to be in the range of few ${\cal O}$ (MeV - 1 GeV), a reasonably broad mass parameter space survives all DM constraints. Moreover, with the considered range of triplet \textit{vev} and a highly fine-tuned Yukawa coupling, the neutrino masses can be generated within the correct ballpark region. We give two concrete benchmark points as examples, and show that the DM observables are most sensitive to the product of the triplet Yukawa coupling of the fermion DM doublet with the triplet \textit{vev}. Note that this scenario connects the existence of the two-component DM to a model which generates neutrino masses. While our analysis requires a triplet scalar mass in the TeV range, it allows for light DM scalar and fermion candidates. 

The presence of both DM fermionic and scalar particles with sub-TeV masses,  which could be produced at the LHC,  would provide a signal in support of this scenario.  Additionally, we looked at the possibility that the doubly charged triplet scalar can decay into the charged scalar or fermion components of the DM doublets, shifting the branching ratios (from almost 100\% into $W^\pm W^\pm$) and lowering the mass limits for this boson at the LHC, thus providing a distinguishing signature for this model.

\vskip 0.5cm
{\large \bf Acknowledgment} \\[3mm]
The work of MF work is funded in part by NSERC under grant number SAP105354. CM acknowledges the Royal Society, UK for support through the Newton International Fellowship with grant number NIF$\backslash$R1$\backslash$221737. SS is partially funded for this work under the US Department of Energy contract DE-SC0011095.



\bibliographystyle{utcaps_mod}
\bibliography{MultiDM}
\end{document}